\documentclass[fleqn,usenatbib]{mnras}

\usepackage{newtxtext,newtxmath}
\usepackage[T1]{fontenc}

\DeclareRobustCommand{\VAN}[3]{#2}
\let\VANthebibliography\thebibliography
\def\thebibliography{\DeclareRobustCommand{\VAN}[3]{##3}\VANthebibliography}

\usepackage{graphicx}	% Including figure files
\usepackage{amsmath}	% Advanced maths commands
\usepackage{rotating}	% sideways figures
\usepackage{caption}
\usepackage{subcaption}
\usepackage{bm}
\usepackage{listings}

\usepackage[british]{babel}

\newcommand{\sTheta}{\bm{\upTheta}} %set Theta
\newcommand{\Lya}{Ly$\alpha$}{}    %set Lyman alpha
\newcommand{\Lyb}{Ly$\beta$}     %set Lyman beta
\usepackage[nameinlink]{cleveref}

\title[Joint constraints on EoR astrophysics]{Rapid and Late Cosmic Reionization Driven by Massive Galaxies: a Joint Analysis of Constraints from 21-cm, Lyman Line \& CMB Data Sets}

\author[P. H. Sims et al.]{Peter H. Sims,$^{1}$\thanks{E-mail: psims3@asu.edu}
Harry T. J. Bevins,$^{2, 3}$
Anastasia Fialkov,$^{3,4}$
Dominic Anstey,$^{2}$
Will J. Handley,$^{2,3}$ \and
Stefan  Heimersheim,$^{4}$
Eloy de Lera Acedo,$^{2,3}$
Rajesh  Mondal,$^{5}$
Rennan Barkana$^{6,7}$
\\
$^{1}$School of Earth and Space Exploration, Arizona State University, Tempe, AZ 85287, USA\\
$^{2}$Cavendish Astrophysics, University of Cambridge, Cambridge, UK\\
$^{3}$Kavli Institute for Cosmology in Cambridge, University of Cambridge, Cambridge, UK\\
$^{4}$Institute of Astronomy, University of Cambridge, Cambridge, UK\\
$^{5}$National Institute of Technology Calicut, Calicut 673601, Kerala, India\\
$^{6}$School of Physics and Astronomy, Tel-Aviv University, Tel-Aviv 69978, Israel \\
$^{7}$School of Natural Sciences, Institute for Advanced Study, 1 Einstein Drive, Princeton, NJ 08540, USA\\
}

\date{Accepted XXX. Received YYY; in original form ZZZ}

\pubyear{2015}

\begin{document}
\label{firstpage}
\pagerange{\pageref{firstpage}--\pageref{lastpage}}
\maketitle

\begin{abstract}
Observations of the Epoch of Reionization (EoR) have the potential to answer long-standing questions of astrophysical interest regarding the nature of the first luminous sources and their effects on the intergalactic medium (IGM). We present astrophysical constraints from a Neural Density Estimation-Accelerated Bayesian joint analysis of constraints deriving from Cosmic Microwave Background power spectrum measurements from Planck and SPT, IGM neutral fraction measurements from Lyman-line-based data sets and 21-cm power spectrum upper limits from HERA, LOFAR and the MWA. In the context of the model employed, the data is found to be consistent with galaxies forming from predominantly atomic-cooled hydrogen gas in dark matter halos, with masses $M_\mathrm{min} \gtrsim 2.6 \times 10^{9}~M_{\odot} ((1+z)/10)^{\frac{1}{2}}$ at 95\% credibility ($V_\mathrm{c} \gtrsim 50~\mathrm{km~s^{-1}}$) being the dominant galactic population driving reionization. These galaxies reionize the neutral hydrogen in the IGM over a narrow redshift interval ($\Delta z_\mathrm{re} < 1.8$ at 95\% credibility), with the midpoint of reionization (when the sky-averaged IGM neutral fraction is 50\%) constrained to $z_{50} = 7.16^{+0.15}_{-0.12}$. Given the parameter posteriors from our joint analysis, we find that the posterior predictive distribution of the global 21-cm signal is reduced in amplitude and shifted to lower redshifts relative to the model prior. We caution, however, that our inferences are model-dependent. Future work incorporating updated, mass-dependent star formation efficiencies in atomic cooling halos, informed by the latest UV luminosity function constraints from the James Webb Space Telescope, promises to refine these inferences further and enhance our understanding of cosmic reionization.
\end{abstract}

\begin{keywords}
% keyword1 -- keyword2 -- keyword3
dark ages, reionization, first stars - cosmology: observations - cosmology: theory - methods: data analysis - methods: statistical
\end{keywords}

%%%%%%%%%%%%%%%%%%%%%%%%%%%%%%%%%%%%%%%%%%%%%%%%%%

%%%%%%%%%%%%%%%%% BODY OF PAPER %%%%%%%%%%%%%%%%%%

%%%%%%%%%%%%%%%%%%%%%%%%%%%%%%%%%%%%%%%%%%%%%%%%%%
\section{Introduction}
\label{Sec:Introduction}
%%%%%%%%%%%%%%%%%%%%%%%%%%%%%%%%%%%%%%%%%%%%%%%%%%

A variety of cosmological probes have provided an outline of how today's astronomical objects came to be. Measurements of the Cosmic Microwave Background (CMB) provide a relatively pristine view of density anisotropies $\sim0.38~\mathrm{Myr}$ after the Big Bang, when the temperature cooled sufficiently for protons and electrons to combine and form neutral hydrogen. At the same time, ground  and space-based telescopes have made detailed measurements of galaxies at epochs following hydrogen reionization (between $1~\mathrm{Gyr}$ after the Big Bang and the present day). However, there remain multiple, as yet unobserved, astrophysical milestones essential for a complete understanding of how the first galaxies evolved and altered the intergalactic medium (IGM) in the intervening $\sim1~\mathrm{Gyr}$ between these two data-rich cosmic periods.

Theoretical considerations paint a rough picture of this time. The first detectable signal is expected from neutral hydrogen during the cosmic Dark Ages (DA; e.g. \citealt{2023NatAs...7.1025M, 2024MNRAS.527.1461M}) when the spin and kinetic temperatures of the gas first decouple from the background radiation field. Subsequently, gravitational collapse in overdense regions of the gas leads to the formation of the first luminous objects  at Cosmic Dawn (CD) and during the Epoch of Reionization (EoR) these objects transformed the IGM from a cold and predominantly neutral gas to a hot, ionized plasma (e.g. \citealt{2006PhR...433..181F, 2012RPPh...75h6901P}). However, the detailed timing of these milestones is only moderately constrained by current analyses, and long-standing astrophysical questions regarding the nature of the sources driving them remain (e.g. \citealt{2019ApJ...879...36F, 2020ApJ...892..109N, 2023MNRAS.520.2757Y, 2024MNRAS.527.6139S}).

Data from a panoply of experiments spanning the electromagnetic spectrum, from radio to ultraviolet (UV) wavelengths, can be used to place constraints on astrophysical parameters of models of the Universe during these early eras (e.g. \citealt{2013MNRAS.432.3340S, 2015MNRAS.449.4246G, 2021ApJ...908..199R, 2022NatAs...6.1473B}). New observations with the James Webb Space Telescope (JWST) are extending the period for which we have detailed measurements of the brightest galaxies further into the first $\mathrm{Gyr}$ after the Big Bang (e.g. \citealt{2022arXiv220802794N, 2022ApJ...940L..55F, 2023ApJ...946L..13F, 2024Natur.633..318C}). Measurements of the CMB power spectrum constrain the midpoint and duration of the EoR ($z_\mathrm{re} = 7.67 \pm 0.73$, assuming an instantaneous reionization model; \citealt{2020A&A...641A...6P} and $\Delta z_\mathrm{re} < 4.1$ at 95\% credibility; \citealt{2021ApJ...908..199R}). Lyman line measurements in high-redshift galaxies and quasars constrain the midpoint and end of the EoR to $z \sim 7$ and $\sim 5.5$, respectively (e.g. \citealt{2022MNRAS.514...55B, 2022MNRAS.517.3263B}). 21-cm power spectrum upper limits constrain the differential brightness between the cosmic radio background temperature and the neutral hydrogen spin temperature at a range of redshifts during CD and the EoR, implying that the IGM must have been heated above the adiabatic cooling limit by $z \sim 8$ and, thus, ruling out "cold reionization" scenarios (e.g. \citealt{2022ApJ...924...51A}).

Where there is statistical consistency between inferences from data sets, a joint analysis allows one to combine information, obtaining more stringent constraints than would be derived through analysis of the data sets individually. Several analyses of this type, for the purpose of constraining EoR and CD astrophysics, have been conducted (e.g. \citealt{2018ApJ...863...11M, 2019ApJ...875...67M, 2019ApJ...879...36F, 2020MNRAS.498.4178M, 2020ApJ...892..109N, 2021MNRAS.500.5322G, 2021MNRAS.507.2405C, 2022ApJ...924...51A, 2024MNRAS.531.1113P, 2024MNRAS.529..519G}). Recently, 21-cm upper limits from the Hydrogen Epoch of Reionization Array (HERA; \citealt{2017PASP..129d5001D}), the LOw Frequency ARray (LOFAR; \citealt{2013A&A...556A...2V}), the Murchison Widefield Array (MWA; \citealt{2018PASA...35...33W}), and the Shaped Antenna measurement of the background RAdio Spectrum (SARAS 3; \citealt{2021arXiv210401756N}) were jointly analysed to place constraints on the astrophysics of galaxies $200~\mathrm{Myr}$ after the Big Bang (\citealt{2024MNRAS.527..813B}; hereafter, B24). In this paper, we build on the constraints from the HERA, LOFAR, and MWA 21-cm data sets derived in B24, combining them with complementary constraints from:
\begin{itemize}
    \item CMB power spectrum data sets, including:
    \begin{itemize}
        \item Planck TT,TE,EE+low$l$+lowE+lensing constraints on the total CMB optical depth, $\tau_\mathrm{CMB}$ (\citealt{2020A&A...641A...6P}; hereafter P20VI);
        \item South Pole Telescope (SPT) patchy kinetic Sunyaev--Zel'dovich (kSZ; \citealt{2021ApJ...908..199R}) constraints on the duration of reionization.
    \end{itemize}
    \item Lyman line data sets, consisting of constraints on the IGM neutral fraction during reionization, $\overline{x}_\mathrm{H_{I}}(z)$, deriving from measurements of:
    \begin{itemize}
        \item the Lyman alpha (\Lya) and Lyman beta (\Lyb) forest dark pixel fraction in high-redshift quasar (\citealt{2015MNRAS.447..499M}),
        \item the clustering of \Lya\ galaxies (\citealt{2015MNRAS.453.1843S}),
        \item bright quasar damping wings (\citealt{2020ApJ...896...23W}), and
        \item \Lya\ equivalent width evolution (\citealt{2018ApJ...856....2M, 2019MNRAS.485.3947M, 2022MNRAS.517.3263B}).
    \end{itemize}
\end{itemize}

Existing analyses which constrain the mass of galaxies driving reionization -- whether faint low-mass galaxies, bright massive galaxies, or a combination of both -- have reached differing conclusions. The assumptions made regarding the fraction of ionizing photons emitted by galaxies that successfully traverses the circumgalactic medium and escapes into the IGM, $f_\mathrm{esc}$, is an important source of these differences. For example, using a model that assumes that bright galaxies do not significantly contribute to the ionizing emissivity (due to having a low $f_\mathrm{esc}$), \citet[hereafter F19]{2019ApJ...879...36F} find that reionization dominated by copious faint low-mass sources completes by $z \sim 6$. However, their model's prediction of an IGM that is approximately $\sim20\%$ neutral at $z\sim7$ is in tension with constraints from \Lya\ data which indicates a neutral fraction of approximately $\sim50\%$ at this redshift (e.g. \citealt{2022MNRAS.517.3263B}). In contrast, fitting a model that assumes that the ionizing photon escape fraction is proportional to the star formation rate surface density, \citet[hereafter N20]{2020ApJ...892..109N} infer that high stellar mass galaxies ($M_{*} \gtrsim 10^{8}~M_{\odot}$) dominate the reionization budget. However, this, in turn, has been found to be in tension with measurements of the ionizing photon mean free path between redshift 5 and 6 (e.g. \citealt{2021ApJ...917L..37C}).

In practice, $f_\mathrm{esc}$ is expected to have large sightline-to-sightline variability and is poorly constrained by observations in the galaxy-mass and redshift range of interest (e.g. \citealt{2023MNRAS.520.2757Y}). Thus, in this work, rather than choosing a specific parametric model for $f_\mathrm{esc}$, we marginalise out its dependence on our conclusions and instead sample directly from the CMB optical depth of the model $\tau_\mathrm{CMB}(f_\mathrm{esc})$. We note, however, that the model used here employs a mass-independent prescription for the star formation efficiency (SFE) in atomic cooling halos. Work conducted during the preparation of this manuscript suggests that models with a mass-dependent SFE combined with a mass- and redshift-dependent escape fraction provide a better fit to JWST UV-luminosity function data (\citealt{2025arXiv250321687D}) and are preferred in fits of \Lya\ opacity constraints from \Lya\ forest data (\citealt{2024arXiv241200799Q}). Future work will extend the analysis presented here by incorporating such models and integrating the JWST and \Lya\ forest data sets into the broader joint analysis employed in this study.

The remainder of this paper is organised as follows. In \Cref{Sec:DataSetsAndMeasurements}, we describe the data sets and observational constraints that inform our posterior parameter estimation. \Cref{Sec:AnalysisMethodology} details the Bayesian joint analysis and information-theoretic frameworks used to derive astrophysical inferences, along with the data model, likelihoods, and priors. In \Cref{Sec:Results}, we present the EoR information content, describing the degree to which each data set contributes information (in nats) to our understanding of EoR astrophysics, as well as the derived astrophysical constraints. Finally, \Cref{Sec:Conclusions} summarises our findings and explores directions for future work.

%%%%%%%%%%%%%%%%%%%%%%%%%%%%%%%%%%%%%%%%%%%%%%%%%%
\section{Data sets and measurements summary}
\label{Sec:DataSetsAndMeasurements}
%%%%%%%%%%%%%%%%%%%%%%%%%%%%%%%%%%%%%%%%%%%%%%%%%%

\Cref{Tab:DataSets} lists the data sets included in our joint analysis according to the observable category that the data falls into (21-cm power spectrum upper limits, Lyman-line-based constraints or CMB-power-spectrum-based constraints), the observables of our model that are constrained by the data ($\Delta^{2}_{21}(k,\bm{z})$, $\overline{x}_\mathrm{H_{I}}$, $\tau_\mathrm{CMB}$, or $\Delta z_\mathrm{re}$), the constraints on those observables associated with the data sets and our likelihood models for the constraints. A brief description of the data sets from which these constraints derive is given in the following subsections.

\begin{table*}
    \caption{
        Summary of the data sets included in our joint analysis. We list, as a function of observable category, the observable constrained by the data and fit with our model, the reference publication, the constraints associated with the data sets and the probability model (likelihood or prior on derived parameters) used to encode the constraint, where these include neural density estimators (NDE), contaminated upper limit (CUL) likelihoods and spline fits of the PDF of a modelled parameter (see \Cref{Sec:DataLikelihoods}). We quote $1 \sigma$ upper-limits and uncertainties, where applicable. The parameter constraints associated with the data sets are illustrated in \Cref{Fig:Priors}.
        }
    \centerline{
    \begin{tabular}{l l l l l }
    \hline
    Observable category & Constrained observable & References & Constraint & Probability model     \\
    \hline
    21-cm power spectrum upper limits & $\Delta^{2}_{21}(k,\bm{z})$ & B24 & See \Cref{Fig:B24Posteriors2} & NDE \\
    & & \citet{2022ApJ...924...51A} & for astrophysical & \\
    & & \citet{2020MNRAS.493.1662M} & parameter constraints & \\
    & & \citet{2020MNRAS.493.4711T} & inferred in B24 & \\
    \hline
    Lyman line constraints & $\overline{x}_\mathrm{H_{I}}$ & \citet{2015MNRAS.447..499M} & $\overline{x}_\mathrm{H_{I}}(z=5.58) < 0.04+0.05(1\sigma)$ & CUL \\
    & & & $\overline{x}_\mathrm{H_{I}}(z=5.87) < 0.06+0.05(1\sigma)$ & CUL \\
    & & & $\overline{x}_\mathrm{H_{I}}(z=6.07) < 0.38+0.20(1\sigma)$ & CUL \\
    & & \citet{2015MNRAS.453.1843S} & $\overline{x}_\mathrm{H_{I}}(z=7.0) < 0.3+0.20(1\sigma)$ & CUL \\
    & & \citet{2018ApJ...856....2M} & $\mathcal{P}(\overline{x}_\mathrm{H_{I}} | z=6.9 \pm 0.5)$ & Spline model of PDF \\
    & & \citet{2019MNRAS.485.3947M} & $\mathcal{P}(\overline{x}_\mathrm{H_{I}} | z=7.9 \pm 0.6)$ & Spline model of PDF \\
    & & \citet{2020ApJ...896...23W} & $\mathcal{P}(\overline{x}_\mathrm{H_{I}} | z=7.0)$ & Spline model of PDF \\
    & & \citet{2022MNRAS.517.3263B} & $\mathcal{P}(\overline{x}_\mathrm{H_{I}} | z=6.7 \pm 0.2)$ & Spline model of PDF \\
    & & & $\mathcal{P}(\overline{x}_\mathrm{H_{I}} | z=7.6 \pm 0.6)$ & Spline model of PDF \\
    \hline
    CMB power spectrum & $\tau_\mathrm{CMB}$ & P20VI & $\tau_\mathrm{CMB} \sim \mathcal{N}(0.054, 0.007)$ & NDE \\
    & $\Delta z_\mathrm{re}$ & R21 & $\mathcal{P}(\Delta z_\mathrm{re} | \mathcal{D}_{l, \mathrm{SPT}})$ & Spline model of PDF \\
    \hline
    \end{tabular}
    }
\label{Tab:DataSets}
\end{table*}

\subsection{21-cm power spectrum upper limits}
\label{Sec:21cmPspecULs}

In B24, five astrophysical parameters of the model used here were constrained using a joint fit to upper limits on the 21-cm power spectrum with HERA at $z=7.9$ and $10.4$ (\citealt{2022ApJ...924...51A}; hereafter, H22), LOFAR at $z=9.1$ (\citealt{2020MNRAS.493.1662M}) and the MWA in the redshift range $6.5 \le z \le 8.7$  (\citealt{2020MNRAS.493.4711T}). The 1D and 2D marginal posterior distributions of the astrophysical parameters derived from that fit are illustrated in \Cref{Fig:B24Posteriors2}. The constrained parameters in that analysis were the minimum circular velocity of star-forming halos ($V_\mathrm{c}$), the star formation, X-ray and radio production efficiencies of early galaxies ($f_*$, $f_\mathrm{X}$ and $f_\mathrm{radio}$, respectively) and the CMB optical depth that corresponds to the reionization history of the model ($\tau_\mathrm{CMB}$). Here, we use the same model (see \Cref{Sec:SemiNumericalSimulations}) as in B24 and use neural density estimation (see \Cref{Sec:NDE,Sec:NDElike} for details) to incorporate in our analysis the constraints from the 21-cm power spectrum upper limits considered in that work.

In addition to 21-cm power spectrum constraints, B24 consider sky-averaged `global' 21-cm signal constraints from SARAS 3 data in their analysis. Given the current tension between available measurements of the global 21-cm signal from SARAS 3 and those reported by the EDGES experiment (\citealt{2018Natur.555...67B}; hereafter B18), however, in this work we conservatively restrict our attention with respect to 21-cm constraints to those deriving from upper limits on the 21-cm power spectrum. Nevertheless, because we have a self-consistent model from which both the 21-cm power spectrum and the global 21-cm signal are derived, our joint analysis results can be used to derive the posterior predictive density (posterior PD; see \Cref{Sec:BayesianInference}) of the global 21-cm signal. In \Cref{Sec:EDGESandSARASImplications}, we compare the posterior PD of the global 21-cm signal, given the model and constraints considered in this work, to the SARAS 3 and EDGES global 21-cm signal constraints.

\subsection{Lyman line constraints}
\label{Sec:LymanLine}

\Cref{Fig:LymanLineConstraints2} shows one-dimensional posterior distributions of $\overline{x}_\mathrm{H_{I}}$ given Lyman line-based constraints from the observations in the redshift range $5.6 \lesssim z \lesssim 8$ that we incorporate in our analysis. These include:
\begin{itemize}
    \item Model-independent upper limits on $\overline{x}_\mathrm{H_{I}}$ in the redshift range $5 \lesssim z \lesssim 6$ estimated from measurements of the \Lya\ and \Lyb\ forest dark pixel fraction derived from spectra of 22 bright high-redshift  quasars, obtained with the Magellan, MMT, and VLT (\citealt{2015MNRAS.447..499M}; orange).
    \item Upper limits on $\overline{x}_\mathrm{H_{I}}$ at $z = 6.6$ from the clustering of \Lya\ emitters (LAEs) in a 1 $\mathrm{deg}^2$ patch of sky in the Subaru/XMM-Newton Deep Survey field (\citealt{2015MNRAS.453.1843S}; red).
    \item Constraints on $\overline{x}_\mathrm{H_{I}}$ at $z = 7.0$, $7.09$ and $7.54$ from the analysis of the \Lya\ damping wings of the bright quasars DES J0252-0503 (\citealt{2020ApJ...896...23W}; green dotted line)
    marginalising out the uncertainty in the quasar lifetimes assuming a uniform prior in the range $10^3$ to $10^8~\mathrm{yr}$.
    \item The evolution of the \Lya\ equivalent width (EW; a measure of the brightness of the emission line relative to the UV continuum) estimated from detections and non-detections of \Lya\ emission from:
    \begin{itemize}
        \item 68 Lyman Break galaxies (LBGs) selected to have a high probability of having redshifts in the range $6.5 \lesssim z \lesssim 7.5$ and with a median redshift $z = 6.9$ (\citealt{2018ApJ...856....2M}; solid cyan curve).
        \item 29 $z \sim 8$ KMOS Lens-Amplified Spectroscopic Survey (KLASS) targets with photometric redshifts consistent with falling in the range $7.2 < z < 8.8$ and with an estimated median and standard deviation photometric redshift of $z = 7.9 \pm 0.6$, as well as 8 Keck/MOSFIRE observations of $z \sim 8$ LBGs from the Brightest of Reionizing Galaxies (BORG) survey (\citealt{2019MNRAS.485.3947M}; dashed cyan curve).
        \item a lensed, intrinsically faint sample of $\sim 200$ LBG candidates in the redshift range $5.5 \lesssim z \lesssim 7$ (\citealt{2022MNRAS.517.3263B}; dotted cyan curve) and 68 low-luminosity candidates in the redshift range $7 \lesssim z \lesssim 8.2$ (\citealt{2022MNRAS.517.3263B}; dot-dashed cyan curve).
    \end{itemize}
\end{itemize}
The approaches we use to incorporate the above constraints in our joint analyses (via forward modelling of the IGM neutral fraction history during the EoR) are described in \Cref{Sec:MHGlike,Sec:SplinePDFlike}.

\subsection{CMB power spectrum constraints}
\label{Sec:CMB}

\begin{figure*}
	\centerline{
        \begin{subfigure}[t]{0.5\textwidth}
        \caption{}
        \includegraphics[width=0.99\textwidth]{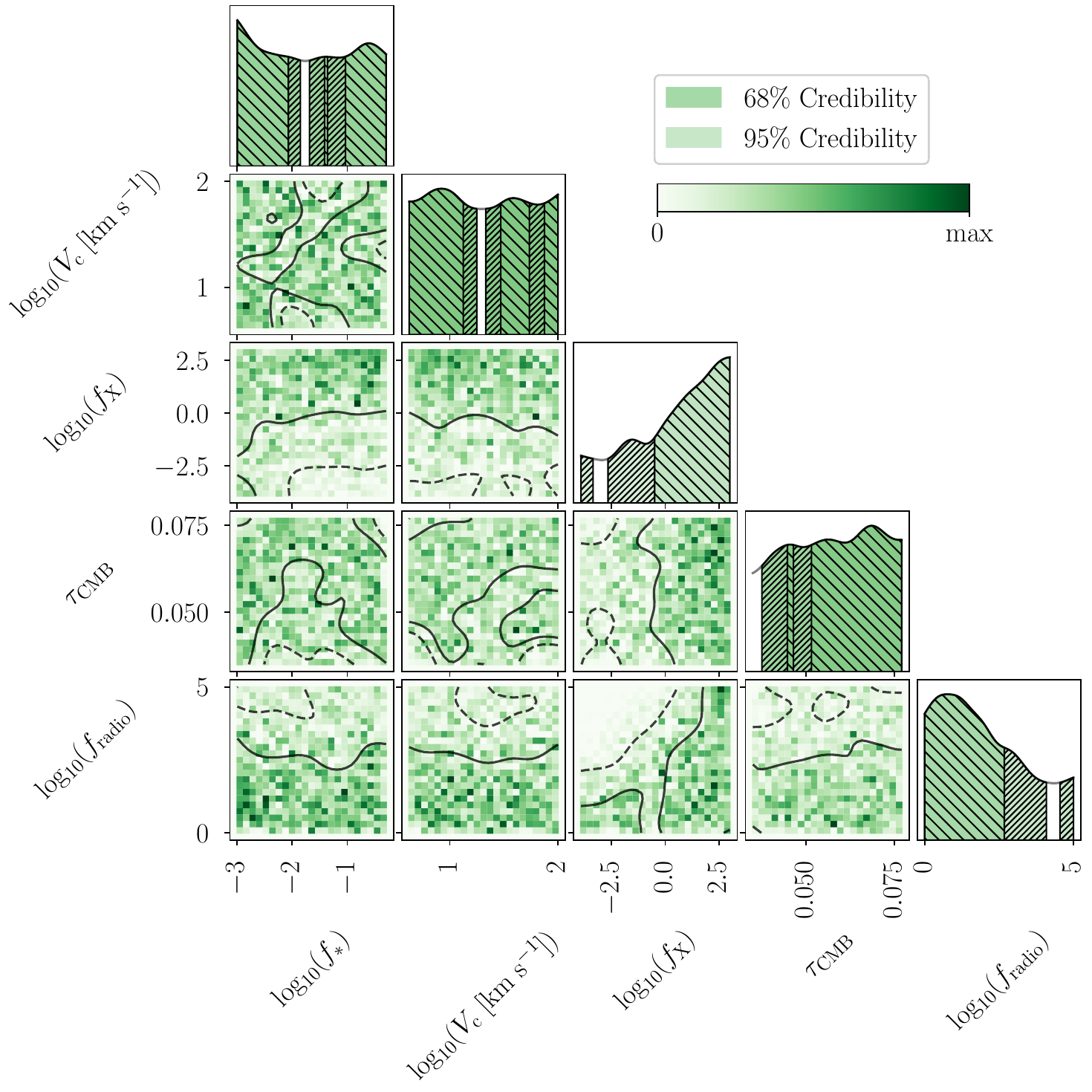}
        \label{Fig:B24Posteriors2}
        \end{subfigure}
        \begin{subfigure}[t]{0.4\textwidth}
        \caption{}
        \includegraphics[width=0.95\textwidth]{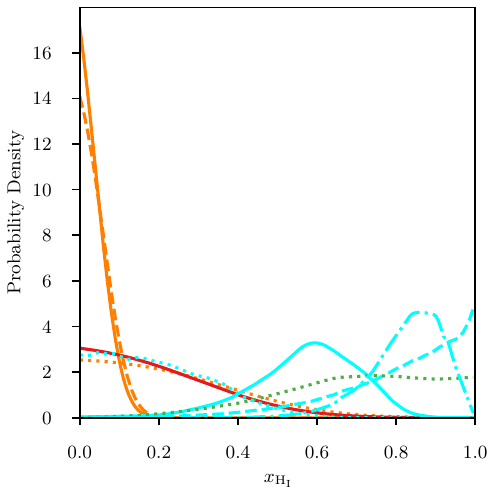}\\
        \includegraphics[width=0.95\textwidth, trim={-1.4cm 0 -0.90cm 0}, clip]{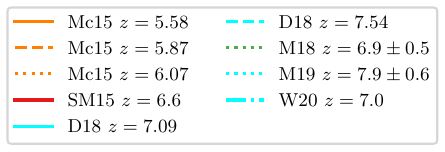}
        \label{Fig:LymanLineConstraints2}
        \end{subfigure}
    }
	\centerline{
        \begin{subfigure}[t]{0.4\textwidth}
        \caption{}
        \includegraphics[width=\textwidth, trim={-1.3cm 0 0.1cm 0}, clip]{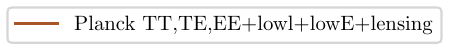}\\
        \includegraphics[width=\textwidth]{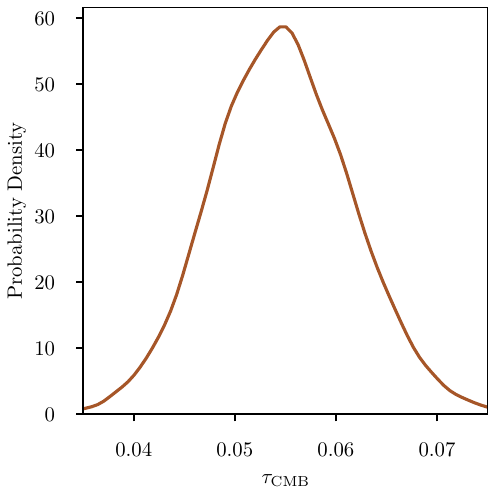}
        \label{Fig:Plancktauposterior2}
        \end{subfigure}
        \begin{subfigure}[t]{0.4\textwidth}
        \caption{}
        \includegraphics[width=0.44\textwidth, trim={-1.9cm 0 0.1cm 0}, clip]{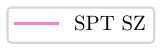}\\
        \includegraphics[width=\textwidth]{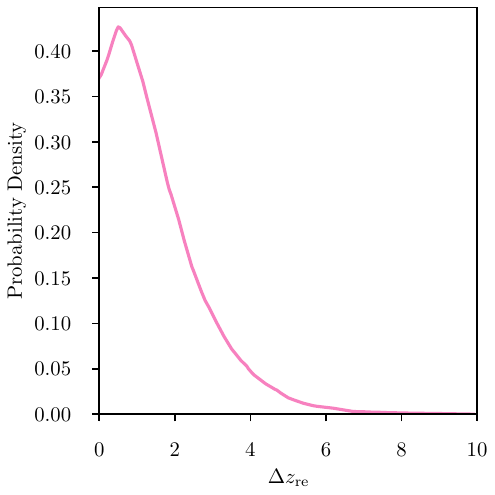}
        \label{Fig:SPTdzrePosterior2}
        \end{subfigure}
    }
    \caption{
        Probability distributions of astrophysical parameters derived with analyses of high-redshift data sets and which form the inputs to the joint analysis carried out in this work. \Cref{Fig:B24Posteriors2} shows the one- and two-dimensional posterior probability distributions of the astrophysical parameters derived from the combination of HERA+MWA+LOFAR analysis in B24. The colorbar displays the peak-normalised binned probability of histogram bins within the 2D joint posterior distributions, where dark green denotes bins that have maximum posterior probability and white bins those with negligible probability. The sparsely and densely hatched regions of the 1D posteriors and areas enclosed by solid and dashed black lines in the 2D posteriors contain 68\% and 95\% of the probability, respectively, in each case. \Cref{Fig:LymanLineConstraints2} shows one-dimensional posterior distributions on $\overline{x}_\mathrm{H_{I}}$ using Lyman line constraints deriving from measurements of the \Lya\ and \Lyb\ forest dark pixel fraction (orange), the clustering of \Lya\ emitters (red), the analysis of the damping wings of bright quasar (green), and the evolution of the \Lya\ equivalent width (cyan). The redshifts of the objects from which the individual PDFs are derived are labelled in the figure legend. The mapping between legend labels and the papers in which the original analyses were carried out is given in \Cref{Tab:DataSets}. \Cref{Fig:Plancktauposterior2} shows the Planck TT,TE,EE+low$l$+lowE+lensing one-dimensional marginal posterior on the total CMB optical depth, $\tau_\mathrm{CMB}$. \Cref{Fig:SPTdzrePosterior2} shows the one-dimensional marginal posterior distribution on $\Delta z_\mathrm{re}$ from \citet{2021ApJ...908..199R}, using SPT measurements of the kSZ effect imprinted on the CMB power spectrum, in combination with a prior on the tSZ bispectrum from \citet{2014ApJ...784..143C} and assuming the \citet{2014JCAP...08..010C} model for the homogeneous kSZ power.
    }
    \label{Fig:Priors}
\end{figure*}

\Cref{Fig:Plancktauposterior2,Fig:SPTdzrePosterior2} illustrate the CMB-power-spectrum-based constraints included in our analysis. \Cref{Fig:Plancktauposterior2} shows the Planck TT,TE,EE+low$l$+lowE+lensing\footnote{This derives from a joint fit of the Planck temperature and polarization E-mode power spectra (TT, EE) and temperature polarization E-mode cross-spectrum (TE), including fits to the large angular scale ($2 \le l \le 29$) temperature and E-mode polarisation power spectra (low$l$+lowE; e.g. P20VI) and CMB lensing measurements (\citealt{2020A&A...641A...8P}).} one-dimensional marginal posterior on the total CMB optical depth. To incorporate this constraint, we use samples\footnote{\url{https://pla.esac.esa.int/}} from the posterior on $\tau_\mathrm{CMB}$ to train a neural density estimator (NDE) which we subsequently employ as a parameter prior in our analysis (see \Cref{Sec:tauLikelihood}).

\Cref{Fig:SPTdzrePosterior2} shows the one-dimensional marginal posterior distribution on $\Delta z_\mathrm{re}$ from \citet{2021ApJ...908..199R}, derived from SPT measurements of the kSZ effect imprinted on the CMB power spectrum, in combination with a prior on the tSZ bispectrum from \citet{2014ApJ...784..143C} and assuming the \citet{2014JCAP...08..010C} model for the homogeneous kSZ power. Here, $\Delta z_\mathrm{re} = z(\overline{x}_\mathrm{H_{I}}=0.75) - z(\overline{x}_\mathrm{H_{I}}=0.25)$, where $z(\overline{x}_\mathrm{H_{I}}=0.25)$ is the redshift at which the sky average IGM neutral fraction reaches $25\%$. We incorporate this constraint in our analysis by forward modelling $\Delta z_\mathrm{re}$, and we employ it as a derived parameter prior (see \Cref{Sec:DeltazReLikelihood}).

%%%%%%%%%%%%%%%%%%%%%%%%%%%%%%%%%%%%%%%%%%%%%%%%%%
\section{Analysis methodology}
\label{Sec:AnalysisMethodology}
%%%%%%%%%%%%%%%%%%%%%%%%%%%%%%%%%%%%%%%%%%%%%%%%%%

In this section, we discuss the Bayesian statistical framework and data modelling that we use to infer parametric constraints on cosmological and astrophysical processes during CD and the EoR via the joint analysis of the high-redshift data sets and constraints described in \Cref{Sec:DataSetsAndMeasurements}.

\subsection{Bayesian inference}
\label{Sec:BayesianInference}

\subsubsection{Bayesian joint analysis}
\label{Sec:BayesianJointAnalysis}

Given $N$ independent data sets\footnote{The measurements used in our joint analysis take the form of constraints on modelled observables, which are inferred from underlying observations. In the context of the Bayesian inference terminology of \Cref{Sec:BayesianInference}, it is these constraints on the modelled observables that constitute our data sets.}, $\bm{D}_{1}$, $\bm{D}_{2}$, $\cdots$, $\bm{D}_{N}$, and a model for the data sets, $M$, described by a set of parameters, $\sTheta$,  Bayesian inference provides a statistically consistent approach for updating one's prior beliefs, given the data. Bayes' theorem states that,
\begin{equation}
\label{Eq:BayesEqnJA}
\mathcal{P}(\sTheta \vert \bm{D}_{1}, \cdots, \bm{D}_{N}, M) = \dfrac{\mathcal{P}(\sTheta \vert M)\ \prod\limits_{i=1}^{N} \mathcal{P}(\bm{D}_{i}\vert\sTheta,M)}{\mathcal{P}(\bm{D}_{1}, \cdots, \bm{D}_{N} \vert M)} \ .
\end{equation}
Here, $\mathcal{P}(\sTheta \vert M) \equiv \mathcal{\pi}(\sTheta)$ describes one's prior beliefs regarding the probability density of the parameters, $\mathcal{P}(\sTheta \vert \bm{D}_{1}, \cdots, \bm{D}_{N}, M)$ is the posterior probability density of the parameters given the data, $\mathcal{P}(\bm{D}_{i} \vert \sTheta,M) \equiv \mathcal{L}_{i}(\sTheta)$ is the likelihood of data set $i$, and the Bayesian evidence $\mathcal{P}(\bm{D}_{1}, \cdots, \bm{D}_{N} \vert M) \equiv \mathcal{Z}$ is the marginal likelihood of the data sets given the model. In what follows, for notational simplicity, we leave the model-dependence implicit and write the set of data sets as $\bm{D} = \{\bm{D}_{1}, \cdots, \bm{D}_{N}\}$. We thus denote, for example, the posterior probability density as $\mathcal{P}(\sTheta \vert \bm{D})$.

Taking the log of both sides of \Cref{Eq:BayesEqnJA}, one can thus write,
\begin{equation}
    \label{Eq:BayesEqnJAlog}
    \log(\mathcal{P}(\sTheta \vert \bm{D})) = -\log(\mathcal{Z}) + \log(\mathcal{\pi}(\sTheta)) + \sum\limits_{i=1}^{N} \log(\mathcal{L}_{i}(\sTheta)) \ .
\end{equation}
This form of Bayes' theorem is generally preferable to work with, for computational reasons, when one or more of the distributions in \Cref{Eq:BayesEqnJA} span many orders of magnitude in probability density. When comparing two models, the relative probability of the data given each can be calculated as the ratio of their Bayesian evidences. Here, we consider parameter inference for a fixed model. For a given model, $\log(\mathcal{Z})$ is a constant with respect to the model parameters and thus can be ignored for this purpose.

In a general Bayesian analysis workflow, posteriors from one analysis form one's priors for subsequent analyses. In this context we note that the joint analysis posterior given in \Cref{Eq:BayesEqnJAlog} is equivalent to the posterior density from a sequential series of analyses of independent data sets,
\begin{align}
    \label{Eq:BayesEqnJALog2}
    \log(\mathcal{P}(\sTheta \vert \bm{D})) = -\log(\mathcal{Z}_{i \ge m}) + \log(\mathcal{\pi}_{m}(\sTheta)) + \sum\limits_{i=m}^{N} \log(\mathcal{L}_{i}(\sTheta)) \ ,
\end{align}
where $\mathcal{Z}_{i \ge m} = \mathcal{P}(\bm{D}_{m}, \cdots, \bm{D}_{N} \vert M)$ and
\begin{align}
    \label{Eq:Priorm}
    \mathcal{\pi}_{m}(\sTheta) &= \mathcal{P}(\sTheta \vert \bm{D}_{i < m}) =  \frac{\mathcal{\pi}(\sTheta)}{\mathcal{Z}_{i < m}}  \prod\limits_{i=1}^{m-1} \mathcal{L}_{i}(\sTheta) \ .
\end{align}
Here, $\mathcal{P}(\sTheta \vert \bm{D}_{i < m})$ is one's posterior on the model parameters derived from the first $(m-1)$ data sets. This constitutes one's state of knowledge, $\mathcal{\pi}_{m}(\sTheta)$, for the joint analysis of the subsequent $i \ge m$ data sets. $\mathcal{L}_{i}(\sTheta)$ is the likelihood of the $i$th data set given the model and parameters, $\sTheta$. The likelihoods of the data sets included in our analysis are described in \Cref{Sec:DataLikelihoods} and summarised in \Cref{Tab:DataSets}.

Several of the measurements discussed in \Cref{Sec:DataSetsAndMeasurements} are of the form $\mathcal{P}(f_{i}(\sTheta) \vert \bm{D}_{i})$. Here, $\bm{D}_{i}$ is the data set on which the PDF of the derived quantity $f_{i}(\sTheta)$ is conditioned; the function $f_{i}(.)$ maps the sampled parameters of our model and the derived quantity constrained by the $i$th data set. In a scenario where one explicitly fits the data set, $\bm{D}_{i}$, from which the constraint was derived, it would be necessary to define a corresponding likelihood and model. However, this can be bypassed\footnote{This requires that the uncertainty deriving from model-choice, on the quantity of interest, is subdominant to other sources of statistical uncertainty. Alternatively, it could be accounted for when estimating the quantity (for example, by estimating it using a Bayesian mixture model; see e.g. \citealt{2023MNRAS.521.3273S}). In \citet{2020MNRAS.499..550Q} it is shown that comparable reionization inference are derived from conditioning the model on the CMB optical depth versus the E-mode polarization power spectra. In future work, validating this assumption for the model and constraints considered here will be valuable for testing the robustness of our conclusions.} by including the derived constraints in \Cref{Eq:BayesEqnJA} as priors, $\mathcal{\pi}_{i}(f_{i}(\sTheta))$, on the derived quantities of our model. In this case, \Cref{Eq:Priorm} generalises to:
\begin{align}
    \label{Eq:Priorm2}
    \mathcal{\pi}_{m}(\sTheta) = k \mathcal{\pi}(\sTheta)  \prod\limits_{i=1}^{n} \mathcal{L}_{i}(\sTheta)   \prod\limits_{j=n}^{m-1} \mathcal{\pi}_{j}(f_{j}(\sTheta)) \ .
\end{align}
Here, $k$ is a constant of proportionality required to normalise $\mathcal{\pi}_{m}(\sTheta)$ to a proper probability distribution\footnote{Using Bayes' theorem in logarithmic form (\Cref{Eq:BayesEqnJALog2}), the log of the constant of proportionality, $k$, in \Cref{Eq:Priorm2}, is a constant and thus can also be ignored for the purpose of parameter inference.}, we have assumed that $n$ of the data sets considered in our joint analysis are jointly fit for explicitly, and we constrain the parameters of our model given the remaining $m-n-1$ data sets via priors on derived quantities of our model $\mathcal{\pi}_{j}(f_{j}(\sTheta)) = \mathcal{P}(f_{i}(\sTheta) \vert \bm{D}_{i})$.

\subsubsection{Information theory}
\label{Sec:InformationTheory}

The Shannon information (\citealt{1949mtc..book.....S})
\begin{equation}
\label{Eq:ShannonInformation}
\mathcal{I}(\sTheta \vert \bm{D}) = \log \left(\frac{\mathcal{P}(\sTheta \vert \bm{D})}{\mathcal{\pi}(\sTheta)}\right) \ ,
\end{equation}
quantifies the information gained, measured in nats (natural bits), about a set of model parameters, $\sTheta$, when moving from the prior to the posterior.
For independent model parameters, such that $\mathcal{P}(\sTheta \vert \bm{D}) = \prod\limits_{i=1}^{n} \mathcal{P}(\theta_{i} \vert \bm{D})$ and $\mathcal{\pi}(\sTheta) = \prod\limits_{i=1}^{n} \mathcal{\pi}(\theta_{i})$, and independent data sets, such that $\mathcal{P}(\sTheta \vert \bm{D}) = \prod\limits_{j=1}^{N} \mathcal{P}(\sTheta \vert \bm{D}_{j})$, the Shannon information is additive:
\begin{equation}
\label{Eq:ShannonInformation2}
\mathcal{I}(\sTheta \vert \bm{D}) = \sum\limits_{i=1}^{n} \mathcal{I}(\theta_{i} \vert \bm{D}) = \sum\limits_{i=1}^{n} \sum\limits_{j=1}^{N} \mathcal{I}(\theta_{i} \vert \bm{D}_{j}) \ .
\end{equation}
Here, $n$ and $N$ are the dimensionalities of the parameter vector and number of data sets considered in the joint analysis, respectively.

The Kullback--Leibler (KL) divergence is defined as the average Shannon information over the posterior (\citealt{10.1214/aoms/1177729694}),
\begin{align}
\label{Eq:KLDivergence}
\mathcal{D}_\mathrm{KL}(\sTheta \vert \bm{D}) &= \int \mathcal{P}(\sTheta \vert \bm{D})\log\left(\frac{\mathcal{P}(\sTheta \vert \bm{D})}{\mathcal{\pi}(\sTheta)}\right) \mathrm{d}^{n}{\theta} \\ \nonumber
&= \left\langle \log \left(\frac{\mathcal{P}(\sTheta \vert \bm{D})}{\mathcal{\pi}(\sTheta)} \right)\right\rangle_{\mathcal{P}(\sTheta \vert \bm{D})} \\ \nonumber
&= \left\langle\mathcal{I}(\sTheta \vert \bm{D})\right\rangle_{\mathcal{P}(\sTheta \vert \bm{D})} \ .
\end{align}
Since $\mathcal{D}_\mathrm{KL}(\sTheta \vert \bm{D})$ is a linear function of the Shannon information, it is also measured in nats and is additive for independent parameters and data sets.

As the data refines one's prior beliefs, $\mathcal{D}_\mathrm{KL}(\sTheta \vert \bm{D})$ accounts for contractions or redistributions in the posterior relative to the prior, including changes in shape or mean. Given this, one can relate it to the fraction of the prior that is consistent with the data via (e.g. B24; \citealt{2024MNRAS.529..519G,2024MNRAS.531.1113P}):
\begin{equation}
    \label{Eq:KLDivergenceFractionalConsistency}
    f_\mathrm{c} = e^{-\mathcal{D}_\mathrm{KL}(\sTheta \vert \bm{D})} \ .
\end{equation}
Here, $0 < f_\mathrm{c} \le 1$. In the limit that the data provides no new information, $\mathcal{D}_\mathrm{KL}(\sTheta \vert \bm{D}) = 0$ and $f_\mathrm{c} = 1$. In contrast, when $\mathcal{D}_\mathrm{KL}(\sTheta \vert \bm{D}) \gg 1$, $f_\mathrm{c} \approx 0$, indicating that the data has provided substantial new information, leading to significant contraction or redistribution of the posterior relative to the prior.

For a top-hat prior and posterior (i.e., uniform distributions over the respective parameter spaces), $f_\mathrm{c}$ corresponds to an intuitive geometric measure of the constraining power of the data set:
\begin{equation}
    \label{Eq:KLDivergenceVolume}
    f_\mathrm{c} = \frac{V_{\mathcal{P}(\sTheta \vert \bm{D})}}{V_{\mathcal{\pi}(\sTheta)}} \ ,
\end{equation}
with $V_{\mathcal{P}(\sTheta \vert \bm{D})}$ and $V_{\mathcal{\pi}(\sTheta)}$ the posterior and prior volumes of the parameter space of the model, respectively.

$\mathcal{D}_\mathrm{KL}(\sTheta \vert \bm{D})$ and $f_\mathrm{c}$ provide intuitive, easily interpretable measures of the information content in individual and combined data sets, which we utilize throughout this work.

\subsubsection{Summary statistics}
\label{Sec:SummaryStatistics}

In \Cref{Sec:Results}, we will provide summary statistics characterising the probability density functions of sampled and derived parameters of our data model. The extent to which data contains information that constrains the PDF of a model parameter can be quantified via the KL divergence between the parameter prior and posterior densities, as discussed in \Cref{Sec:InformationTheory}. Here, we define a marginal posterior on a parameter (or set of parameters) as constrained by a data set under the following condition:
\paragraph*{Definition: constrained posterior distribution.}{\textit{We describe a data set $D$ as providing an appreciable constraint on the marginal posterior distribution of a parameter (or set of parameters) $\upTheta_{i}$ when it has a KL divergence relative to the prior on $\upTheta_{i}$ of $\mathcal{D}_\mathrm{KL}(\upTheta_{i} \vert D) \ge \mathcal{D}_\mathrm{KL}^\mathrm{min}$, where $\mathcal{D}_\mathrm{KL}^\mathrm{min}$ is a minimum-information-content threshold.\\}}

Here, we describe a parameter posterior as appreciably constrained by the data if it has a KL divergence above $\mathcal{D}_\mathrm{KL}^\mathrm{min} \simeq 3\mathcal{D}_\mathrm{KL}^\mathrm{noise}$, where $\mathcal{D}_\mathrm{KL}^\mathrm{noise} \sim 0.0033~\mathrm{nats}$ is an estimate of the KL divergence between prior and posterior distributions of our sampled astrophysical parameters due to sampling noise\footnote{We estimate this by dividing the astrophysical parameters of our model into two subsets, $\bm{\theta} = \{\bm{\theta}_{1}, \bm{\theta}_{2}\}$, of which only the 1D and 2D marginal posterior distributions of the parameters in $\bm{\theta}_{1}$ are constrained by the data. We train an NDE on samples from the posterior of $\bm{\theta}_{1}$. Using the NDE as a likelihood that is independent of $\bm{\theta}_{2}$ we re-sample from the full set of parameters of our model and use the maximum KL divergence between the marginal 1D and 2D posteriors of the unconstrained parameters as our estimate of $\mathcal{D}_\mathrm{KL}^\mathrm{noise}$.}. In the remainder of this paper we quote summary statistics for posterior distributions only when they are appreciably constrained by the data\footnote{The statistics of unconstrained parameter posteriors can optionally be calculated from the parameter priors, where they are of interest.}, as determined by the above metric.

For the purpose of providing explicative summary statistics of the posterior PDFs of constrained parameters, it is useful to subdivide them into two groups: those with and without prior distributions with compact support (with or without PDFs characterised by non-zero probability densities only in closed, bounded regions). The prior distributions of the sampled astrophysical parameters considered in this work all have compact support (see \Cref{Sec:SemiNumericalSimulations}). Parameters with compact support can be further subdivided into two subclasses, within which we consider different statistics to provide summaries of the distributions: those which have posteriors that are or are not prior-limited. We use the following definition of prior-limited posterior PDFs:
\paragraph*{Definition: prior-limited posterior distribution.}{\textit{We define a posterior distribution as prior-limited when the highest probability density interval (HPDI; e.g. \citealt{b7f71c99-f621-3c7e-a9dd-9d152d4822a4}) of the posterior PDF of a parameter (or group of parameters) has an upper or lower limit equal to a boundary of the parameter's prior PDF.\\}}

Here, we use one of two summary statistics to characterise the posterior PDFs of parameters with constrained distributions and priors with compact support, depending on whether they have prior-limited posterior distributions or not:
\begin{itemize}
    \item \textit{Non-prior-limited posterior distributions}, we characterise via their highest probability density estimate (HPDE; e.g. \citealt{b7f71c99-f621-3c7e-a9dd-9d152d4822a4}) and 68\% HPDI, $X_\mathrm{HPDE}|^{+\sigma_{+}}_{-\sigma_{-}}$. Here, $X_\mathrm{HPDE}$ is the highest probability density value of the PDF of a parameter (or set of parameters), $X$, and $\sigma_{\pm} = |X_\mathrm{HPDI\pm} - X_\mathrm{HPDE}|$ characterises its width, with $X_\mathrm{HPDI+}$ and $X_\mathrm{HPDI-}$, the upper and lower bound of the HPDI, respectively.
    \item \textit{Prior-limited posterior distributions}, we characterise via their 68\% or 95\% credibility upper limits when the HPDI is in contact with the maximum of the corresponding prior and by the equivalent lower limits when the HPDI is in contact with the minimum of the prior. We note that the 68\% credibility limits we use as summary statistics for prior-limited posterior distributions are equal to the 68\% HPDI when those distributions are unimodal (as they are for the prior-limited posterior distributions considered in this work). In these cases, the specification that the quoted parameter interval is an upper or lower limit provides a succinct way of conveying that the PDF is prior limited and indicating whether it is the largest or smallest values in the parameter prior that are preferred in the posterior PDF.
\end{itemize}

\subsubsection{Computational methods}
\label{Sec:ComputationalMethods}

Throughout our analysis in \Cref{Sec:Results}, we sample directly from $\log(\mathcal{P}(\sTheta \vert \bm{D}))$ using nested sampling as implemented by \textsc{PolyChord} \citep{2015MNRAS.450L..61H, 2015MNRAS.453.4384H}. Given samples from the posterior distribution of the parameters, $\mathrm{Pr}(\upTheta\vert\bm{D},M)$, one can estimate the posterior PD $\mathrm{Pr}(y\vert z,\upTheta,\bm{D},M)$, of a variable or function $y = f(z,\upTheta)$, by calculating the corresponding set of samples from $\mathrm{Pr}(y\vert z,\upTheta,\bm{D},M)$. We derive contour plots of prior and posterior PDs using the \textsc{fgivenx} software package (\citealt{2018JOSS....3..849H}). We calculate $\mathcal{D}_\mathrm{KL}(\sTheta_{\mathrm{s}_{i}} \vert \bm{D}_{\mathrm{s}_{j}})$ using marginal density estimation as implemented by \textsc{margarine} (\citealt{2022arXiv220711457B, 2023MNRAS.526.4613B}). Here, $\sTheta_{\mathrm{s}_{i}}$ are 1D and 2D subsets of the model parameters, and we use $\bm{D}_{\mathrm{s}_{j}}$ to denote observation-technique specific subsets of the data (or, in the case of the final joint analysis, the full range of data sets considered). We use kernel density estimation in \textsc{getdist} (\citealt{2019arXiv191013970L}) to generate plots of exclusively well constrained marginalised 2D and 1D parameter PDFs. Histogramming of the posterior samples provides a higher fidelity representation of unconstrained 2D distributions, therefore we use this approach as implemented in \textsc{anesthetic} (\citealt{2019JOSS....4.1414H}) to generate plots containing a mix of unconstrained and well constrained marginalised 2D PDFs.

\subsection{21cmSPACE semi-numerical simulations}
\label{Sec:SemiNumericalSimulations}

The simulation framework employed in this work, recently named 21cmSPACE (21-cm Semi-numerical Predictions Across Cosmic Epochs; e.g., \citealt{2024MNRAS.529..519G}), is described in detail in \citet{2019MNRAS.486.1763F}, \citet{2021MNRAS.506.5479R}, and \citealt{2023MNRAS.526.4262G}, as well as references therein.. In brief, each simulation is initialized with cubes, $384~\mathrm{Mpc}$ on a side, of density, temperature and relative velocity between dark matter and baryons. The density and velocity fields are evolved using linear perturbation theory. The number of dark matter halos per voxel is determined based on the values of the local density and relative velocity and is derived at each redshift using a modified Press-Schechter model (we refer the reader to \citealt{2023MNRAS.526.4262G} for a detailed overview of the model and B24 and references therein for specifics of the model used in this work).

\Cref{Tab:Priors} lists the prior distributions assumed for the astrophysical parameters of the 21cmSPACE semi-analytic simulations, which form the foundational level of our data model: $\sTheta = [f_*, V_\mathrm{c}, f_\mathrm{X}, \tau_\mathrm{CMB}, f_\mathrm{radio}, R_\mathrm{MFP}]$. To combine the interferometric constraints derived in B24 with those from the additional data sets described in \Cref{Sec:DataSetsAndMeasurements}, we follow B24 in fixing the mean free path of ionizing photons -- a parameter to which the data is relatively insensitive -- to $R_\mathrm{MFP} = 40~\mathrm{Mpc}$.

\Cref{Fig:SensitivityPlotxHI_T21_21cmPS} illustrates the sensitivity of the three modelled summary statistics considered in our analysis -- (i) the IGM neutral fraction (left column), (ii) the global 21-cm signal (middle column\footnote{See \citealt{2019ApJ...875...67M} for an equivalent plot of the astrophysical parameter sensitivity of the global 21-cm signal in an excess-radio-background-free 21cmSPACE model.}), and (iii) the 21-cm power spectrum (right column) -- to variations in the five free astrophysical parameters. Brief definitions of these parameters are provided below.

\begin{table}
    \caption{
        Astrophysical priors and parameter descriptions. We select broad priors designed to encompass the large theoretical uncertainty in the properties of the high-redshift Universe in the absence of the constraints from the data sets considered in this work. Parameters in the upper section of the table are sampled over in our analysis; $R_\mathrm{MFP}$ in the lower section is fixed to match B24 in order to use NDEs to consistently combine the interferometric constraints derived in that work with constraints from the additional data sets considered here.
        }
    \centerline{
    \begin{tabular}{l|ccc}
        \hline
        Parameter & Prior  & Description     \\
        \hline
        $f_*$ & $\mathrm{log}U(0.001, 0.5)$  &  Star formation efficiency \\
        $V_\mathrm{c}$ & $\mathrm{log}U(4.2, 100)$ $\mathrm{km~s^{-1}}$  & Minimum circular velocity\\
        $f_\mathrm{X}$ &  $\mathrm{log}U(10^{-4}, 10^{3})$  & X-ray production efficiency\\
        $\tau_\mathrm{CMB}$ &  $U(0.04, 0.1)$  & CMB optical depth\\
        $f_\mathrm{radio}$ &  $\mathrm{log}U(1.0, 99500)$  & Radio production efficiency\\
        \hline
        $R_\mathrm{MFP}$ &  $40~\mathrm{Mpc}$  & Mean free path of ionizing photons\\
        \hline
        \end{tabular}
    }
\label{Tab:Priors}
\end{table}

\begin{enumerate}
    \item $V_\mathrm{c}$: the minimum circular velocity of dark matter halos hosting star forming galaxies. The corresponding redshift-dependent minimum halo mass threshold, $M_\mathrm{min}(z)$, scales as a cubic function of the circular velocity, modulated by the redshift (e.g. \citealt{2020MNRAS.499.5993R}):
    \begin{equation}
        \label{Eq:Vc}
        M_\mathrm{min} \propto \frac{V_\mathrm{c}^{3}}{(1+z)^{\frac{3}{2}}} \ .
    \end{equation}
    We explore a broad range of star formation thresholds by sampling $V_\mathrm{c}$ from a log-uniform prior, $\mathcal{\pi}(V_\mathrm{c}) = \mathrm{log}U(4.2, 100)$ $\mathrm{km~s^{-1}}$, such that the prior is uniform in the variable $\log(V_\mathrm{c})$. The lower bound, $V_\mathrm{c} = 4.2~\mathrm{km~s^{-1}}$, corresponds to the threshold for molecular-hydrogen-mediated cooling. Assuming spherical collapse, this translates to a minimum halo mass of $M_\mathrm{min}^\mathrm{molecular}(z=10) \simeq 1 \times 10^{6}~M_{\odot}$. For $V_\mathrm{c} \geq 16.5~\mathrm{kms^{-1}}$, atomic-hydrogen-mediated cooling becomes efficient, corresponding to a halo mass of $M_\mathrm{min}^\mathrm{atomic}(z=10) \simeq 8 \times 10^{7}M_{\odot}$. Finally, the upper bound of the prior, $V_\mathrm{c} = 100~\mathrm{kms^{-1}}$, corresponds to a halo mass of $M_\mathrm{min}(z=10) \simeq 2 \times 10^{10}~M_{\odot}$. In addition to the minimum circular velocity cut-off, star formation is further suppressed due to environmental effects, which boost the minimum mass of star-forming halos (e.g. \citealt{2020MNRAS.499.5993R} and references therein). These effects include: (1) the relative velocity between dark matter and baryons (\citealt{2012MNRAS.424.1335F}), and (2) the destruction of molecular hydrogen by Lyman-Werner radiation (\citealt{2013MNRAS.432.2909F}) in molecular cooling halos. As reionization proceeds (3) photoheating feedback becomes significant in atomic cooling halos, limiting further accretion of gas on to lower mass halos (e.g. \citealt{2013MNRAS.432.3340S,2016MNRAS.459L..90C}).
    \begin{figure*}
        \includegraphics[width=1.0\textwidth]{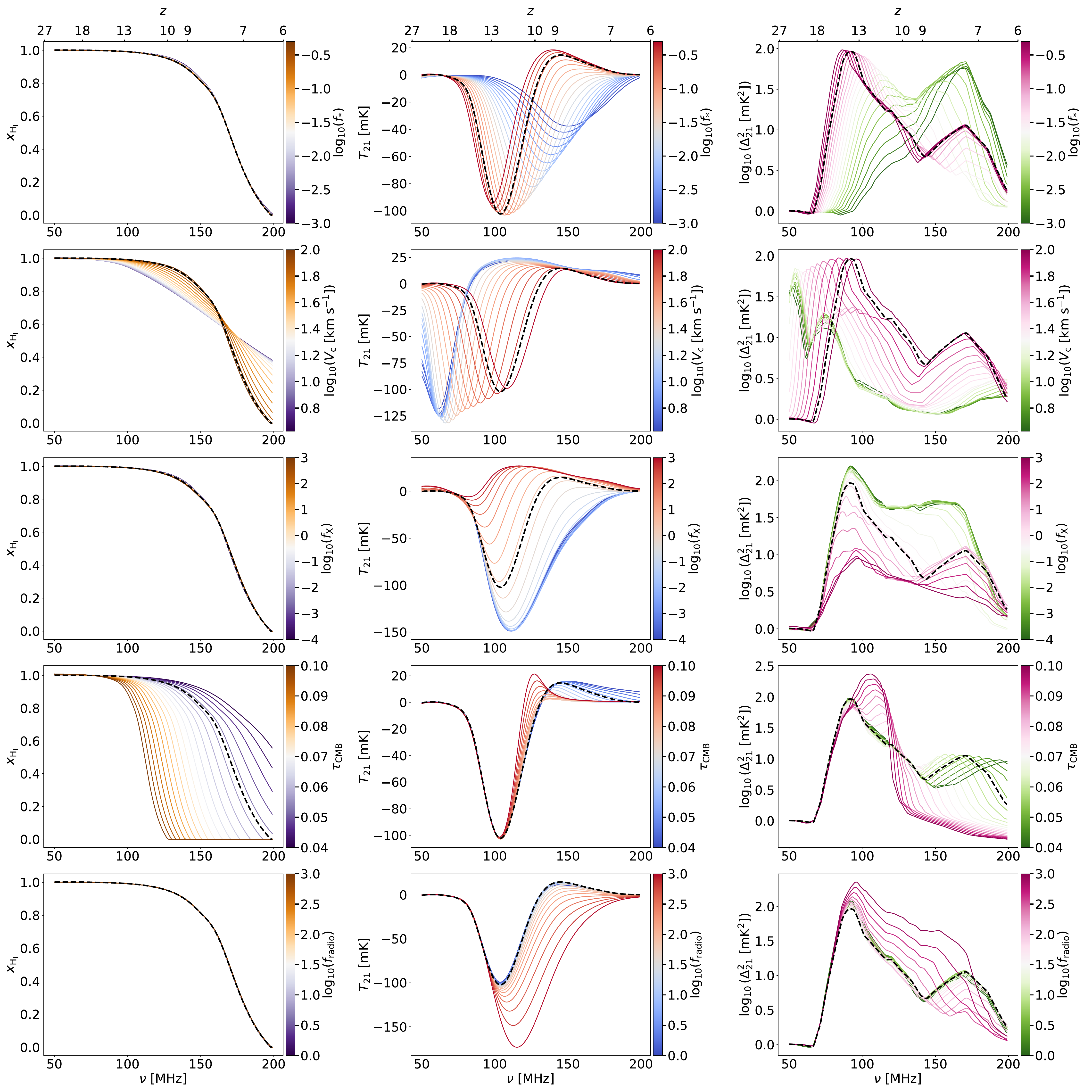}
        \caption{
            Sensitivity plot illustrating the impact of variations in astrophysical parameters, relative to a fiducial set of values, on: (i) the redshift-dependent IGM neutral fraction (left column), (ii) the global 21-cm signal (middle column), and (iii) the redshift- and spatial-scale-dependent dimensionless 21-cm power spectrum, $\Delta^{2}_{21}(k, z)$, evaluated at $k = 0.1h\mathrm{Mpc^{-1}}$ (right column). Each parameter is sampled in 20 uniform steps, in linear or logarithmic space according to its prior distribution, over the prior ranges listed in \Cref{Tab:Priors}. Among the five free parameters in our model, variations in the simulated global 21-cm signal and 21-cm power spectrum are dominated by changes in $\log_{10}(f_\mathrm{radio})$ when sampling its full prior range, $0 \leq \log_{10}(f_\mathrm{radio}) \lesssim 5$. To better illustrate the relative sensitivity to the remaining parameters, we display a compressed range for $\log_{10}(f_\mathrm{radio})$, sampling log-uniformly over $0 \leq \log_{10}(f_\mathrm{radio}) \lesssim 3$. The fiducial parameter values used are: $f_* = 0.2$, $V_\mathrm{c} = 90.0~\mathrm{km~s^{-1}}$, $f_\mathrm{X} = 2.0$, $\tau_\mathrm{CMB} = 0.054$, and $f_\mathrm{radio} = 10.0$. The corresponding signals for this fiducial model are shown as dashed black lines.
            }
        \label{Fig:SensitivityPlotxHI_T21_21cmPS}
    \end{figure*}
    \item $f_{*}$: The SFE in halos heavier than the atomic cooling mass describes the mass fraction of gas that is converted into stars in halos with a circular velocity $V_\mathrm{c} \ge 16.5~\mathrm{km~s^{-1}}$:
    \begin{equation}
        \label{Eq:fstar}
        f_{*} = \frac{M_{*}}{M_\mathrm{g}} \ .
    \end{equation}
    Here, $M_\mathrm{g}$ and $M_{*}$ are the initial gas mass (e.g. \citealt{2012MNRAS.424.1335F}) and final stellar mass, respectively. We model the SFE as constant in halos heavier than the atomic cooling mass (hereafter, atomic cooling halos) and assume a logarithmic cut-off in the SFE in halos with masses in the range $M_{\mathrm{min}}<M<M^{\mathrm{atomic}}_{\mathrm{min}}$. The general SFE of a halo with mass $M_\mathrm{h}$ in our model is thus given by (e.g. \citealt{2020MNRAS.495.4845C}):
    \begin{equation}
        \label{Eq:fstar2}
        \mathrm{SFE(M_\mathrm{h})}  =
        \begin{cases}
            f_{*} , & M^{\mathrm{atomic}}_{\mathrm{min}}<M \ , \\
            f_{*}\frac{\log(M/M_{\mathrm{min}})}{\log(M^{\mathrm{atomic}}_{\mathrm{min}}/M_{\mathrm{min}})} , & M_{\mathrm{min}}<M<M^{\mathrm{atomic}}_{\mathrm{min}} \ , \\
            0, & \text{otherwise} \ .
        \end{cases}
    \end{equation}
    We sample star formation efficiencies between a minimum and maximum of 0.1\% and 50\%, respectively, from a log-uniform prior $\mathcal{\pi}(f_{*}) = \mathrm{log}U(0.001, 0.5)$.
    \item $f_\mathrm{X}$: Observations of X-ray binaries\footnote{Population synthesis simulations calibrated to low-redshift observations of X-ray binaries suggest high-mass X-ray binaries (HMXB) dominate the total X-ray budget at redshifts above $z \sim 6$ (e.g. \citealt{2013ApJ...776L..31F}); here, we assume an X-ray spectral energy distribution (SED) that would be typical for a population of these sources at high redshifts.} (XRBs) in the local Universe find a strong correlation between the bolometric X-ray luminosity, $L_\mathrm{X}$, and the star formation rate (SFR; e.g. \citealt{2006PhR...433..181F,  2014Natur.506..197F}):
    \begin{equation}
        L_\mathrm{X}(z)=3\times10^{33}f_\mathrm{X} \left(\frac{\mathrm{SFR}(z)}{M_{\odot} \rm{yr}^{-1}}\right) ~\mathrm{W} \ ,
    \end{equation}
    where $f_\mathrm{X}$ is the X-ray efficiency. Observations of nearby XRBs yield $f_\mathrm{X} = 1.0$; however, larger values of $f_\mathrm{X}$ are predicted in low metallicity high-redshift galaxies. We sample X-ray efficiency from a log-uniform prior $\mathcal{\pi}(f_\mathrm{X}) = \mathrm{log}U(10^{-4}, 10^{3})$.

    \item $\tau_\mathrm{CMB}$: The CMB optical depth is proportional to the sky-averaged integrated column density of free electrons along the line of sight\footnote{Given an EoR endpoint (e.g. $z \lesssim 5.5$), the contribution to $\tau_\mathrm{CMB}$ from free electrons following hydrogen and first helium reionization, as well as second helium reionization by quasars, is well constrained (e.g., P20VI and references therein). The remaining contribution to the measured value of $\tau_\mathrm{CMB}$ arises from free electrons in the IGM during the EoR ($z \gtrsim 5.5$). Consequently, measurements of $\tau_\mathrm{CMB}$ provide constraints on the timeline of reionization: larger measured values correspond to an earlier or extended EoR, while smaller values suggest a shorter reionization period.},
    \begin{equation}
        \label{Eq:tauCMB}
        \tau_\mathrm{CMB}(z) = \sigma_\mathrm{T}\int_{0}^{l(z)} (1-\overline{x}_\mathrm{H_{I}}(z)) \overline{n}_\mathrm{e}(z) \mathrm{d}l^{\prime} \ .
    \end{equation}
    Here, $\overline{n}_\mathrm{e}$ is the average number density of free electrons in ionized regions accounting for hydrogen ionisation and first helium ionisation, $\sigma_\mathrm{T}$ is the Thomson cross-section and $\mathrm{d}l^{\prime}$ is the line of sight proper distance element. We sample CMB optical depth from a uniform prior $\mathcal{\pi}(\tau_\mathrm{CMB}) = U(0.04, 0.1)$.
    \item $f_\mathrm{radio}$: We model the galactic radio production efficiency as proportional to the ratio of the galactic radio luminosity per unit frequency, $L_\mathrm{radio}$, and the star formation rate,
    \begin{equation}
        \label{Eq:Fradio}
        L_\mathrm{radio} (\nu, z ) = f_\mathrm{radio} 10^{22} \left(\frac{\nu}{150 \mathrm{MHz}}\right)^{-\alpha_\mathrm{radio}} \left(\frac{\mathrm{SFR}(z)}{M_{\odot} \mathrm{yr^{-1}}}\right)~\mathrm{W~Hz^{-1}} \ ,
    \end{equation}
    with the average radio production efficiency for present-day star-forming galaxies corresponding to $f_\mathrm{radio} = 1$. We set $\alpha_\mathrm{radio} = 0.7$, consistent with synchrotron emission (e.g. \citealt{2020MNRAS.499.5993R}). We sample galactic radio production efficiency from a log-uniform prior $\mathcal{\pi}(f_\mathrm{radio}) = \mathrm{log}U(1.0, 99500)$.
\end{enumerate}

\subsection{Emulators}
\label{Sec:Emulators}

For a given vector of input parameters, $\sTheta$, the output of the 21cmSPACE simulations described in \Cref{Sec:SemiNumericalSimulations} is a set of ionisation, density and 21-cm temperature cubes at a discrete set of redshifts. From these simulation cubes, we derive the summary statistics necessary to jointly analyse the data sets and constraints described in \Cref{Sec:DataSetsAndMeasurements}. The two summary statistics required for this purpose are:
\begin{enumerate}
    \item the 1D `dimensionless' 21-cm power spectrum:
    \begin{equation}
        \Delta^{2}_{21}(k,\bm{z}, \sTheta) = \frac{2\pi^{2}}{k^{3}} P(k,\bm{z}, \sTheta) \ .
    \end{equation}
    Here, $k$ is the wavenumber and $\Delta^{2}_{21}(k,\bm{z}, \sTheta)$ is formed by averaging the 3D `dimensionless' 21-cm power spectrum, $\Delta^{2}_{21}(\bm{k},\bm{z}, \sTheta)$, in spherical annuli in $k$-space. $\Delta^{2}_{21}(\bm{k},\bm{z}, \sTheta)$ quantifies the variance of 21-cm brightness temperature fluctuations and has units of $\mathrm{mK^{2}}$. The 21-cm power spectrum $P(\bm{k},\bm{z}, \sTheta)$ has units $\mathrm{mK^{2}Mpc^{3}}$ and is defined as: $\left\langle \widetilde{\delta T_\mathrm{b}}(\bm{k},\bm{z}, \sTheta)\widetilde{\delta T_\mathrm{b}}^{*}(\bm{k}^\prime,\bm{z}) \right\rangle \equiv (2\pi)^3\delta_D(\bm{k}-\bm{k}^\prime)P(\bm{k},\bm{z}, \sTheta)$. Here, $\delta_D$ is the Dirac delta function and $\widetilde{\delta T_\mathrm{b}}(\bm{k})$ is the Fourier transform of $\delta T_\mathrm{b}(\bm{x},\bm{z}, \sTheta)$ which describes deviations of the 21-cm brightness temperature from the background radiation temperature (e.g. \citealt{2006PhR...433..181F}).
    \item the redshift-dependent sky averaged IGM neutral fraction:
    \begin{equation}
        \overline{x}_\mathrm{H_{I}}(\bm{z}, \sTheta) = \langle x_\mathrm{H_{I}}(\bm{x}, \bm{z}, \sTheta) \rangle \ ,
    \end{equation}
    where $\langle.\rangle$ is a spatial average and $\overline{x}_\mathrm{H_{I}}(z_{i}, \sTheta)$ is the IGM neutral fraction cube at redshift $z_{i}$, with $z_{i}$ the $i$th element of $\bm{z}$.
\end{enumerate}

Each 21cmSPACE simulation, from which the above summary statistics are calculated, requires several hours to run on a desktop computer (e.g., \citealt{2019ApJ...875...67M}). While this is significantly faster than full hydrodynamical simulations, directly using the 21cmSPACE simulator to robustly constrain astrophysical parameters through the joint analysis of the datasets discussed in \Cref{Sec:DataSetsAndMeasurements} would be computationally prohibitive.

To overcome this challenge, we employ the \textsc{globalemu} framework (\citealt{2021MNRAS.508.2923B}) to construct an emulator for the redshift-dependent IGM neutral fraction, $\overline{x}_\mathrm{H_{I}}(\bm{z})$. This emulator is trained on the neutral fraction histories produced by the 21cmSPACE simulations used as training data in B24. By doing so, we achieve a normalized root mean square error (RMSE) emulation accuracy of $\sim 1\%$ for $\overline{x}_\mathrm{H_{I}}(\bm{z})$, comparable to the performance demonstrated in \citet{2021MNRAS.508.2923B}.

To derive the global 21-cm signal sensitivity plots in \Cref{Fig:SensitivityPlotxHI_T21_21cmPS} and the posterior PD of the global 21-cm signal given the posterior density of the astrophysical parameters conditioned on the data sets described in \Cref{Sec:DataSetsAndMeasurements}, we also use the \textsc{globalemu} framework. Our $T_{21}$ emulator matches the one used to infer astrophysical constraints given upper limits on the global 21-cm signal in B24.

In addition, we use a 21-cm power spectrum emulator\footnote{For the purpose of imposing the astrophysical constraints derived from upper limits on the 21-cm power spectrum measured by HERA, LOFAR and the MWA, we circumvent the need to fit these data directly, in our joint analysis, by training an NDE on posterior samples from the analysis of the interferometric data sets performed in B24 (see \Cref{Sec:NDE} for details).} to derive the 21-cm power spectrum sensitivity plots in \Cref{Fig:SensitivityPlotxHI_T21_21cmPS} and the posterior PD of the 21-cm power spectrum. This emulator matches the one used to infer astrophysical constraints from the upper limits on the 21-cm power spectrum from HERA in H22.

The evaluation time for each of these emulators, on a Macbook Pro with an M1 Pro chip, is of order a few milliseconds. This results in a roughly six order of magnitude improvement in the computational efficiency of our analysis.

\subsection{Data likelihoods \& priors on derived parameters}
\label{Sec:DataLikelihoods}

In this section, we outline the likelihood functions and priors on derived parameters. These are used to incorporate each of the data sets or constraints discussed in \Cref{Sec:DataSetsAndMeasurements} into our joint analysis. This follows the approach detailed in \Cref{Sec:BayesianInference}. A summary of the data sets, along with the corresponding likelihood functions and prior distributions, is provided in \Cref{Tab:DataSets}.

\subsubsection{Notation}
\label{Sec:Notation}

For any given redshift-dependent observable, say $o(z)$, throughout this section,  we use $o^\mathrm{d}(z)$ to denote a data-derived estimate of the observable at redshift $z$ and $o(z, \sTheta)$ to denote our model for the observable at redshift $z$, given sampled astrophysical parameters, $\sTheta$.

\subsubsection{Neural density estimators}
\label{Sec:NDElike}

\label{Sec:NDE}

For an in depth description of density estimators, including NDEs, see \citet{2022arXiv220711457B, 2023MNRAS.526.4613B}. In brief, NDEs can be used to model probability distributions given a set of representative samples. Once trained, an NDE can be used to evaluate the logarithm of the probability of a set of samples on the learned distribution. As probability emulators, NDEs can be used to perform efficient joint analysis of constraints from multiple different experiments probing the same core science with different nuisance
parameters.

We use this property to circumvent the need to directly fit upper limits on the 21-cm power spectrum measured by HERA, LOFAR and the MWA by instead using an equivalent joint likelihood provided by an NDE trained on the posterior distribution of the parameters given these data sets, derived in B24. Specifically, we train the NDE using the \textsc{margarine} software package (\citealt{2022arXiv220711457B, 2023MNRAS.526.4613B}) on posterior samples from the joint analysis of the interferometric data sets performed in B24. We use the trained NDE to emulate the joint loglikelihood of the 21-cm power spectrum upper limit given our model:
\begin{multline}
\log(L_\mathrm{NDE, \Delta^{2}_{21}}(\bm{\theta})) = \\ \log(L_\mathrm{HERA}(\bm{\theta})) + \log(L_\mathrm{LOFAR}(\bm{\theta})) + \log(L_\mathrm{MWA}(\bm{\theta})) \ ,
\end{multline}
where $\bm{\theta} = [V_\mathrm{c}, f_{*}, f_\mathrm{X}, \tau_\mathrm{CMB}, f_\mathrm{radio}]$ is a vector of the astrophysical parameters of the model. $\log(L_\mathrm{HERA})(\bm{\theta})$, $\log(L_\mathrm{LOFAR})(\bm{\theta})$ and $\log(L_\mathrm{MWA})(\bm{\theta})$, are the log-likelihoods of the HERA, LOFAR and MWA upper limit data sets, respectively, given our model for the data evaluated with the set of parameters, $\bm{\theta}$.

Use of the NDE likelihood facilitates a factor-of-several improvement in computational efficiency relative to explicit evaluation of the applicable 21-cm power spectrum upper limit likelihoods for each of the data sets. Furthermore, it simplifies the data analysis by replacing the three likelihoods with a single joint likelihood.

Additionally, we use an equivalent procedure to train an NDE on samples\footnote{\url{https://pla.esac.esa.int/}} from the Planck TT,TE,EE+low$l$+lowE+lensing posterior. Marginal parameter posteriors deriving from the NDE likelihood models for these two data set combinations are illustrated in Figures \ref{Fig:B24Posteriors2} and \ref{Fig:Plancktauposterior2}, respectively.

\subsubsection{LAE upper limits likelihoods}
\label{Sec:MHGlike}

As discussed in \Cref{Sec:LymanLine}, the fraction of \Lya\ and \Lyb\ forest pixels in the spectra of bright high-redshift quasars that is dark, due to absorption of the quasar emission by neutral hydrogen, measures the presence of neutral hydrogen along the line of sight between us and the quasar and provides an upper limit on the sky-averaged IGM neutral fraction. \citet{2015MNRAS.447..499M} derive redshift-dependent estimates of the dark pixel fraction, $f_\mathrm{dp}$, with $1 \sigma$ errors, $\sigma_{f_\mathrm{dp}}$, from a sample of 20 high-S/N quasar spectra. For the $N=3$ independent estimates of $f_\mathrm{dp}$, we incorporate these constraints in our joint analysis using a contaminated upper limit (CUL) likelihood of the form (see \Cref{Sec:UpperLimitLikelihood}):
\begin{multline}
    \mathcal{L}_\mathrm{CUL,dp}(\sTheta) = \prod\limits_{i=1}^{N} \frac{1}{2} \Bigg[ \mathrm{erf}\left(\frac{(1 + \overline{x}_\mathrm{H_{I}}(\sTheta, z_{i}) - f_\mathrm{dp}(z_{i}))}{\sqrt{2}\sigma_{f_\mathrm{dp}}}\right) \\
    - \mathrm{erf}\left(\frac{(\overline{x}_\mathrm{H_{I}}(\sTheta, z_{i}) - f_\mathrm{dp}(z_{i}))}{\sqrt{2}\sigma_{f_\mathrm{dp}}}\right) \Bigg] \ ,
\label{Eq:CULMarginalLikelihoodDP}
\end{multline}
where $\overline{x}_\mathrm{H_{I}}(\sTheta, z_{i})$ is an emulator estimate of the sky averaged IGM neutral fraction at the redshifts $z_{i}$. The corresponding three upper limit constraints are illustrated in orange in \Cref{Fig:LymanLineConstraints2}.

\citet{2015MNRAS.453.1843S} derive an estimate of the IGM neutral fraction from measurements of the LAE ACF at $z = 6.6$ of $\hat{\overline{x}}_\mathrm{H_{I}, UL}(z=6.6) = 0.3$, with an estimated uncertainty of $\sigma_{\hat{\overline{x}}_\mathrm{H_{I}, UL}}(z=6.6) = 0.2$. The corresponding $1 \sigma$ upper limit is thus $\hat{\overline{x}}_\mathrm{H_{I}, UL}(z=6.6) = 0.5$. We incorporate this constraint in our joint analysis, using a contaminated upper limit likelihood of the form:
\begin{multline}
    \mathcal{L}_\mathrm{CUL,ACF}(\sTheta) = \frac{1}{2} \Bigg[ \mathrm{erf}\left(\frac{(1 + \overline{x}_\mathrm{H_{I}}(\sTheta, z=6.6) - \hat{\overline{x}}_\mathrm{H_{I}, UL})}{\sqrt{2}\sigma_{\hat{\overline{x}}_\mathrm{H_{I}, UL}}}\right) \\
    - \mathrm{erf}\left(\frac{(m(\sTheta) - \hat{\overline{x}}_\mathrm{H_{I}, UL})}{\sqrt{2}\sigma_{\hat{\overline{x}}_\mathrm{H_{I}, UL}}}\right) \Bigg] \ .
\label{Eq:MarginalLikelihoodACF}
\end{multline}
The corresponding upper limit constraint is illustrated in red in \Cref{Fig:LymanLineConstraints2}.

\subsubsection{IGM neutral fraction PDFs}
\label{Sec:SplinePDFlike}

We use the full $\overline{x}_\mathrm{H_{I}}^\mathrm{d}$ PDFs for Lyman line analyses, where available. We assume the redshift uncertainties associated with the LAE-derived $\overline{x}_\mathrm{H_{I}}^\mathrm{d}(z)$ PDF are sufficiently small to neglect for our purposes when point estimates of the relevant redshift, alone, are provided in the original analyses. We use the IGM neutral fraction PDFs in this set as priors on the IGM neutral fraction, $\overline{x}_\mathrm{H_{I}}(z_{i})$, of our model, where $z_{i}$ is the central redshift of the LAE-derived constraint. For the $i$th PDF, we define the corresponding prior as follows:
\begin{equation}
    \label{Eq:PriorxHILAE}
    \pi_{\mathrm{LAE}_i}(\overline{x}_\mathrm{H_{I}}(z_{i}, \sTheta)) = \phi_{i}(\mathcal{P}(\overline{x}_\mathrm{H_{I}}^\mathrm{d}(z_{i}) \vert \bm{D}_{i})) \ .
\end{equation}
Here, $\phi_{i}(\mathcal{P}(\overline{x}_\mathrm{H_{I}}^\mathrm{d}(z_{i}) \vert \bm{D}_{i}))$ is a spline fit to the posterior PDF of $\overline{x}_\mathrm{H_{I}}^\mathrm{d}(z_{i})$, given the $i$th LAE data set $\bm{D}_{i}$, and $\overline{x}_\mathrm{H_{I}}(z_{i}, \sTheta)$ is our emulator estimate of the modelled sky averaged IGM neutral fraction at redshift $z_{i}$, given astrophysical parameters $\sTheta$.

Where the redshift uncertainty associated with the LAE-derived $\overline{x}_\mathrm{H_{I}}^\mathrm{d}(z)$ PDF is reported, we add the redshift of the constraint as an additional free parameter of our model and computationally marginalise over the redshift uncertainty (see \Cref{Sec:LyaRedshiftMarginalisation} for details). Throughout the remainder of the paper, we use the marginal posteriors on the parameters marginalised over redshift uncertainty in these cases, in our joint analysis.

\subsubsection{$\tau_\mathrm{CMB}$ PDF}
\label{Sec:tauLikelihood}

Since $\tau_\mathrm{CMB}$ is a sampled parameter of our model, the Planck TT,TE,EE+low$l$+lowE+lensing one-dimensional marginal constraint can be trivially included in our analysis as a parameter prior. When including this constraint, we use the Planck TT,TE,EE+low$l$+lowE+lensing one-dimensional marginal posterior on $\tau_\mathrm{CMB}$ as our prior in place of the uniform $\tau_\mathrm{CMB}$ PDF listed in \Cref{Tab:Priors}.

\subsubsection{$\Delta z_\mathrm{re}$ PDF}
\label{Sec:DeltazReLikelihood}

We use the R21 posterior on $\Delta z_\mathrm{re}$ given the SPT measurements of the patchy kSZ power spectrum as a constraint on the duration of reionization in our joint analysis (see \Cref{Sec:CMB}). We use a spline fit to $\mathcal{P}(\Delta z_\mathrm{re}^\mathrm{d} \vert D^\mathrm{p-kSZ}_{3000})$ (see \Cref{Fig:SPTdzrePosterior2}) as a prior on the value of $\Delta z_\mathrm{re}(\sTheta)$ predicted by our model, such that:
\begin{equation}
    \label{Eq:PriorDzre}
    \pi_{\Delta z_\mathrm{re}}(\Delta z_\mathrm{re}(\sTheta)) = \phi(\mathcal{P}(\Delta z_\mathrm{re}^\mathrm{d} \vert D^\mathrm{p-kSZ}_{3000})) \ .
\end{equation}
Here, $D^\mathrm{p-kSZ}_{3000}$ is the measured constraint on the patchy kSZ power at an angular multipole scale of $l=3000$, and,
\begin{equation}
    \label{Eq:Dzre}
    \Delta z_\mathrm{re}(\sTheta) = z_{75}(\sTheta) - z_{25}(\sTheta) \ ,
\end{equation}
with $z_{\alpha}$ the redshift at which the sky averaged IGM neutral fraction is $\alpha\%$, defined such that:
\begin{equation}
    \label{Eq:z_percentile}
    \overline{x}_\mathrm{H_{I}}(z_{\alpha}, \sTheta) = \frac{\alpha}{100} \ .
\end{equation}

%%%%%%%%%%%%%%%%%%%%%%%%%%%%%%%%%%%%%%%%%%%%%%%%%%
\section{Results}
\label{Sec:Results}
%%%%%%%%%%%%%%%%%%%%%%%%%%%%%%%%%%%%%%%%%%%%%%%%%%

In this section, we present the results of our individual and joint analyses of the data sets and measurements detailed in \Cref{Sec:DataSetsAndMeasurements}, using the analysis methodology described in \Cref{Sec:AnalysisMethodology}.

\subsection{Information content}
\label{Sec:DKL}

\begin{figure*}
	\centerline{
        \hspace*{1.0cm}
        \begin{subfigure}[t]{0.55\textwidth}
        \caption{\Large{CMB}}
        \includegraphics[height=9.5cm, trim={+0.2cm -0.0cm 0cm -0.3cm}, clip]{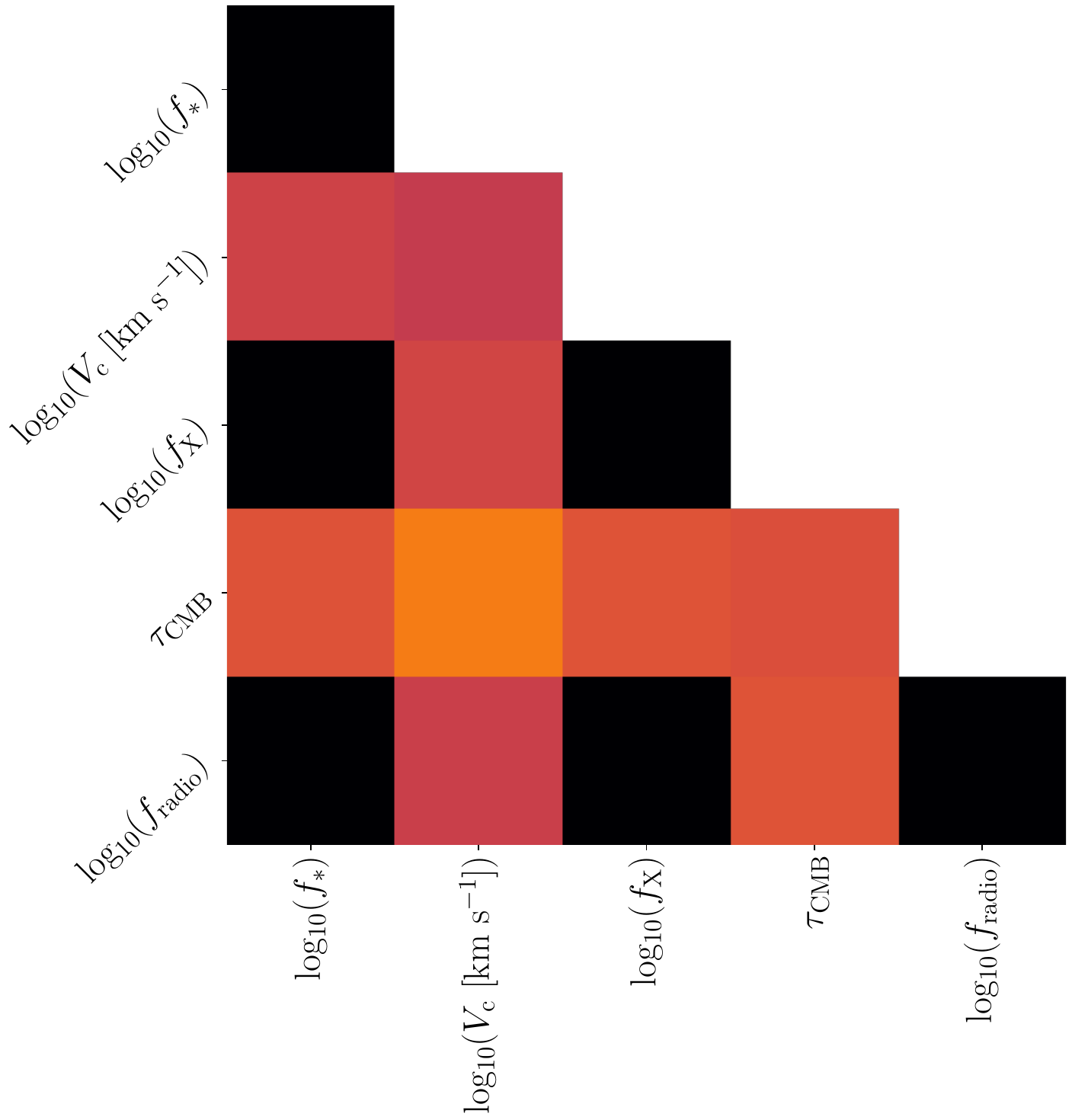}
        \label{Fig:InformationContent2DSummaryCMB}
        \end{subfigure}
        \begin{subfigure}[t]{0.55\textwidth}
        \caption{\Large{Lyman line}}
        \includegraphics[height=9.5cm, trim={0.0cm -0.3cm 0.0cm 0.0cm}, clip]{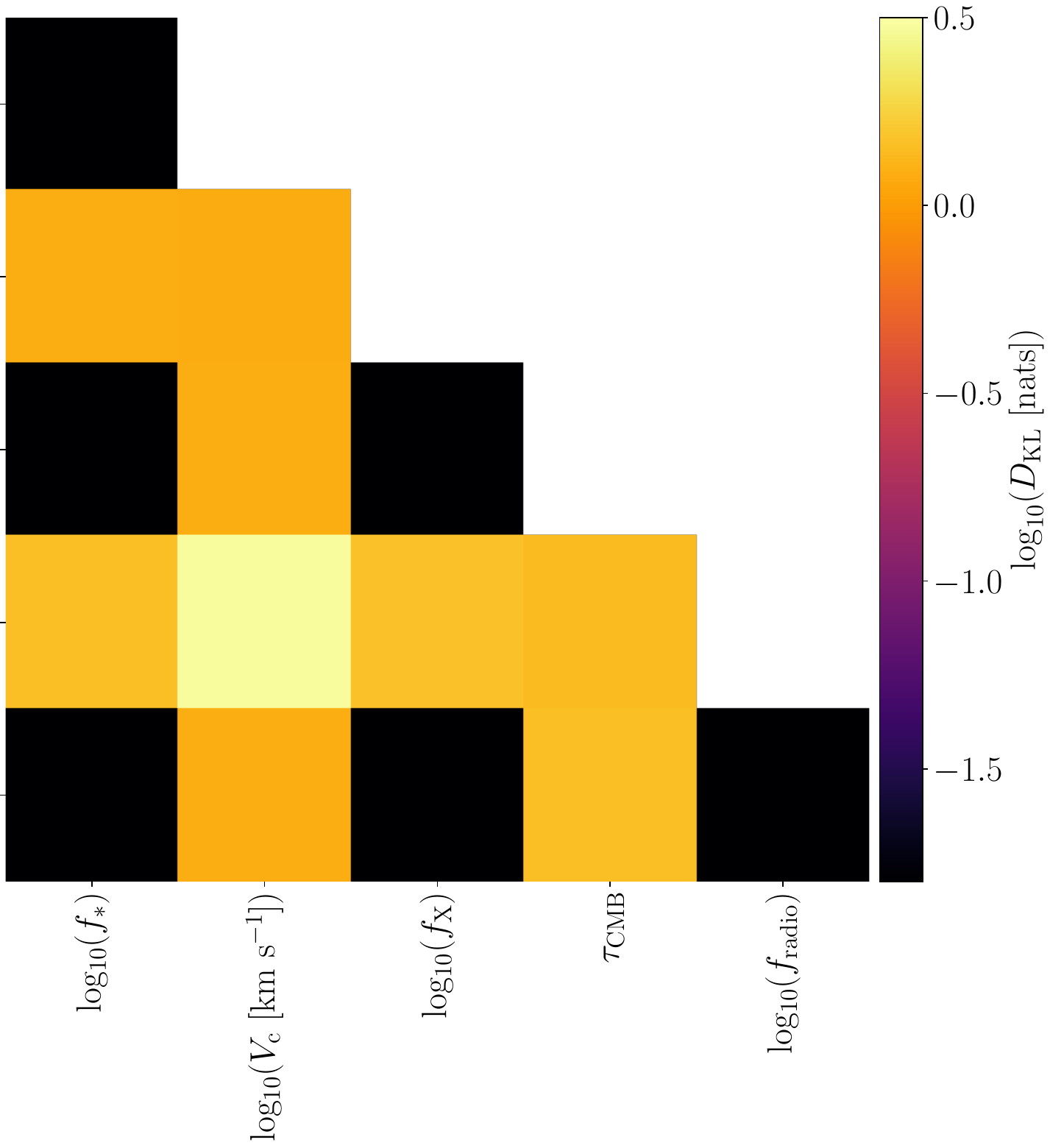}
        \label{Fig:InformationContent2DSummaryLl}
        \end{subfigure}
    }
	\centerline{
        \hspace*{1.0cm}
        \begin{subfigure}[t]{0.55\textwidth}
        \caption{\Large{21-cm}}
        \includegraphics[height=9.5cm, trim={+0.2cm -0.0cm 0cm -0.3cm}, clip]{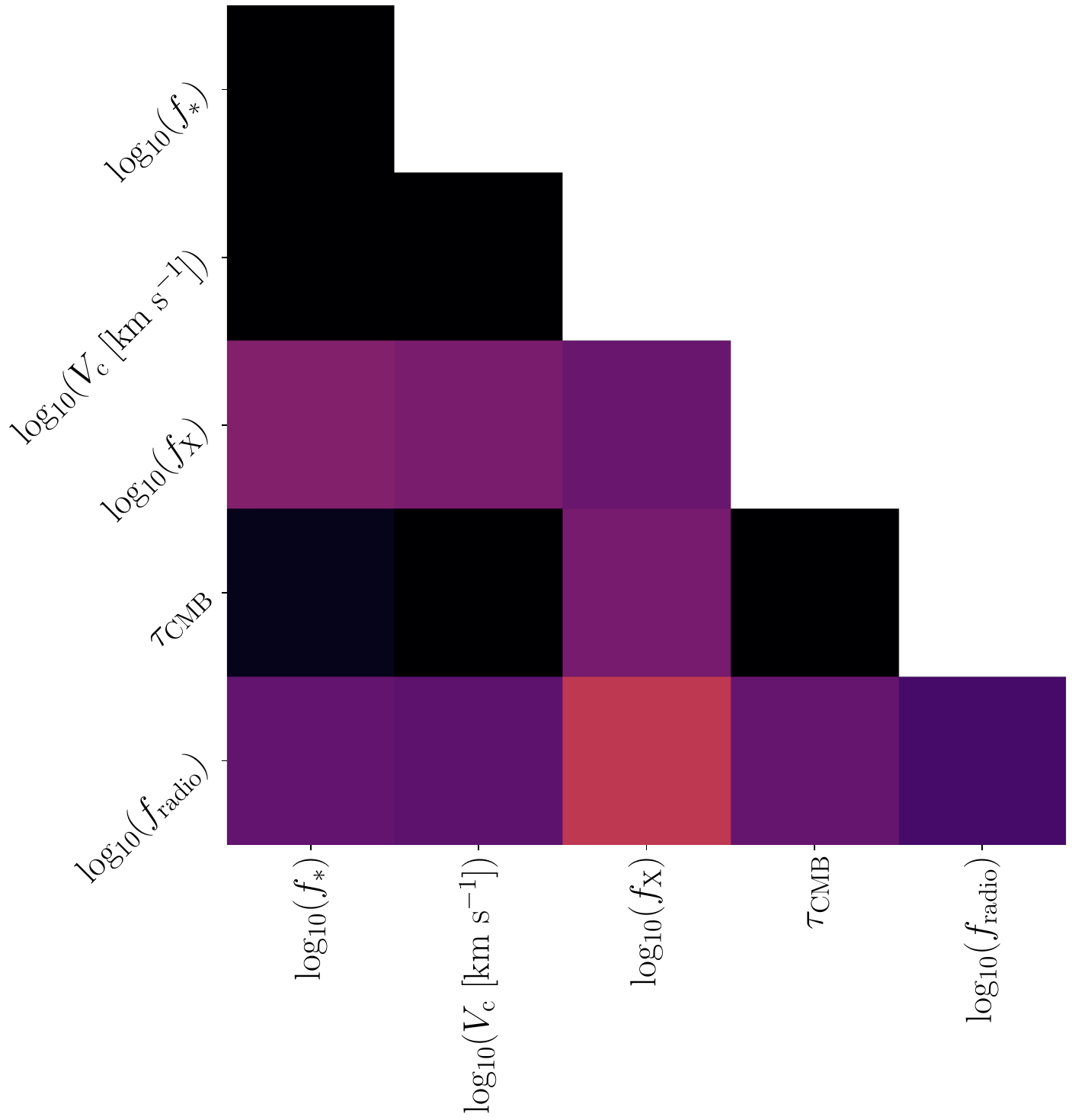}
        \label{Fig:InformationContent2DSummary21cm}
        \end{subfigure}        %
        \begin{subfigure}[t]{0.55\textwidth}
        \caption{\Large{Joint}}
        \includegraphics[height=9.5cm, trim={0.0cm -0.3cm 0.0cm 0.0cm}, clip]{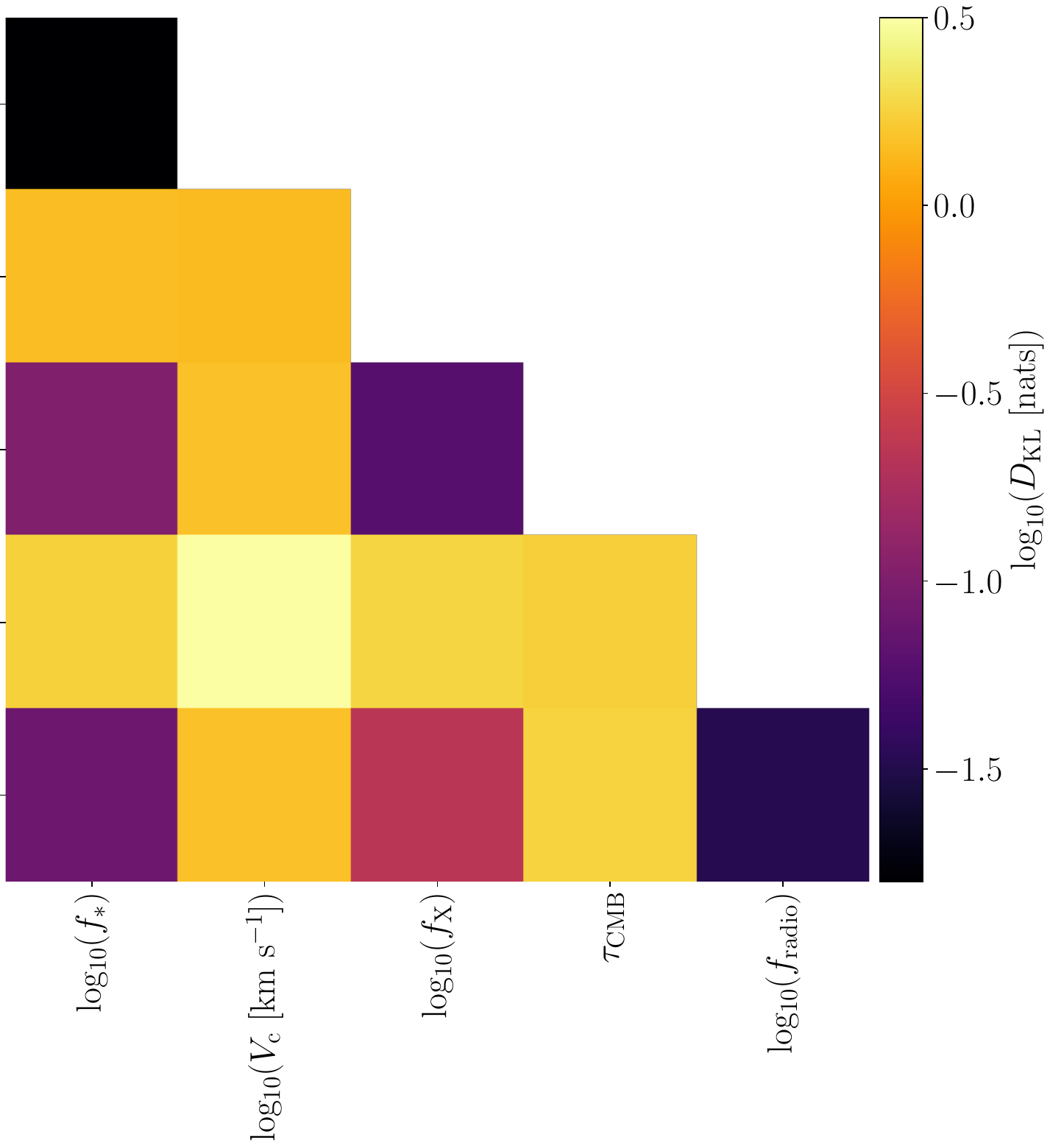}
        \label{Fig:InformationContent2DSummaryJoint}
        \end{subfigure}
    }
    \caption{
        EoR information content provided by the data sets included in our analysis, as measured by the marginal KL divergence ($\mathcal{D}_\mathrm{KL}(\sTheta \vert \bm{D})$) between the priors and posteriors of the one- and two-dimensional probability densities of the astrophysical parameters of the model. \Cref{Fig:InformationContent2DSummaryCMB} shows $\mathcal{D}_\mathrm{KL}(\sTheta \vert \bm{D}_\mathrm{CMB})$, where $\bm{D}_\mathrm{CMB}$ are the Planck 2018 constraints on $\tau_\mathrm{CMB}$ and R21 constraints on $\Delta z_\mathrm{re}$ from SPT data. \Cref{Fig:InformationContent2DSummaryLl} shows $\mathcal{D}_\mathrm{KL}(\sTheta \vert \bm{D}_\mathrm{Ll})$, where $\bm{D}_\mathrm{Ll}$ are the Lyman line constraints on the high-redshift IGM neutral fraction illustrated in \Cref{Fig:LymanLineConstraints2}. \Cref{Fig:InformationContent2DSummary21cm} shows $\mathcal{D}_\mathrm{KL}(\sTheta \vert \bm{D}_\mathrm{21-cm})$, where $\bm{D}_\mathrm{21-cm}$ are the HERA+MWA+LOFAR power spectrum upper limits analysed in B24. \Cref{Fig:InformationContent2DSummaryJoint} shows $\mathcal{D}_\mathrm{KL}(\sTheta \vert \bm{D}_\mathrm{Joint})$, where $\bm{D}_\mathrm{Joint}$ is the combination of the aforementioned constraints. Squares that are black indicate that the corresponding data set contains no information about those parameters (or, equivalently, there is not a statistically significant difference between their prior and posterior distributions). Conversely, the light squares indicate that the data set contains a greater quantity of information that constrains those parameters, as quantified by a significant difference in the posterior distribution of the parameters relative to their priors.
    }
    \label{Fig:InformationContent2DSummary}
\end{figure*}

\begin{figure}
    \centerline{
        \includegraphics[width=0.5\textwidth]{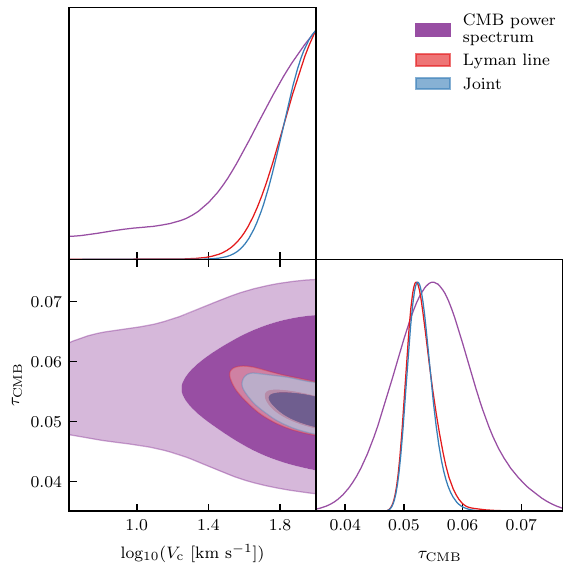}
    }
    \caption{
        One- and two-dimensional posterior probability densities of the circular velocity, $\log_{10}(V_\mathrm{c})$, and the CMB optical depth, $\tau_\mathrm{CMB}$. Posteriors in purple derive from CMB power spectrum constraints (Planck TT,TE,EE+low$l$+lowE+lensing constraint on $\tau_\mathrm{CMB}$ and R21 constraint on $\Delta z_\mathrm{re}$ from SPT data). Those in red derive from Lyman line constraints on the sky averaged IGM neutral fraction (see \Cref{Tab:DataSets}). Those in blue derive from our full joint analysis. The solid and transparent shaded contours plotted in the 2D posteriors contain 68\% and 95\% of the probability, respectively.
        }
    \label{Fig:tauVcPosteriors}
\end{figure}

\begin{figure}
    \centerline{
        \includegraphics[width=0.5\textwidth]{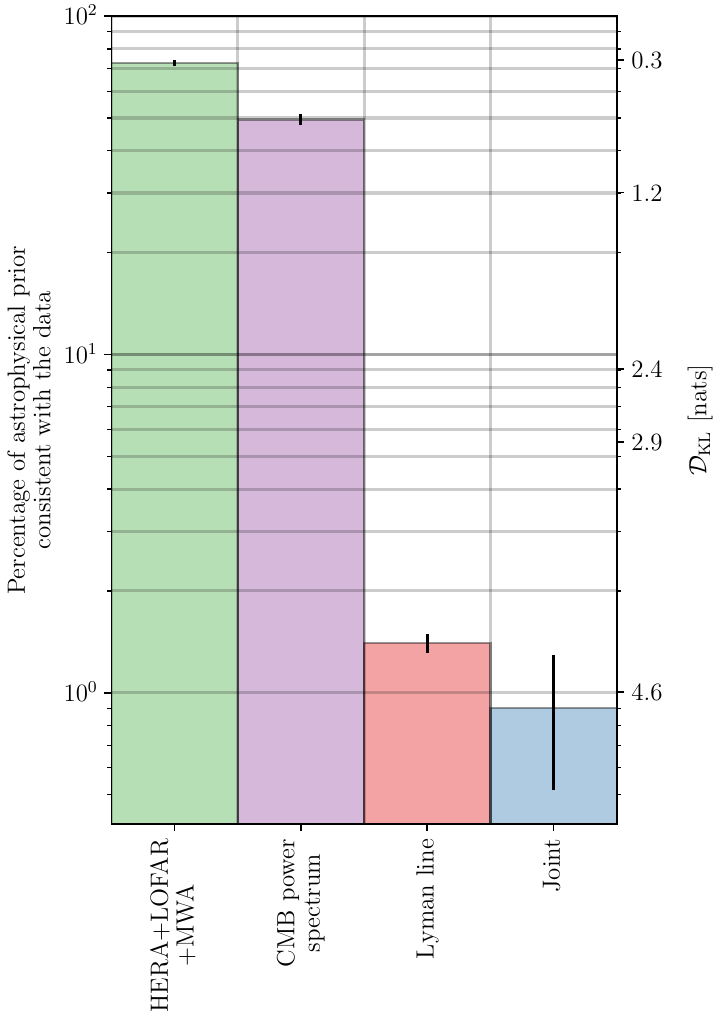}
    }
    \caption{
        The percentage ($100f_\mathrm{c}$; see \Cref{Eq:KLDivergenceFractionalConsistency}) of the five dimensional prior probability density of the astrophysical parameters of our model that is consistent with each data set individually and in combination, as well as their associated $1 \sigma$ uncertainties accounting for sampling and model error. The smaller the percentage of the astrophysical prior consistent with the data, the more constraining that data set is. The corresponding KL divergences between the five dimensional prior and posterior probability densities, from which the percentile constraints are calculated, can be read from the twin y-axis.
    }
    \label{Fig:InformationContentNDSummary}
\end{figure}

In \Cref{Fig:InformationContent2DSummary}, we show triangle plots of the marginal KL divergence (see \Cref{Sec:BayesianInference}) between the priors and posteriors of the one- and two-dimensional probability densities of the astrophysical parameters of our model ($V_\mathrm{c}, f_{*}, f_\mathrm{X}, \tau_\mathrm{CMB}, f_\mathrm{radio}$) given Lyman line constraints on the high-redshift IGM neutral fraction, the P20VI TT,TE,EE+low$l$+lowE+lensing constraint on $\tau_\mathrm{CMB}$ and SPT constraint on $\Delta z_\mathrm{re}$, the HERA+MWA+LOFAR upper limits on the 21-cm power spectrum, and the combination of the aforementioned constraints. Henceforth, we refer to plots of this type as information triangle plots.

Comparing the Lyman line, CMB and 21-cm information triangle plots, one sees that the information pertaining to the astrophysical parameters of CD and the EoR in our model of the high-redshift Universe (hereafter, EoR information content) is not uniformly distributed among the data sets included in our analysis. The Lyman line constraints on the IGM neutral fraction, in aggregate, are the most EoR-information-rich data set in our analysis, with an EoR information content of order several nats. In comparison, CMB power spectrum constraints provide approximately an order of magnitude lower EoR information (a few tenths of a nat), and redshifted 21-cm power spectrum constraints provide a factor of a few lower EoR information than the CMB based constraints (approximately a tenth of a nat).

In addition to the total information content being unevenly distributed between data sets, so too are the specific parameters constrained by the data. The 21-cm power spectrum upper limits constrain $\log_{10}(f_\mathrm{X})$ and $\log_{10}(f_\mathrm{radio})$, providing the most information about the $\log_{10}(f_\mathrm{X})$ - $\log_{10}(f_\mathrm{radio})$ joint PDF. In contrast, the Lyman line and CMB power spectrum data sets both constrain $\log_{10}(V_\mathrm{c})$ and $\tau_\mathrm{CMB}$, providing the most information about the $\log_{10}(V_\mathrm{c})$ - $\tau_\mathrm{CMB}$ joint PDF. The respective data sets thus provide independent information. The significantly different volumes of EoR information content in the Lyman line and CMB power spectrum data sets imply that one should anticipate only a minor improvement on the Lyman line constraints when these two data sets are analysed jointly. However, the qualitative similarity of the information content distribution across parameters means that the CMB power spectrum-based constraints provide a consistency check on the Lyman-line-derived constraints (see \Cref{Fig:tauVcPosteriors}).

In totality, the full range of data sets and measurements considered in this work contains EoR information associated with four of the five sampled parameters of our astrophysical model: $\log_{10}(V_\mathrm{c})$, $\log_{10}(f_\mathrm{X})$, $\tau_\mathrm{CMB}$ and $\log_{10}(f_\mathrm{radio})$. Future inclusion of data sets or measurements that directly or indirectly constrain $f_{*}$, such as measurements of the UV-luminosity function of high-redshift galaxies (e.g. \citealt{2017MNRAS.464.1365M, 2019MNRAS.484..933P, 2025arXiv250321687D}), is thus of interest. Furthermore, incorporating estimates of the cosmic radio and X-ray background (e.g. \citealt{2024MNRAS.531.1113P}), measurements of ionizing photon mean free path constraints from \Lya\ forest data (e.g. \citealt{2021ApJ...917L..37C,2024MNRAS.531.1951C}) and future global 21-cm, 21-cm power spectrum measurements will provide additional EoR information content.

In addition to using information triangle plots to analyse the distribution of EoR information among the parameters of our model, one can also calculate the total EoR information content as the KL divergence between the full 5D parameter prior and posterior probability densities. In \Cref{Fig:InformationContentNDSummary}, we plot these KL divergences and their $1 \sigma$ uncertainties, as well as the approximate corresponding percentage of the astrophysical prior consistent with the data ($100f_\mathrm{c}$; see \Cref{Eq:KLDivergenceFractionalConsistency}). We find that $72.8 \pm 1.6 \%$, $49.5 \pm 1.7 \%$, $1.4 \pm 0.1\%$ and $0.9 \pm 0.4\%$ of our astrophysical prior volume is consistent with the posterior given the HERA+MWA+LOFAR upper limits, CMB power spectrum measurements and constraints, Lyman line measurements and constraints, and their combination, respectively. When calculating the uncertainties on the KL divergences, $\sigma_{\mathcal{D}_\mathrm{KL}} = \sqrt{\sigma_{\mathcal{D}_\mathrm{KL, sample}}^{2} + \sigma_{\mathcal{D}_\mathrm{KL, model}}^{2}}$, we account both for the uncertainty associated with the finite sampling of the probability distributions by our nested sampling analysis, $\sigma_{\mathcal{D}_\mathrm{KL, sample}}$, which we calculate using \textsc{anesthetic} (\citealt{2019JOSS....4.1414H}), and, for data sets including NDE constraints, the error associated with the NDE modelling of the posteriors, $\sigma_{\mathcal{D}_\mathrm{KL, model}}$, which we estimate with \textsc{margarine} (\citealt{2022arXiv220711457B, 2023MNRAS.526.4613B}).

\subsection{Rapid and late reionization driven by massive galaxies}
\label{Sec:tauVcConstraints}

\begin{figure*}
    \centerline{
        \includegraphics[width=1.0\textwidth]{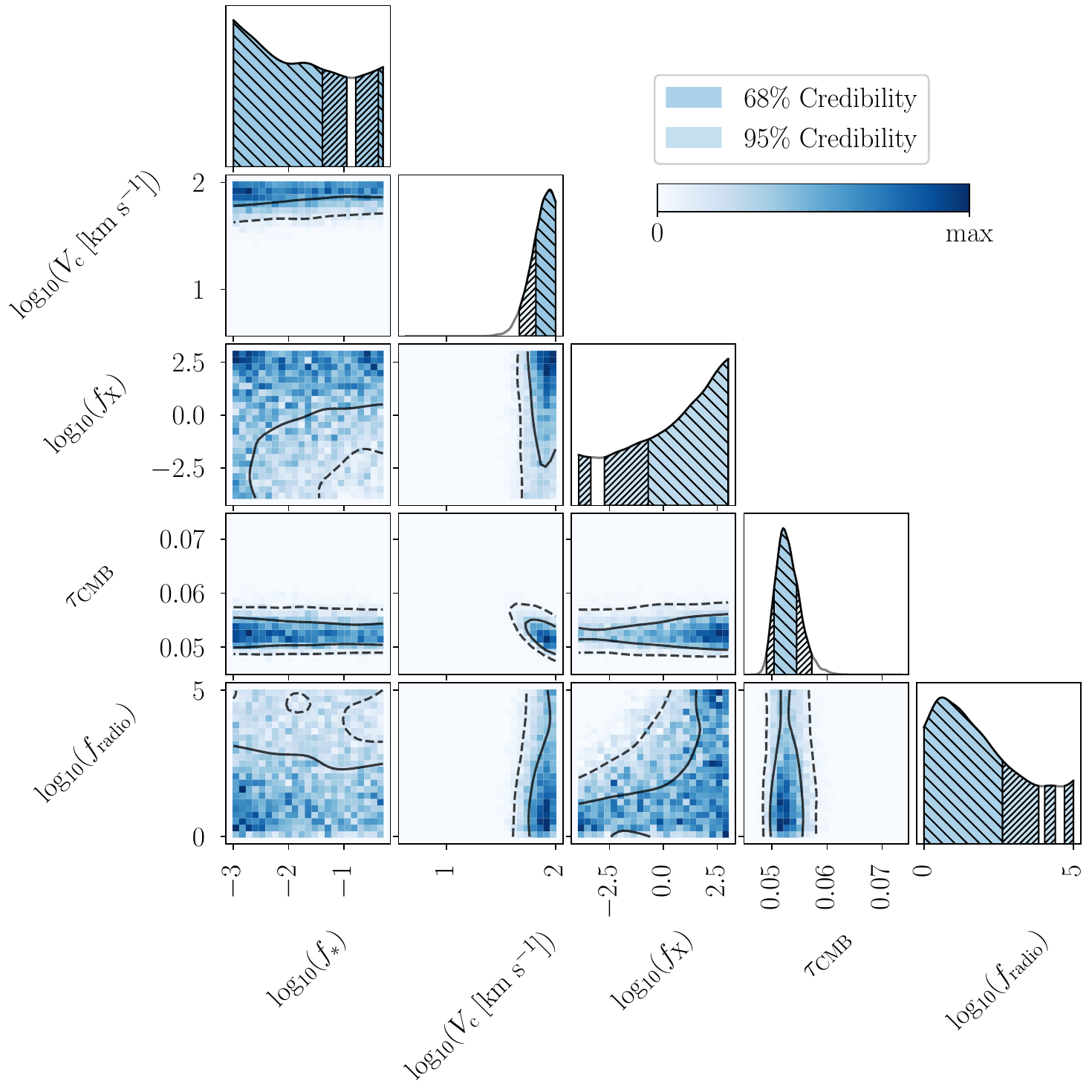}
        }
    \caption{
        One- and two-dimensional marginal posterior probability densities of the astrophysical parameters derived from the combination of upper limits on the power spectrum of redshifted 21-cm emission from the EoR with HERA, the MWA and LOFAR, high-redshift IGM neutral fraction measurements inferred from Lyman line data, and CMB power spectrum constraints on $\tau_\mathrm{CMB}$ from Planck and on $\Delta z_\mathrm{re}$ from SPT. The colorbar displays the peak-normalised, binned probability of histogram bins within the 2D joint posterior densities, where dark blue denotes bins that have maximum posterior probability and white bins those with negligible probability. The sparsely and densely hatched regions of the 1D posteriors and areas enclosed by solid and dashed black lines in the 2D posteriors contain 68\% and 95\% of the probability, respectively, in each case.
        }
    \label{Fig:JointAnalysisPosteriors}
\end{figure*}

\begin{figure*}
    \centerline{
        \includegraphics[width=1.0\textwidth, trim={-2.9cm 0cm 0.0cm 0}, clip]{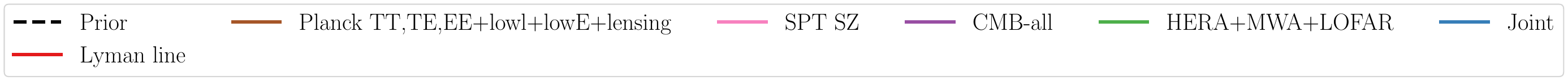}
            }
    \centerline{
        \includegraphics[width=0.33\textwidth]{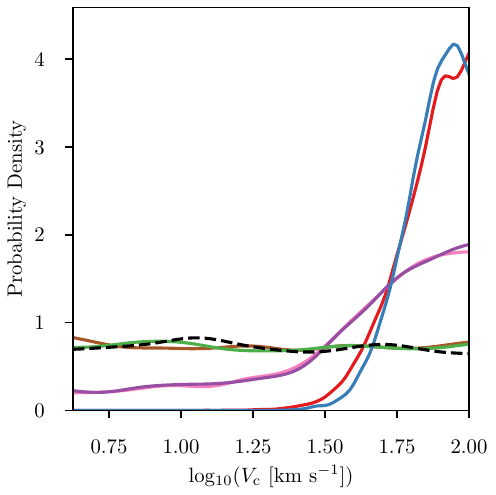}
        \includegraphics[width=0.33\textwidth]{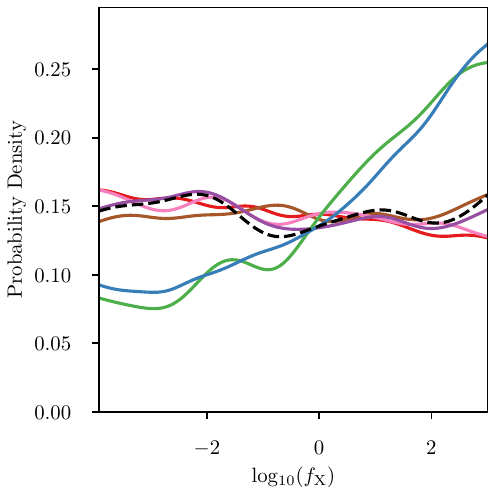}
        \includegraphics[width=0.33\textwidth]{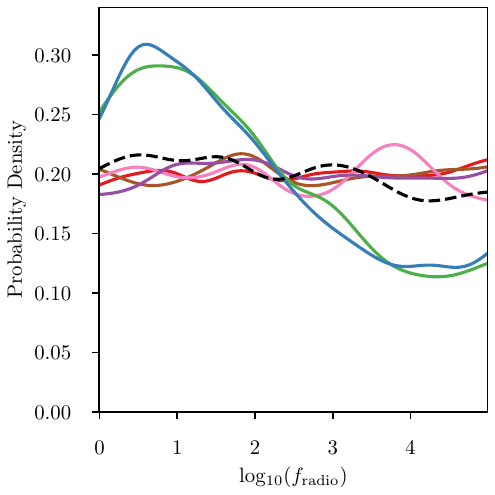}
    }
    \centerline{
        \includegraphics[width=0.33\textwidth]{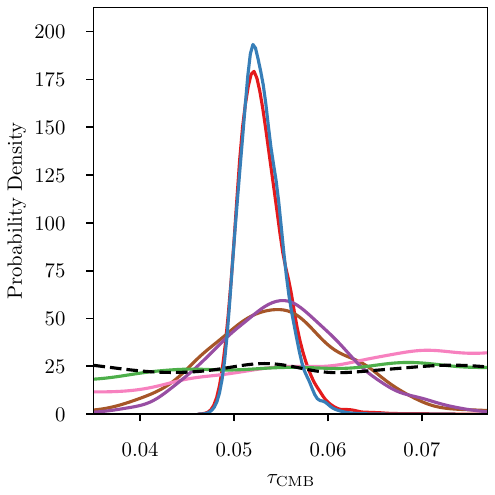}
        \includegraphics[width=0.33\textwidth]{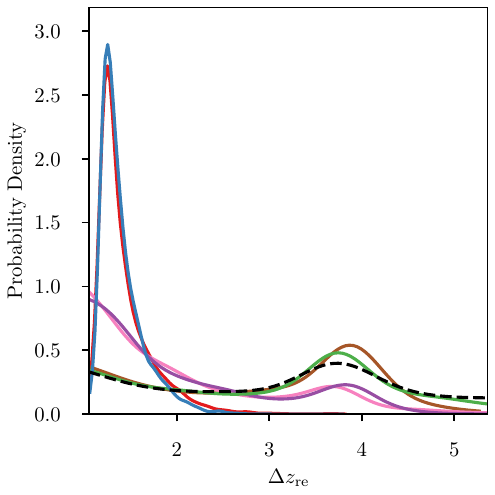}
    }
    \caption{
        Prior and marginal one-dimensional posterior probability densities of $\log_{10}(V_\mathrm{c})$, $\log_{10}(f_\mathrm{X})$, and $\log_{10}(f_\mathrm{radio})$ (top left, middle and right, respectively) and $\tau_\mathrm{CMB}$ and $\Delta z_\mathrm{re}$ (bottom left and right, respectively), given individual and joint constraints from high-redshift data sets as listed in the figure legend. In the context of our model for the data (see \Cref{Sec:SemiNumericalSimulations}) $\log_{10}(V_\mathrm{c})$, $\log_{10}(f_\mathrm{X})$,  $\log_{10}(f_\mathrm{radio})$ and $\tau_\mathrm{CMB}$ are sampled parameters and $\Delta z_\mathrm{re}$ is a derived parameter. $\log_{10}(f_{*})$ is not plotted because it is unconstrained by the data sets considered in this work. Individual or combinations of data sets that do not include the HERA+MWA+LOFAR 21-cm power spectrum upper limits contain minimal information regarding $\log_{10}(f_\mathrm{X})$ and $\log_{10}(f_\mathrm{radio})$ (see \Cref{Sec:DKL}). Resultantly, in all but the HERA+MWA+LOFAR and joint analyses, structure in the marginal one-dimensional posterior probability densities of these two parameters is dominated by the effects of sampling noise. Conversely, the HERA+MWA+LOFAR 21-cm power spectrum upper limits themselves contain minimal information regarding $\log_{10}(V_\mathrm{c})$ and $\tau_\mathrm{CMB}$; thus, the posteriors on these parameters, given this data set, are also dominated by the effects of sampling noise. Nevertheless, for completeness, here we plot marginal posteriors from each analysis for each of the aforementioned parameters.
        }
    \label{Fig:DerivedParamPosteriors}
\end{figure*}

\begin{table*}
    \caption{
        Summary statistics describing the 1D marginal probability densities of the sampled astrophysical parameters of our model given joint constraints from CMB (CMB joint analysis), Lyman line (Lyman line joint analysis) and 21-cm (21-cm joint analysis) data sets and measurements individually and in combination (Full joint analysis). We omit $\log_{10}(f_{*})$, the 1D marginal posterior density of which is unconstrained by the data sets analysed in this work (see \Cref{Sec:DKL}). We list the HPD parameter estimates and uncertainties for non-prior-limited distributions constrained by the data ($\tau_\mathrm{CMB}$). We list 68\% and 95\% credibility upper or lower limits of parameters with prior-limited posteriors (see \Cref{Sec:SummaryStatistics}), in which the probability density is concentrated at the upper or lower bound of the prior, respectively ($\log_{10}(V_\mathrm{c})$, $\log_{10}(f_\mathrm{X})$ and $\log_{10}(f_\mathrm{radio})$).
    }
    \centerline{
    \begin{tabular}{ll|lllll}
        \hline
        Analysis type & Parameter & Parameter estimate & Limit type  & 68\% credibility limit  & 95\% credibility limit     \\
        \hline
        CMB joint analysis             & $\log_{10}(V_\mathrm{c}\ [\mathrm{km~s^{-1}}])$ & - & Lower & $ > 1.5$ & $ > 0.8$ \\
                                       & $\tau_\mathrm{CMB}$ & $0.056^{+0.0055}_{-0.0076}$  & - &  -  & - \\
                                       & $\Delta z_\mathrm{re}$ & - &  Upper &  $ < 2.5$  &  $ < 4.0$ \\
        \hline
        Lyman line joint analysis      & $\log_{10}(V_\mathrm{c}\ [\mathrm{km~s^{-1}}])$ & - & Lower & $ > 1.8$ & $ > 1.7$ \\
                                       & $\tau_\mathrm{CMB}$ & $0.052^{+0.0016}_{-0.0018}$  & - &  -  & - \\
                                       & $\Delta z_\mathrm{re}$ & - &  Upper &  $ < 1.4$  &  $ < 1.8$ \\
        \hline
        21-cm joint analysis           & $\log_{10}(f_\mathrm{X})$ & - &  Lower  & $ > -0.5$  & $ > -3.3$ \\
                                       & $\log_{10}(f_\mathrm{radio})$ & - &  Upper &  $ < 2.6$  &  $ < 4.6$ \\
        \hline
        Full joint analysis            & $\log_{10}(V_\mathrm{c}\ [\mathrm{km~s^{-1}}])$ & - & Lower & $ > 1.8$ & $ > 1.7$ \\
                                       & $\log_{10}(f_\mathrm{X})$ & - &  Lower  & $ > -0.5$  & $ > -3.3$ \\
                                       & $\tau_\mathrm{CMB}$ & $0.052^{+0.0016}_{-0.0018}$  & - &  -  & - \\
                                       & $\log_{10}(f_\mathrm{radio})$ & - &  Upper &  $ < 2.6$  &  $ < 4.5$ \\
                                       & $\Delta z_\mathrm{re}$ & - &  Upper &  $ < 1.4$  &  $ < 1.8$ \\
        \hline
        \end{tabular}
    }
\label{Tab:CredibilityIntervals}
\end{table*}

\Cref{Fig:JointAnalysisPosteriors} shows the one- and two-dimensional marginal posterior probability densities of the astrophysical parameters of our model, derived from the joint analysis of the full range of data sets and measurements discussed in \Cref{Sec:DataSetsAndMeasurements}. \Cref{Fig:DerivedParamPosteriors} shows the one-dimensional marginal prior and posterior probability densities of the sampled and derived parameters of our model as a function of the data set, or combination of data sets, under analysis.

The relative strengths with which parameters are constrained by the data is in qualitative agreement with expectations from the EoR information analysis in \Cref{Sec:DKL}. Specifically, we derive strong constraints on $\tau_\mathrm{CMB}$ and $\log_{10}(V_\mathrm{c})$, moderate constraints on $\log_{10}(f_\mathrm{X})$ and $\log_{10}(f_\mathrm{radio})$ and the marginal posterior on $\log_{10}(f_{*})$ is unconstrained by our joint analysis.

\Cref{Tab:CredibilityIntervals} lists the HPD parameter estimates, including 68\% symmetric HPD credibility intervals for the astrophysical parameters that are well constrained by the data. Additionally, it provides the 68\% and 95\% credibility limits for prior-limited 1D marginal posteriors, given joint constraints from the CMB, Lyman line, and 21-cm datasets individually, as well as from their combination (full joint analysis).

The concentration of posterior density at large \(\log_{10}(V_\mathrm{c})\) in the CMB-power-spectrum-only joint analysis (shown in purple in \Cref{Fig:tauVcPosteriors}) is primarily driven by the SPT constraint on \(\Delta z_\mathrm{re}\) (see \Cref{Fig:DerivedParamPosteriors}, bottom right). When combined with the strong constraints on \(\tau_\mathrm{CMB}\), the joint posterior on \(\log_{10}(V_\mathrm{c})\) and \(\tau_\mathrm{CMB}\) indicates a preference for EoR histories characterised by later and more rapid reionization. Similar conclusions were reached in the analysis of SPT data by \citet{2023MNRAS.526.3170N}.

As discussed in \Cref{Sec:DKL}, the primary information content on \(\tau_\mathrm{CMB}\) and \(\log_{10}(V_\mathrm{c})\) arises from the Lyman line datasets, with only a subdominant contribution from the weaker CMB power-spectrum-based constraints. \Cref{Fig:tauVcPosteriors} confirms that the results of Lyman-line-only analyses are consistent with, but more precise than, those inferred from the CMB-only joint analysis.

However, we note that the constraint on \(\log_{10}(V_\mathrm{c})\) is sensitive to our model assumptions - particularly the assumption of a constant SFE in atomic-cooled halos. We explore this in greater detail in the following section.

\subsubsection{Masses of dark matter halos hosting star forming galaxies}
\label{Sec:HaloMasses}

The redshift-dependent mass of the dark matter halos in which a galaxy forms is proportional to the cube of its circular velocity (see \Cref{Eq:Vc}). The tightly constrained posterior density at high $V_\mathrm{c}$ recovered in our joint analysis ($V_\mathrm{c} \gtrsim 50~\mathrm{km~s^{-1}}$ at $95\%$ credibility; see \Cref{Fig:JointAnalysisPosteriors}) implies that in the considered model, with mass-independent SFE galaxies forming from atomic-cooled hydrogen gas in dark matter halos with masses $M_\mathrm{min} \gtrsim 2.6 \times 10^{9}~M_{\odot} ((1+z)/10)^{\frac{1}{2}}$ are the dominant galactic population required to explain the full set of constraints derived from the CMB, Lyman line, and 21-cm datasets included in our joint analysis.

We note that this constraint aligns with the findings of the \textsc{thesan} large-volume cosmological radiation-magneto-hydrodynamic simulations of the Epoch of Reionization, regarding the galaxies driving the later stages of reionization (e.g., \citealt{2022MNRAS.511.4005K, 2022MNRAS.512.4909G, 2022MNRAS.512.3243S}).
% ---
% Quoting from the conclusions of Yeh et al. 2024:
% ---
These simulations suggest that high-mass halos ($M_\mathrm{h} \gtrsim 10^{10}~M_{\odot}$) dominate the photon budget at $z \lesssim 7$, where the majority of the neutral fraction constraints driving our $V_\mathrm{c}$ inference are found. Furthermore, the most massive galaxies ($M_{*} \ge 10^{9}~M_{\odot}$) play a particularly significant role in the later stages of reionization due to their disproportionately high intrinsic luminosities (\citealt{2023MNRAS.520.2757Y}).

This halo mass is several orders of magnitude larger than the minimum mass required for star formation via molecular-hydrogen-mediated cooling under the assumption of spherical collapse (see \Cref{Sec:SemiNumericalSimulations}). Within the EoR mass and redshift range of interest, assuming that galaxies reside in halos approximately 1--3 orders of magnitude more massive than their stellar masses (e.g., \citealt{2021ApJ...922...29S, 2023MNRAS.520.2757Y}), it is comparable to the expected halo masses of the lowest-mass galaxies detected by JWST in highly magnified gravitationally lensed observations. Examples include the Cosmic Gems arc galaxy at $z_\mathrm{phot} \approx 10.2$, with a lensing-corrected stellar mass in the range of $2.4$--$5.6 \times 10^{7}~M_{\odot}$ (\citealt{2024Natur.632..513A}), and the Firefly Sparkle galaxy at $z_\mathrm{spec} = 8.269 \pm 0.001$, with a lensing-corrected stellar mass of $\log(M_{*}/M_{\odot}) = 7.0^{+1.0}_{-0.3}$ (\citealt{2024arXiv240208696M}). Additionally, it remains 2--4 orders of magnitude smaller than the expected halo masses of the largest galaxies observed at $7 \lesssim z \lesssim 9$ with JWST, which have stellar masses of $\sim 10^{10}~M_{\odot}$ (\citealt{2023Natur.616..266L}).

The tight constraint on $\log_{10}(V_\mathrm{c})$ and $\tau_\mathrm{CMB}$ we derive from our joint analysis constitutes a precise but model-dependent answer to the question: how massive were the halos in which the dominant population of galaxies driving reionization were formed? The confidence placed in these conclusions should therefore be weighted by the credibility of the assumption that the astrophysical processes one anticipates could affect these results have been accurately captured by the model. We note the following approximations that require consideration when drawing conclusions from the $\log_{10}(V_\mathrm{c})$ posterior derived in our analysis. The model assumes:
\begin{enumerate}
    \item a constant SFE as a function of halo mass in atomic-cooled halos, with a logarithmically declining SFE with mass in hydrogen molecular cooling halos (see \Cref{Eq:fstar2}),
    \item the population II and III SFEs are not differentiated\footnote{During the writing of this manuscript, a version of the 21cmSPACE simulation suite was updated to remove this limitation (\citealt{2024MNRAS.531.1113P,2024MNRAS.529..519G}). We refer the interested reader to these papers for additional details.}, and
    \item the galaxy SFE and the escape fraction of ionizing radiation are constant with redshift.
\end{enumerate}

To understand the impact of these approximations on the conclusions that can be drawn from our parameter constraints, we first consider the underlying astrophysical mechanism by which our constraint on $\log_{10}(V_\mathrm{c})$ comes about. \Cref{Fig:SensitivityPlotxHI_T21_21cmPS} illustrates the impact on the IGM neutral fraction history of changing $\log_{10}(V_\mathrm{c})$, given otherwise fixed galaxy properties. Increasing $\log_{10}(V_\mathrm{c})$ steepens the redshift evolution of $\overline{x}_\mathrm{H_{I}}$, shortening the modelled duration of reionization. In models with higher $\log_{10}(V_\mathrm{c})$ and fixed $\tau_\mathrm{CMB}$, this occurs due to later onset and more rapid reionization driven by smaller numbers of high mass galaxies. Such a rapid reionization is preferred both by the Lyman line and the kSZ data fit in this work (\Cref{Fig:xHIContour}, bottom panel, and \Cref{Fig:SPTdzrePosterior2}, respectively). In contrast, in lower $\log_{10}(V_\mathrm{c})$ models, the contributions of a larger number of smaller, earlier forming galaxies leads to a more gradual reionization of the IGM that begins at higher redshift, which is disfavoured by the data.

While the conclusion of a preference for a late onset and rapid reionization epoch is data driven, the inferences one can draw from it are model dependent. This is because, for a fixed SFE as per assumption (i), the formation of galaxies in lower-mass halos would lead to an extended reionization that begins at higher redshift. However, if, in contrast, the SFE was modelled as declining in lower-mass atomic cooling halos, a non-zero stellar contribution from lower-mass galaxies could be consistent with a similarly rapid reionization of the IGM and the greater the drop-off in SFE with mass, the greater the number of such low mass galaxies that can be accommodated by the data.

Such an anticorrelation between the SFE and halo mass is theoretically well motivated. Feedback from supernova-driven winds, radiation pressure, and cosmic rays can expel large amounts of gas from less massive halos with low escape velocities, reducing the availability of fuel for star formation. In the highest mass halos, feedback from active galactic nuclei can heat the gas, preventing it from cooling enough to form stars. In combination, these effects are expected to result in a peaked SFE curve characterised by increasing SFE with halo mass up to a maximum ($M_\mathrm{halo} \sim 10^{12}~M_{\odot}$, at high redshift), where the combination of negative feedback effects is minimised, and a subsequent decline in the most massive halos (\citealt{2018MNRAS.477.1822M}). Additionally, \citet{2017MNRAS.464.1365M} found that this form of SFE evolution was preferred in high-redshift galaxy models when requiring that the model self-consistently reproduced the available ultraviolet luminosity function (UVLF) data. However, in contrast, we also note that the high-resolution, cosmological hydrodynamical simulation from the Feedback in Realistic Environments (FIRE) project by \citet{2017MNRAS.464.1365M} successfully reproduces the observed cosmic UV luminosity density at $z \sim 6-14$ and finds that the SFE-halo mass relation for intermediate mass halos ($M_\mathrm{halo} \sim 10^{9} - 10^{11}~M_{\odot}$) does not significantly evolve with redshift and is only weakly mass-dependent. An update to the 21cmSPACE SFE evolution with redshift given currently available UVLF data (including new JWST observations) is currently in progress (\citealt{2025arXiv250321687D}).

Furthermore, one can expect to reproduce a similar effect by relaxing assumption (ii), modelling population II and population III SFE independently, and accounting for potential supernovae-feedback-induced delay between population III and population II star formation. In this latter case, one can allow for early population III star formation in lower-mass halos, while still expecting to recover a similarly rapid reionization history to that required by Lyman line and preferred by kSZ data, as long as such galaxies are characterised by a lower SFE.

Finally, assumption (iii) means that the increasingly stochastic, bursty, nature of star formation at high redshifts implied by JWST survey data (e.g. \citealt{2023arXiv230602470L, 2023arXiv231210152C}), and proposed as a solution to account for the higher number density of high redshift galaxies at the bright end of the UVLF (e.g. \citealt{2023MNRAS.521..497M}), is not included in our model. This stochasticity will plausibly impact the expected power spectrum of 21-cm fluctuations; however, we leave more detailed investigation of this to future work. Additionally, recent work suggests that models including a mass-dependent SFE combined with a mass- and redshift-dependent escape fraction can provide a better fit to JWST UV-luminosity function data (\citealt{2025arXiv250321687D}) and to \Lya\ opacity constraints from \Lya\ forest data relative to equivalent models in which the escape fraction is redshift independent (\citealt{2024arXiv241200799Q}). This provides motivation for future work jointly analysing the data sets considered in this work in combination with UVLF and \Lya\ opacity constraints to refine the conclusions drawn here.

\subsubsection{IGM neutral fraction history}
\label{Sec:IGMneutralFractionConstraints}

\begin{figure}
    \centerline{
        \includegraphics[width=0.5\textwidth, trim={0 1.3cm 0 0}, clip]{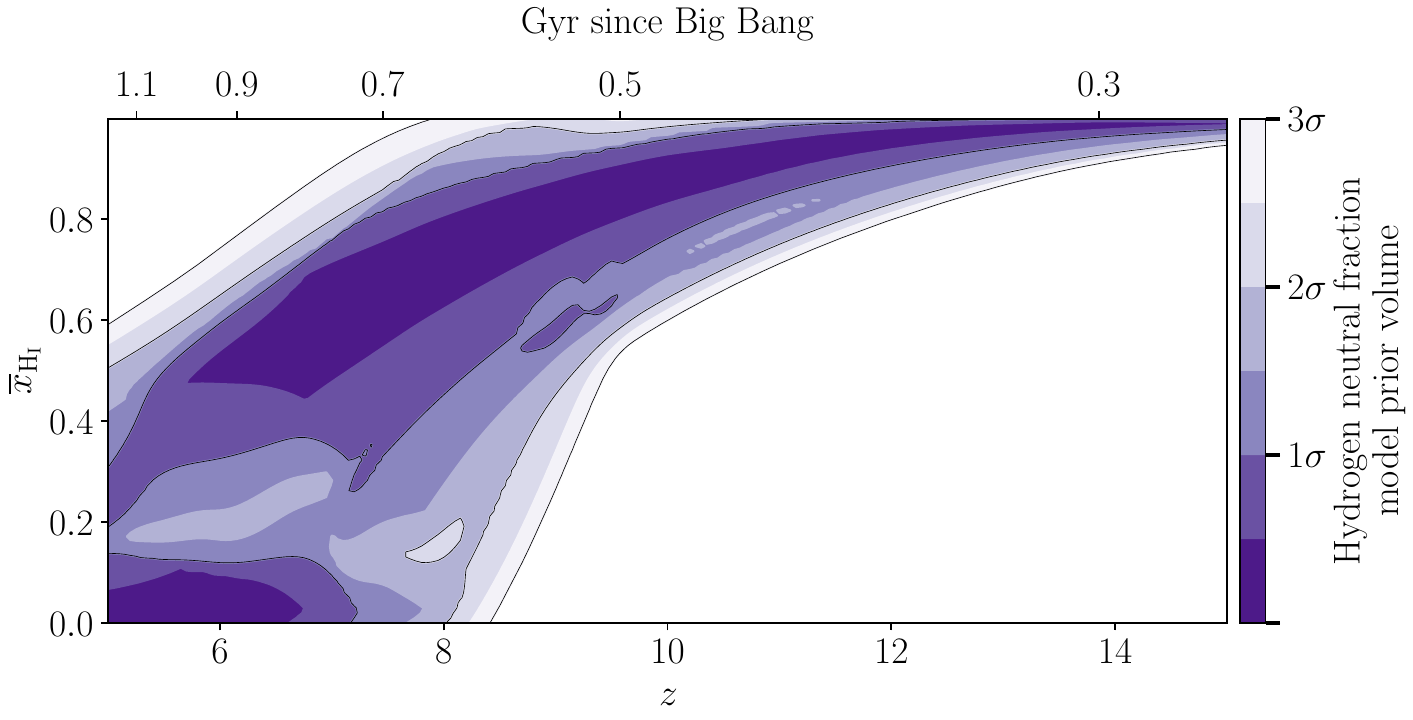}
    }
    \centerline{
        \includegraphics[width=0.5\textwidth, trim={0 1.3cm 0.2cm 1.5cm}, clip]{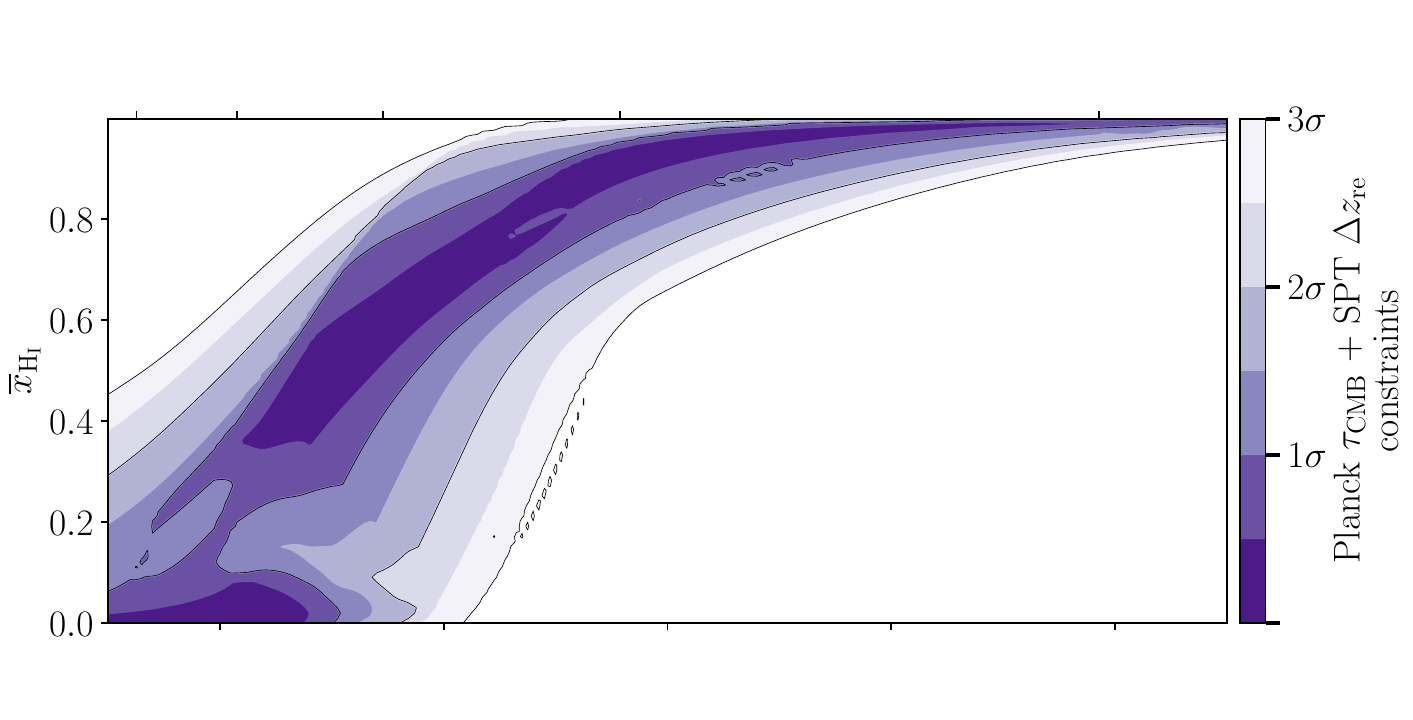}
    }
    \centerline{
        \includegraphics[width=0.5\textwidth, trim={0 0  0.2cm 1.5cm}, clip]{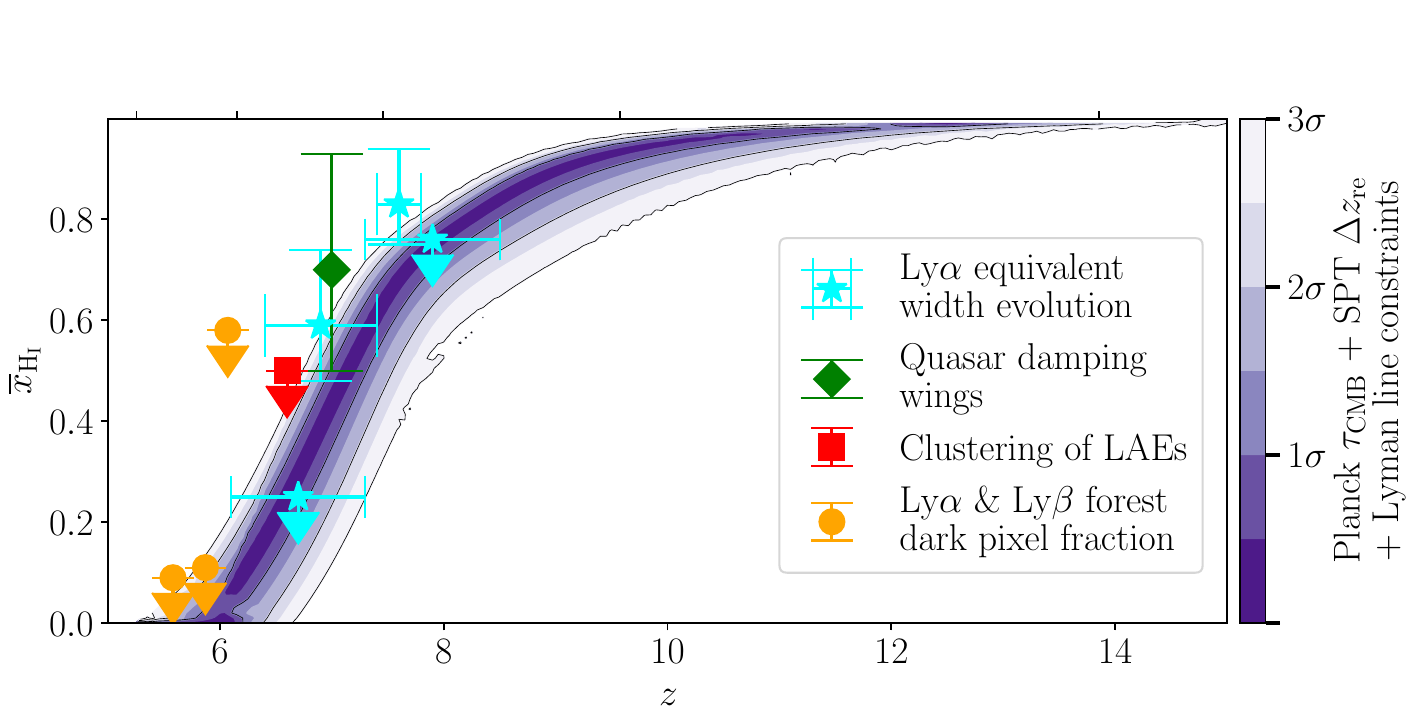}
    }
    \caption{
        Prior (top) and posterior (middle and bottom) PDs of the redshift-dependent IGM neutral fraction. Middle: The posterior PD incorporates constraints on $\tau_\mathrm{CMB}$ from Planck TT,TE,EE+low$l$+lowE+lensing. It also includes inferred constraints on the duration of reionization from patchy kSZ power spectrum measurements by SPT.
        Bottom: The posterior PD combines these constraints with high-redshift IGM neutral fraction measurements inferred from Lyman line data. Points with $1 \sigma$ error bars display constraints on the IGM neutral fraction from the Lyman line analyses indicated in the figure legend. The 21-cm-power-spectrum-based datasets considered in this work place negligible constraints on $x_\mathrm{H_{I}}(z)$, yielding a posterior PD indistinguishable from the prior PD. Therefore, we do not include it here.
    }
    \label{Fig:xHIContour}
\end{figure}

\begin{figure}
    \centerline{
        \includegraphics[width=0.5\textwidth, trim={0 0 0.1cm 0}, clip]{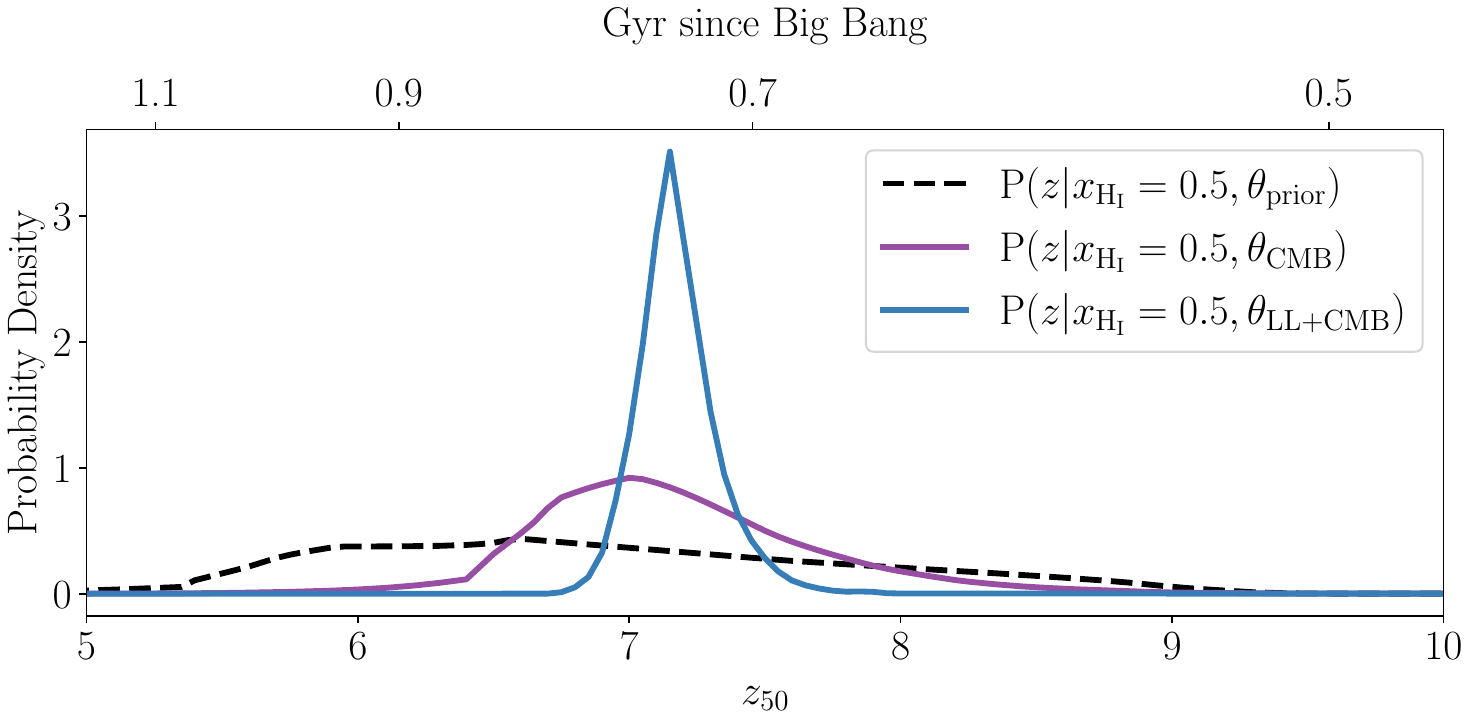}
        }
    \caption{
        Marginal one-dimensional posterior PD of the mid-point of reionization assuming probability density functions for the model parameters that are given by: (i) our model priors (black dashed), (ii) the posteriors on the parameters given the Planck TT,TE,EE+low$l$+lowE+lensing constraint on $\tau_\mathrm{CMB}$ and the R21 constraints on $\Delta z_\mathrm{re}$ from SPT data (purple), (iii) the posteriors on the parameters given the joint constraint from measurements of the IGM neutral fraction from Lyman continuum observations, the Planck TT,TE,EE+low$l$+lowE+lensing constraint on $\tau_\mathrm{CMB}$ and the R21 constraints on $\Delta z_\mathrm{re}$ from SPT data (blue). The 21-cm-power-spectrum-based datasets considered in this work place negligible constraints on the mid-point of reionization, yielding a posterior indistinguishable from the prior. Therefore, we do not include it here.
        }
    \label{Fig:zreConstraints}
\end{figure}

\Cref{Fig:xHIContour} shows the prior (top) and posterior PD of $x_\mathrm{H_{I}}(z)$ given CMB-based constraints (measurement of $\tau_\mathrm{CMB}$ inferred from Planck TT,TE,EE+low$l$+lowE+lensing data and constraints on the duration of reionization inferred from measurements of the patchy kSZ power spectrum with SPT; middle) and constraints derived from the analysis of the combination of these data sets and Lyman-line-data-based high-redshift IGM neutral fraction measurements (bottom). We overlay the individual Lyman-line-based constraints listed in \Cref{Tab:DataSets}, categorised by technique, on the posterior PD of $x_\mathrm{H_{I}}(z)$ from the joint analysis, illustrating the mutual consistency between the respective data points and the model.

With the model parametrisation considered in this work, the IGM neutral fraction history has a strong dependence on $\log_{10}(V_\mathrm{c})$ and on $\tau_\mathrm{CMB}$ and is approximately independent of our remaining three sampled astrophysical parameters (see \Cref{Fig:SensitivityPlotxHI_T21_21cmPS}). The 21-cm-power-spectrum-based datasets considered in this work contain negligible information about $\log_{10}(V_\mathrm{c})$ and $\tau_\mathrm{CMB}$ (see \Cref{Fig:InformationContent2DSummary}). Consequently, they provide limited insight into the IGM neutral fraction history and we do not consider them further in this section. Additionally, the $\log_{10}(V_\mathrm{c})$ and $\tau_\mathrm{CMB}$ Lyman line + CMB information content and constraints are dominated by the Lyman line constraints (\Cref{Fig:InformationContent2DSummary,Fig:DerivedParamPosteriors}). The posterior PD of $x_\mathrm{H_{I}}(z)$ given the Lyman line constraints and that given the Lyman line + CMB constraints are also comparable. Therefore, going forward we focus on the latter constraint, which is marginally stronger.

Comparing the prior and posterior PDs of $x_\mathrm{H_{I}}(z)$ given CMB power spectrum measurements, one sees that the enhanced constraints on $\tau_\mathrm{CMB}$ and $\log_{10}(V_\mathrm{c})$ that the CMB power spectrum measurements yield (see \Cref{Fig:tauVcPosteriors} and \Cref{Tab:CredibilityIntervals}) induce a moderate concentration of the posterior PD of $x_\mathrm{H_{I}}(z)$, decreasing uncertainty in $x_\mathrm{H_{I}}(z)$ throughout the EoR and shifting the midpoint of reionization to higher redshifts in \Cref{Fig:xHIContour}.
In \Cref{Fig:zreConstraints} we show the corresponding one-dimensional PDF of the mid-point of reionization, $\mathrm{P}(z_{50} | \theta)$, which clearly illustrates these effects (here, $\mathrm{P}(z_{50} | \theta)$ is defined such that $x_\mathrm{H_{I}}(z_{50}, \theta) = 0.5$). In particular, in the prior PD we find that $z_{50} = 6.81^{+1.13}_{-0.86}$ and in the posterior given CMB-based constraints, we find $z_{50} = 7.11^{+0.54}_{-0.4}$.

Fitting the Lyman-line-based constraints jointly with the CMB-based constraints yields tighter constraints on $\tau_\mathrm{CMB}$ and $\log_{10}(V_\mathrm{c})$. This leads to a further, more significant, contraction in the posterior PD of $x_\mathrm{H_{I}}(z)$ and an additional shift in the midpoint of reionization to higher redshift, with an HPD estimate of $z_{50} = 7.16^{+0.15}_{-0.12}$. The posterior PD is visually indistinguishable to that from our full joint analysis and yields a constrained reionization history with reionization occurring rapidly and relatively late.

Quantitatively, our joint analysis predicts the IGM was predominantly neutral during the first $0.5~\mathrm{Gyr}$ following recombination ($x_\mathrm{H_{I}}>0.75$ at $95\%$ credibility at $z \gtrsim 8$). Then, driven by galaxies forming in massive dark matter halos, reionization occurred rapidly, over a redshift interval $\Delta z_\mathrm{re} < 1.8$ at 95\% credibility (see \Cref{Fig:DerivedParamPosteriors}, right), with a midpoint of reionization at $z_{50} = 7.16^{+0.15}_{-0.12}$ and an IGM that was predominantly ionised ($x_\mathrm{H_{I}}<0.25$ at 95\% credibility) by $z \approx 6.2$.

\subsection{21-cm power spectrum \& global 21-cm signal posterior predictive densities}
\label{Sec:21cmConstraints}

\begin{figure*}
    \centerline{
        \includegraphics[width=0.5\textwidth, trim={0 1.3cm 0 0}, clip]{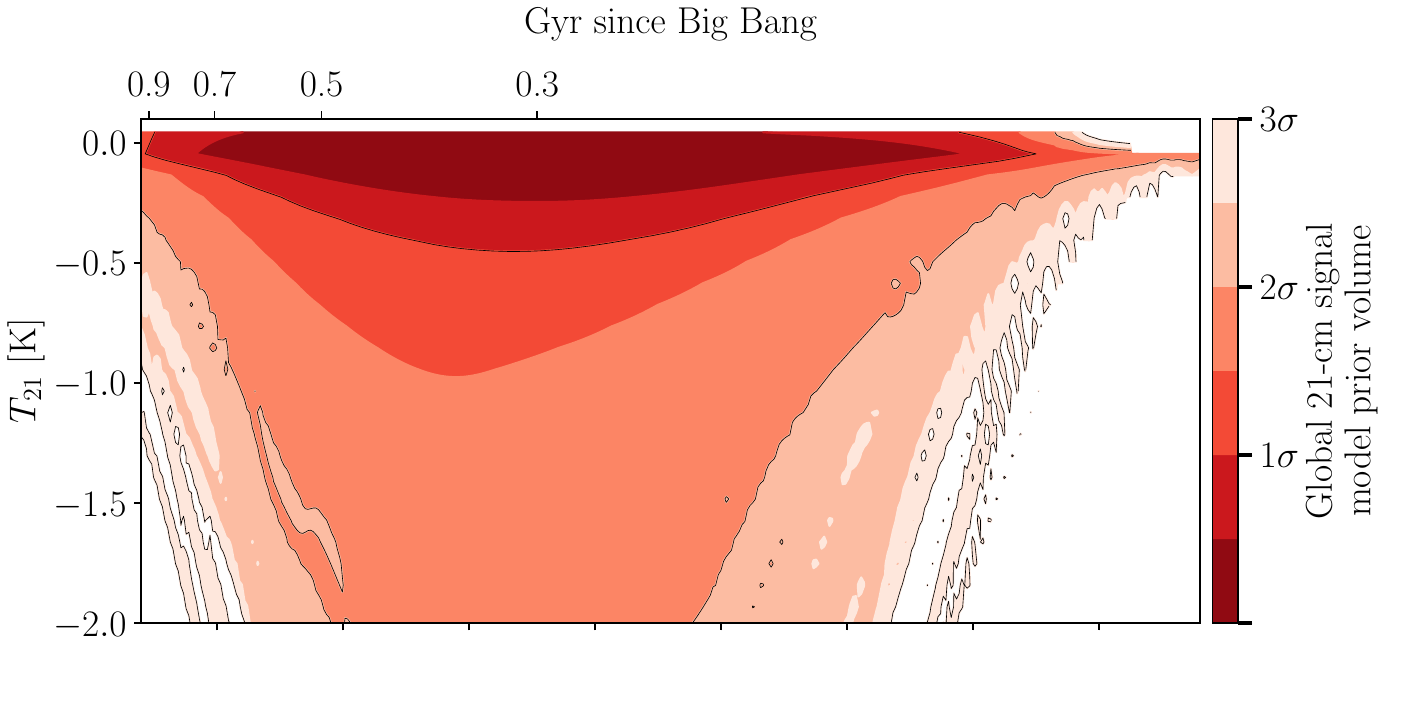}
        \includegraphics[width=0.5\textwidth, trim={0 1.3cm 0 0}, clip]{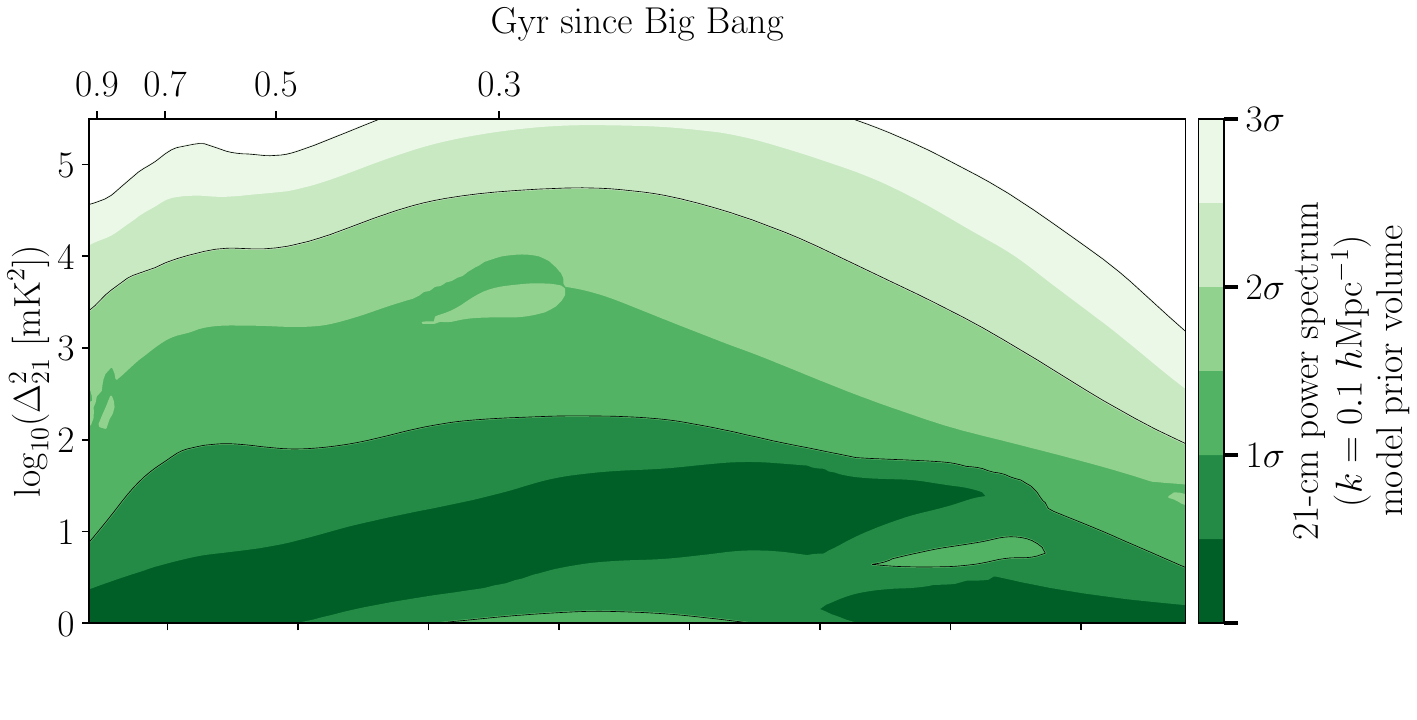}
    }
    \centerline{
        \includegraphics[width=0.5\textwidth, trim={0 1.3cm 0.2cm 1.5cm}, clip]{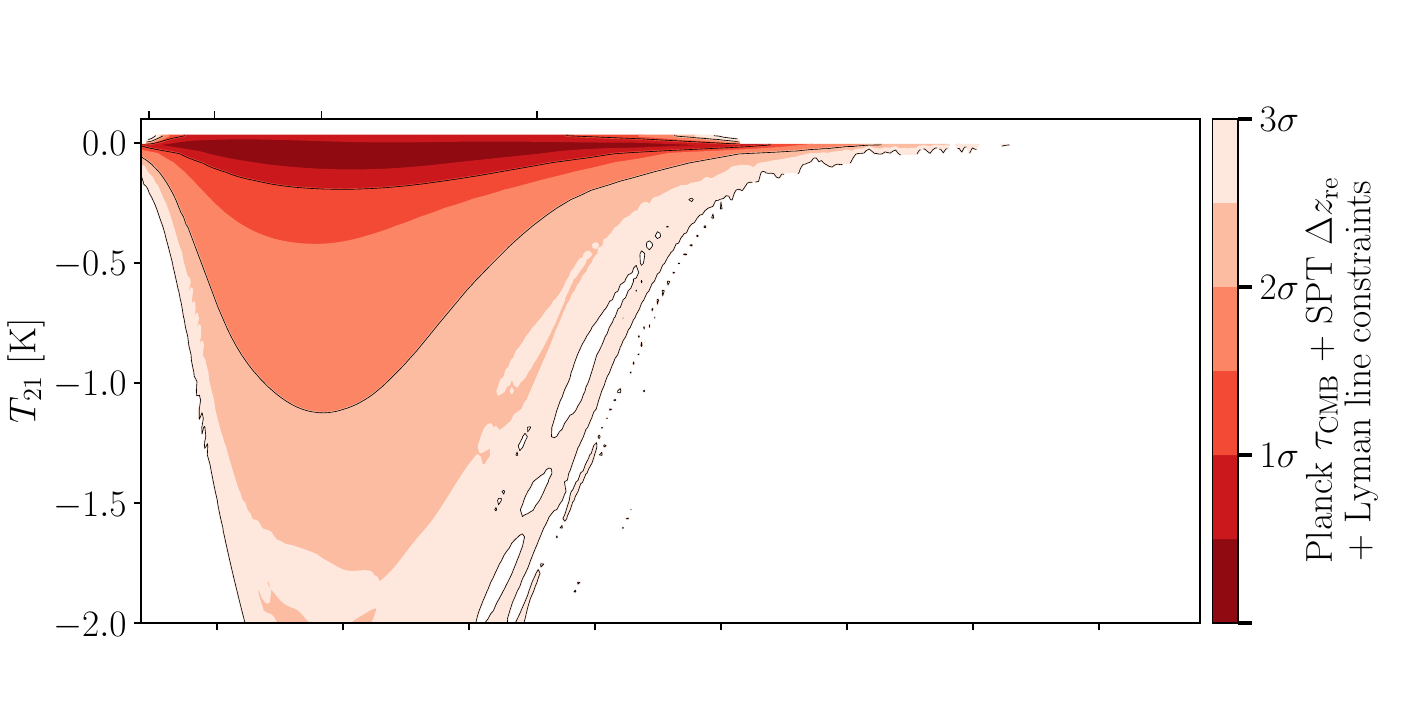}
        \includegraphics[width=0.5\textwidth, trim={0 1.3cm 0.2cm 1.5cm}, clip]{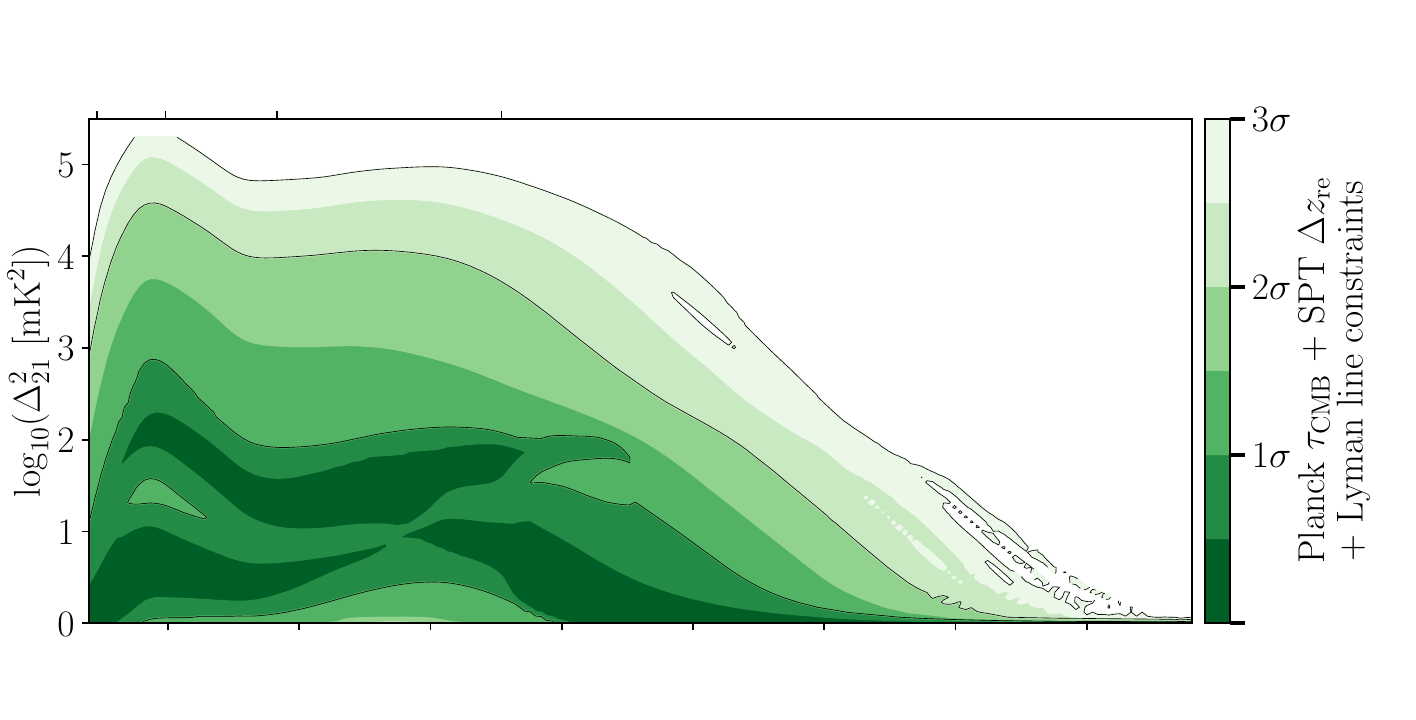}
    }
    \centerline{
        \includegraphics[width=0.5\textwidth, trim={0 0  0.2cm 1.5cm}, clip]{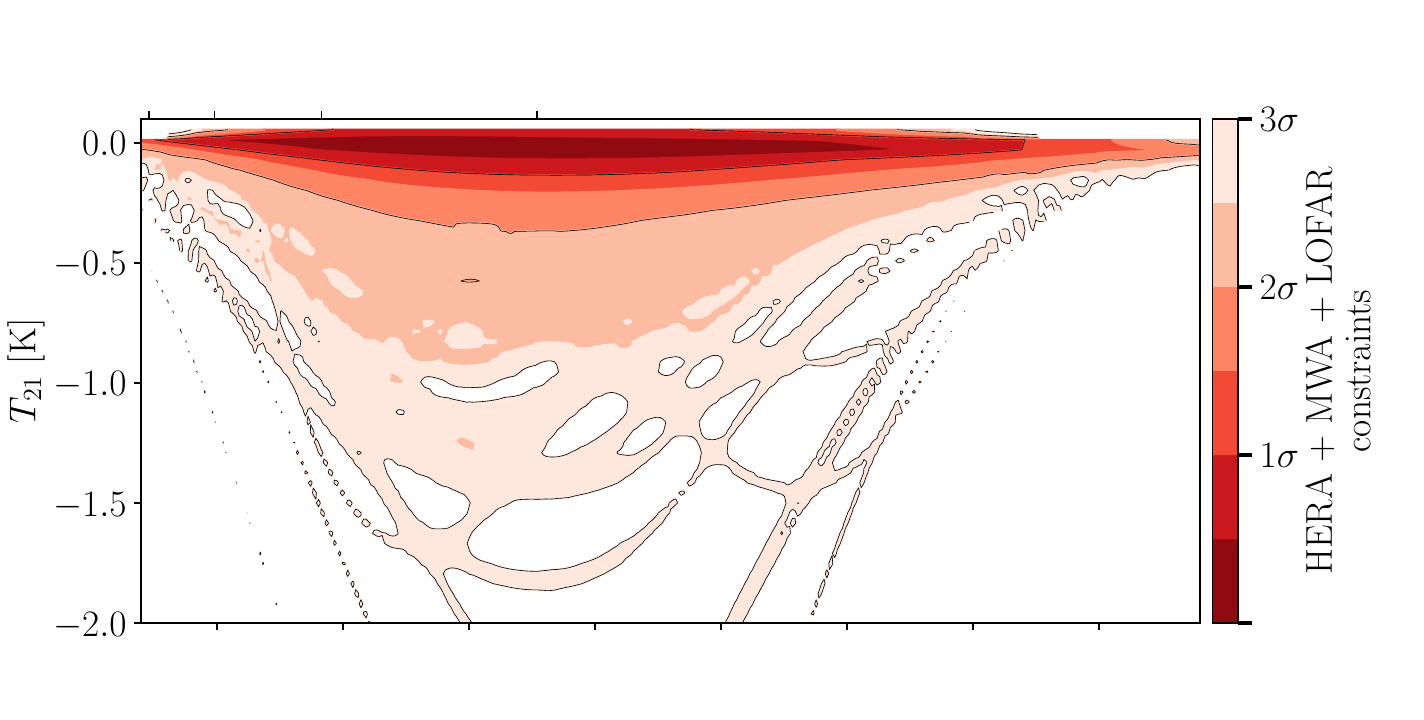}
        \includegraphics[width=0.5\textwidth, trim={0 0  0.2cm 1.5cm}, clip]{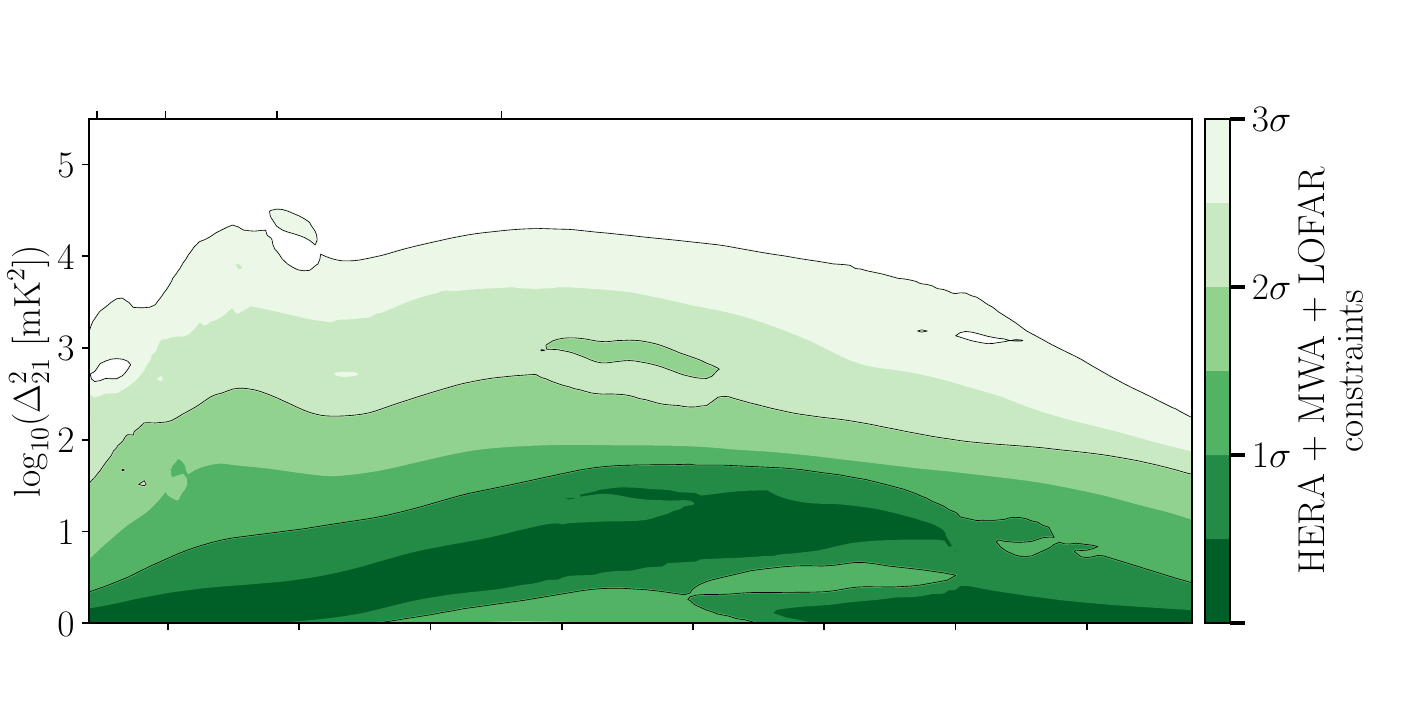}
    }
    \centerline{
        \includegraphics[width=0.5\textwidth, trim={0 0  0.2cm 1.5cm}, clip]{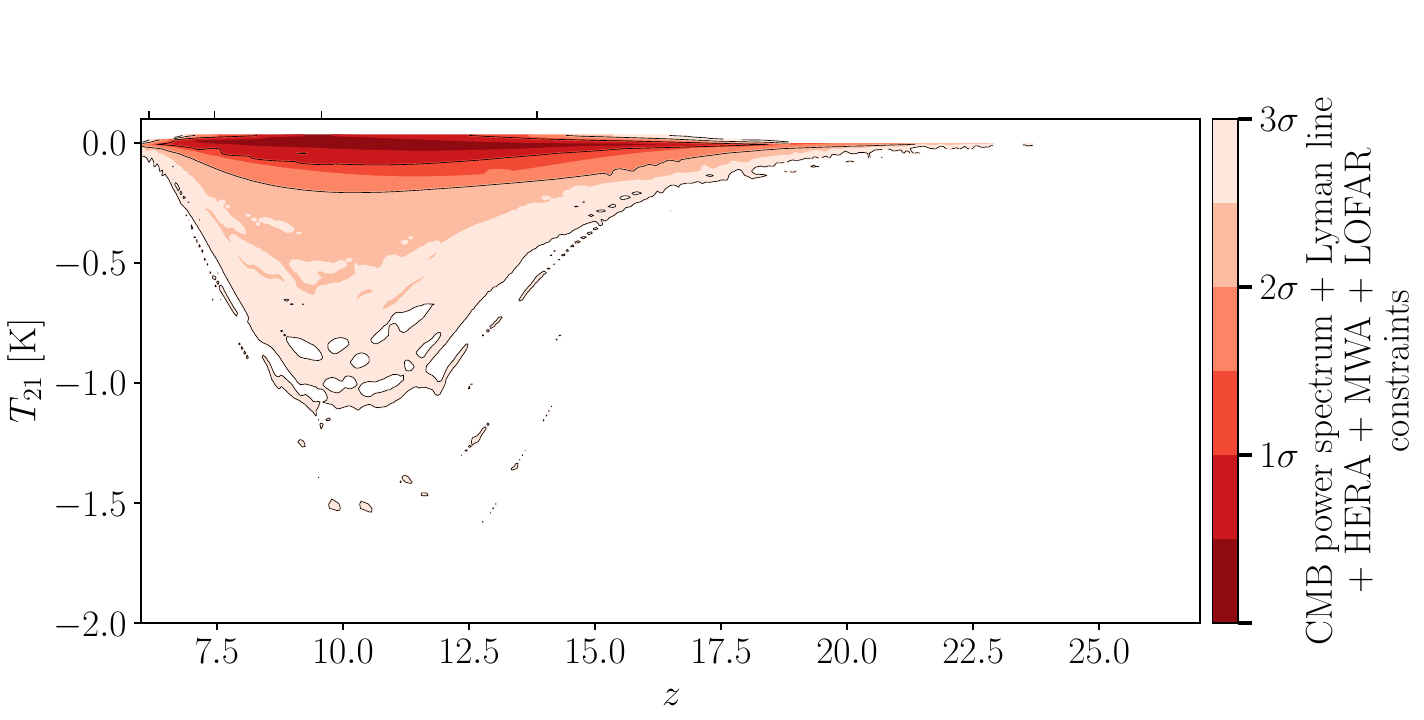}
        \includegraphics[width=0.5\textwidth, trim={0 0  0.2cm 1.5cm}, clip]{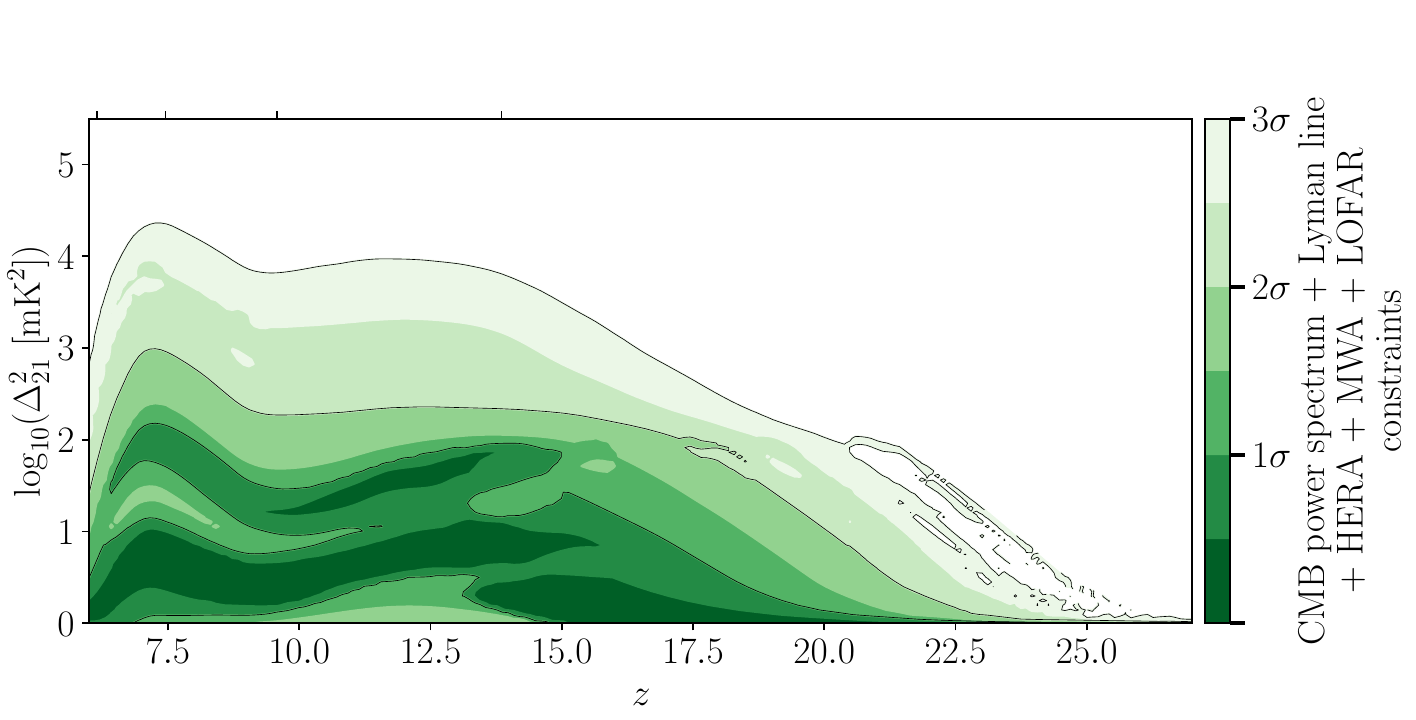}
    }
    \caption{
        Prior (top row) and posterior (rows 2-4) PDs of the global 21-cm signal (left) and dimensionless 21-cm power spectrum at $k=0.1~h\mathrm{Mpc^{-1}}$ (right). For both modelled observables, the data sets constraining the posterior PDs are indicated in the colorbar labels. They include constraints on the IGM neutral fraction inferred from Lyman line data, constraints on $\tau_\mathrm{CMB}$ derived from Planck TT,TE,EE+low$l$+lowE+lensing data and a constraint on the duration of reionization from the measurement of the patchy kSZ power at high $l$ in the CMB power spectrum by SPT (second row), constraints from upper limits on the power spectrum of redshifted 21-cm emission from the EoR with HERA, the MWA and LOFAR (third row) and the combination of these constraints (bottom row).
        }
    \label{Fig:T21andPspecContour}
\end{figure*}

Before discussing the impact of the data sets on the 21-cm signal posterior PDs, we caution that, as with the $x_\mathrm{H_{I}}(z)$ posteriors discussed in \Cref{Sec:IGMneutralFractionConstraints}, the $T_{21}(z)$ and $\Delta^{2}_{21}(z)$ posterior PDs are model-dependent. In particular, replacing the assumed constant SFE as a function of halo mass in atomic-cooled halos with a more flexible parametrisation (e.g., \citealt{2025arXiv250321687D}) is expected to result in less stringent constraints on the high-redshift 21-cm signals from the data sets considered in this work.

Nevertheless, we emphasise that the posterior PDs of summary statistics provide a valuable metric for directly comparing models with different astrophysical parametrisations and assumptions. Thus, we present the posterior PDs here as a reference set of predictions for the 21-cm signal, based on the model assumptions considered in this work, against which future analyses can be compared.

\Cref{Fig:T21andPspecContour} shows the prior (top) and posterior PDs of the redshift-dependent global 21-cm signal ($T_{21}$; left) and 21-cm power spectrum at $k=0.1~h\mathrm{Mpc^{-1}}$ ($\Delta^{2}_{21}(k = 0.1 h~\mathrm{Mpc^{-1}})$; right) given,
\begin{enumerate}
    \item CMB- and Lyman-line-based constraints (measurements of $\tau_\mathrm{CMB}$
     inferred from Planck TT,TE,EE+low$l$+lowE+lensing data, constraints on the duration of reionization inferred from measurements of the patchy kSZ power spectrum with SPT and high-redshift IGM neutral fraction measurements inferred from Lyman line data; row two)
    \item upper limits on the 21-cm power spectrum measured by HERA+LOFAR+MWA (row three) and
    \item the combination of these data sets (bottom row).
\end{enumerate}

Comparing the prior and posterior PDs of $T_{21}$ and $\Delta^{2}_{21}(k = 0.1 h~\mathrm{Mpc^{-1}})$ given CMB + Lyman-line-based constraints, one sees that the enhanced constraints on $\tau_\mathrm{CMB}$ and $\log_{10}(V_\mathrm{c})$ resulting from our joint analysis of these data sets (see \Cref{Fig:tauVcPosteriors} and \Cref{Tab:CredibilityIntervals}) lead to a concentration of the posterior PD of the 21-cm signals at lower redshifts. This shifts the most probable redshift of peak $T_{21}$ and $\Delta^{2}_{21}(k = 0.1 h~\mathrm{Mpc^{-1}})$ from $z \approx 13$ and $16$ ($t \approx 0.33$ and $0.24~\mathrm{Gyr}$, where $t$ is time since the Big Bang) in the prior PDs to $z \approx 10$ and $7$ ($t \approx 0.47$ and $0.76~\mathrm{Gyr}$), respectively, in the posterior PDs.

In the model fit here, the shift to lower redshift (later times) results from star formation being delayed until high-mass atomic-cooling halos form. This leaves models in which reionization, \Lya\ coupling of the neutral hydrogen spin and kinetic temperatures, and reheating of the IGM occur later and more rapidly, on average, than in our prior volume. Models with contemporaneous \Lya\ coupling of the neutral hydrogen spin and kinetic temperatures and reheating of the IGM, on average, are more prominent in the posterior, leading to a modest decline in the a posteriori peak depth of the global 21-cm signal in absorption. Quantifying this impact of the data, we find that signals deeper than $2~\mathrm{K}$ in absorption are disfavoured at $95\%$ credibility in the prior PD of $T_{21}$, and this tightens to $1~\mathrm{K}$, at $95\%$ credibility, in the posterior PD. We also note that the most probable a posteriori redshift of the peak of $\Delta^{2}_{21}(k = 0.1 h~\mathrm{Mpc^{-1}})$ approximately coincides with our $z_{50} = 7.11^{+0.54}_{-0.4}$ a posteriori HPD estimate of the midpoint of reionization, given the same data sets. This is consistent with the inference from our joint analysis that reionization is driven by massive galaxies (see \Cref{Sec:tauVcConstraints}), which, at the midpoint of reionization, are surrounded by large ionized bubbles that contrast strongly with adjoining, more neutral, regions, leading to large spatial fluctuations in the $x_\mathrm{H_{I}}$ field and a correspondingly large 21-cm power spectrum amplitude.

In the absence of the modelling approximations discussed in \Cref{Sec:tauVcConstraints}, one may expect models in which star formation in low-mass halos is non-zero but makes only a subdominant contribution to reionization (for example, due to a reduced SFE) to also be consistent with the data. This would require the increased contribution to $\tau_\mathrm{CMB}$ from these early sources to be balanced by a smaller contribution to $\tau_\mathrm{CMB}$ from star formation at lower redshift. Such a scenario is possible while remaining consistent with the Lyman line data sets considered in our analysis (which provide limited constraint in the redshift range $8 \le z \le 10$; see \Cref{Fig:xHIContour}, bottom panel) if the transition to rapid reionization beginning around $z \sim 10$ in our model was further delayed to $z \sim 9$ and then occurred even more rapidly in the redshift range $8 \lesssim z \lesssim 9$.

In contrast to the impact of Lyman-line- and CMB-power-spectrum-based constraints on the 21-cm posterior densities (which are principally timing-based), the constraints on $\log_{10}(f_\mathrm{X})$ and $\log_{10}(f_\mathrm{radio})$ from our joint analysis of HERA+LOFAR+MWA 21-cm power spectrum upper limits principally translate to a constraint on the amplitudes of the 21-cm posterior densities. In particular, the $95\%$ credibility upper limit on the peak depth of $T_{21}$ in absorption and amplitude of $\Delta^{2}_{21}(k = 0.1 h~\mathrm{Mpc^{-1}})$ drops from approximately $2~\mathrm{K}$ and $10^{4}~\mathrm{mK}^{2}$ at $z \approx 13$ and $ 17$ ($t \approx 0.33$ and $0.22~\mathrm{Gyr}$), respectively, in the prior, to $0.35~\mathrm{K}$ and $3 \times 10^{2}~\mathrm{mK}^{2}$ at $z \approx 12$ and $14$ ($t \approx 0.37$ and $0.30~\mathrm{Gyr}$), respectively, in the posterior.

This comes as a result of models described by the combination of low values of $\log_{10}(f_\mathrm{X})$ and large values of $\log_{10}(f_\mathrm{radio})$ being disfavoured by the HERA+LOFAR+MWA 21-cm power spectrum upper limits (see the $\log_{10}(f_\mathrm{X})$ - $\log_{10}(f_\mathrm{radio})$ 2D marginal posterior in \Cref{Fig:JointAnalysisPosteriors}). Models with low values of $\log_{10}(f_\mathrm{X})$ have reduced reheating of the IGM by X-ray emission and those with large values of $\log_{10}(f_\mathrm{radio})$ have a correspondingly high radio background temperature ($T_\mathrm{b,r} \propto L_\mathrm{radio}(\nu, z)$, where $T_\mathrm{b,r}$ is the contribution to the radio background temperature from high redshift galactic radio luminosity). Thus, it is models with a combination of these effects that, on average, have the largest differential brightness temperatures between the 21-cm spin temperature and background radiation temperature and are therefore the first to be ruled out by existing 21-cm power spectrum upper limits\footnote{During the writing of this manuscript, \citet{2024MNRAS.534..738C} demonstrated that soft photon heating may influence this conclusion. For instance, in the presence of an intense radio background with a sufficiently steep spectrum, it could suppress the resultant 21-cm signal amplitude. Further consideration of this effect is left to future work.}.

The posterior PD of the 21-cm signal given the parameter posteriors from our full joint analysis exhibits a combination of the aforementioned effects, concentrating posterior density at lower amplitude and lower redshift. In particular, we find that in the posterior the $95\%$ credibility upper limit on the peak amplitude of $T_{21}$ and $\Delta^{2}_{21}(k = 0.1 h~\mathrm{Mpc^{-1}})$ drops to $0.18~\mathrm{K}$ and $10^{2.8}~\mathrm{mK}^{2}$ at $z \approx 11$ and $7$ ($t \approx 0.41$ and $0.76~\mathrm{Gyr}$), respectively.

\subsubsection{Comparison with EDGES and SARAS 3 global signal measurements}
\label{Sec:EDGESandSARASImplications}

In 2018, the EDGES experiment reported the first detection of the sky-averaged `global' 21-cm absorption signal (B18) that is expected to accompany the onset of CD. The best fitting flat-bottomed 21-cm signal recovered in absorption in B18 is centred at redshift $z = 17.2 \pm 0.2$ and has a depth of $A = 500^{+500}_{-200}~\mathrm{mK}$, where the uncertainties correspond to 99 percent confidence intervals, accounting for both thermal and systematic errors. However, the flattened shape and large depth of the signal reported in B18 was unexpected and a recent analysis of observations by SARAS 3, in a similar but narrower redshift range, disfavours the presence of a signal with the position, shape and amplitude found in B18, finding a 95.3\% credibility upper limit on the amplitude of such a signal $A < 500~\mathrm{mK}$ (\citealt{2022NatAs...6..607S}; hereafter, S22). Furthermore, the presence in the SARAS 3 data of signals with the amplitude found by B18 but marginalised over astrophysically motivated shapes is also found to be disfavoured (e.g. \citealt{2022NatAs...6.1473B}, B24). This implies either one or both of the EDGES and SARAS measurements are contaminated by systematics\footnote{A number of reanalyses of the EDGES data have also suggested the possible existence of unaccounted for systematics, which complicate drawing firm conclusions regarding a detection of a cosmological signal in the data (e.g. \citealt{2018Natur.564E..32H, 2019ApJ...874..153B, 2019ApJ...880...26S, 2020MNRAS.492...22S, 2021MNRAS.502.4405B, 2024arXiv241108134C}). The impact of systematic structure sourced by incompletely accounted for receiver- or antenna-based chromaticity is currently under investigation (e.g. \citealt{2022MNRAS.517.2264M, 2023MNRAS.521.3273S}).}.

The relative credibility of these constraints can be assessed, in the context of our model and the data analysed in this work, using the posterior PD of $T_{21}$. The posterior PD of $T_{21}$ from our joint analysis yields a model-dependent upper limit on the absorption depth of the global 21-cm signal at redshift $z = 17.2$ of $A < 62~\mathrm{mK}$ at 95\% credibility.

This limit disfavours a deep absorption trough at high redshift, consistent with the conclusions drawn from the SARAS 3 data in S22 and B24. However, we caution that the modelling caveats discussed in \Cref{Sec:tauVcConstraints} could influence this conclusion. In particular, as illustrated in \Cref{Fig:SensitivityPlotxHI_T21_21cmPS}, model realisations with large $V_\mathrm{c}$ exhibit delayed star formation, shifting the peak of the global 21-cm absorption signal and translating the power in 21-cm signal fluctuations to lower redshifts. As a result, our inferred constraint on $V_\mathrm{c}$ suppresses both the posterior PD of the amplitude of $T_{21}$ and the power in 21-cm fluctuations at high redshifts.

In contrast, introducing a more sophisticated, mass-dependent SFE model in place of the current approximation of a constant SFE in atomic cooling halos could lead to less stringent constraints on the high-redshift 21-cm signal amplitude. For example, star formation at high redshift could still be consistent with the data in a model where the SFE declines with mass, assuming that this star formation occurs in low-mass, low-SFE halos that contribute minimally to early reionization. In such cases, additional high-redshift star formation could result in less stringent limits on the high-redshift absorption amplitude of $T_{21}$. Similar effects may arise from addressing other modelling approximations discussed in \Cref{Sec:tauVcConstraints}. A detailed exploration of these effects is left to future work.

\subsubsection{Implications for 21-cm power spectrum \& global 21-cm signal detectability}
\label{Sec:21cmImplications}

The primary challenge to recovering unbiased estimates of the redshifted 21-cm signal is its extraction from data also containing bright foreground emission\footnote{Here, foreground emission refers to all sources of radio emission within the observing band of interest for recovery of the 21-cm signal, other than the signal itself.} (e.g. \citealt{2006PhR...433..181F, 2016MNRAS.462.3069S, 2019MNRAS.484.4152S, 2019MNRAS.488.2904S, 2020PASP..132f2001L, 2023MNRAS.520.4443B}). The 21-cm foregrounds are dominated by synchrotron emission from the Milky Way and extragalactic sources that is spectrally well characterised by a spatially dependent power law distribution (e.g. \citealt{1999A&A...345..380S, 2023MNRAS.521.3273S}) with a mean temperature spectral index of $\beta \sim 2.5$ (where $T \propto \nu^{-\beta}$; e.g. \citealt{2017MNRAS.464.4995M, 2019MNRAS.483.4411M}), in the $45 \lesssim \nu \lesssim 200~\mathrm{MHz}$ ($30 \lesssim z \lesssim 6$)\footnote{The observed frequency and redshift at which the restframe 21-cm hyperfine line is emitted are related by $\nu_\mathrm{obs} = \nu_{21} / (1+z)$, where $\nu_{21} \approx 1420~\mathrm{MHz}$ is the rest-frame emission frequency of the 21-cm hyperfine line of neutral hydrogen.} band of interest for CD and EoR science.

While this emission is intrinsically spectrally smooth, it propagates through an instrumental transfer function that can have structure on the narrower spectral scales on which the 21-cm signal is intrinsically dominant. Consequently, to recover unbiased estimates of the 21-cm signal with robust error estimates, the experiment must be exquisitely calibrated, with the associated uncertainties in the instrument calibration robustly propagated to the signal estimates (e.g. \citealt{2021MNRAS.505.2638R, 2022MNRAS.517..910S, 2022MNRAS.517..935S, 2022MNRAS.517.2264M, 2024arXiv241214023K}).

As such, while model-dependent, the shift to lower redshifts in the a posteriori average peak amplitudes of the global 21-cm signal and of the 21-cm power spectrum implied by our analysis is encouraging for their detectability. Assuming a sky-averaged temperature spectral index $\beta = 2.5$, the foreground brightness temperature decreases by a factor of $(\nu_{2}/\nu_{1})^{\beta} = [(1 + z_{1})/(1 + z_{2})]^{\beta} \approx 1.5$ between $z_{1} \approx 13$ and $z_{2} \approx 11$ ($\nu_{1} \approx 101~\mathrm{MHz}$ and $\nu_{2} \approx 118~\mathrm{MHz}$). These redshifts correspond to the maxima of the $95\%$ credibility iso-probability contours of the prior and joint analysis posterior PDs of $T_{21}$, respectively. Furthermore, the temperature drops by a factor of 2.8 when comparing the $z \approx 17$ central redshift of the EDGES signal with the $z \approx 11$ maximum of the joint analysis posterior PD of $T_{21}$.

This effect is even more pronounced for the 21-cm power spectrum. Specifically, the foreground brightness temperature decreases by approximately a factor of 6.6 between the $z \approx 16$ and $z \approx 7$ redshifts, corresponding to the maxima of the $95\%$ credibility iso-probability contours of $\Delta^{2}_{21}(k = 0.1 h~\mathrm{Mpc^{-1}})$ in our prior and joint analysis posterior PDs, respectively. In both cases, this reduction in foreground brightness at lower redshifts decreases the amplitude of foreground-coupled calibration systematics for a fixed level of calibration accuracy, enhancing the prospects for successful detection and characterization of the 21-cm signal.

Multiple interferometric EoR experiments have sensitivity to the power spectrum of redshifted 21-cm emission in the $70 \lesssim \nu \lesssim 200~\mathrm{MHz}$ spectral band in which our full joint analysis posterior PD of $\Delta^{2}_{21}(k = 0.1 h~\mathrm{Mpc^{-1}})$ is concentrated. These include, for example, HERA phase II ($50$ -- $250~\mathrm{MHz}$; \citealt{2024arXiv240104304B}), LOFAR ($10$ -- $240~\mathrm{MHz}$; \citealt{2013A&A...556A...2V}), the MWA ($80$ -- $300~\mathrm{MHz}$; \citealt{2013PASA...30....7T}) and NenuFAR ($40$ -- $85~\mathrm{MHz}$; \citealt{2021sf2a.conf..211M}). In particular, the $\Delta^{2}_{21}(k = 0.36 h~\mathrm{Mpc^{-1}}) \le 3.5\times10^{3}~\mathrm{mK^{2}}$ most recent \citet{2023ApJ...945..124H} upper limits on the 21-cm power spectrum at $z=7.9$ are within approximately an order of magnitude\footnote{The power spectrum near the midpoint of reionization is typically approximately flat (to within a factor of a few; e.g. \citealt{2014ApJ...782...66P}) across the spatial scale range relevant for comparing the $k = 0.1 h~\mathrm{Mpc^{-1}}$ posterior PD of the 21-cm power spectrum plotted in \Cref{Fig:T21andPspecContour} and the \citet{2023ApJ...945..124H} upper limits on the 21-cm power spectrum at $k = 0.36 h~\mathrm{Mpc^{-1}}$ and $z=7.9$. The posterior PD of the 21-cm power spectrum plotted in \Cref{Fig:T21andPspecContour} thus provides a reasonable guide to the credibility of 21-cm power spectrum estimates at this redshift, across this spatial scale range. } of the $\Delta^{2}_{21}(k = 0.1 h~\mathrm{Mpc^{-1}}) \sim 10^{2}~\mathrm{mK^{2}}$ upper limit on the HPD $68\%$ credibility interval of the posterior PD of the 21-cm power spectrum, given our full joint analysis.

Furthermore, multiple radiometric EoR experiments have significant sensitivity to the global 21-cm signal in the $70 \lesssim \nu \lesssim 200~\mathrm{MHz}$ spectral band in which our full joint analysis posterior PD of $T_{21}$ is concentrated. These include, for example, EDGES high-band ($90$ -- $190~\mathrm{MHz}$; \citealt{2017ApJ...847...64M}), REACH ($50$ -- $170~\mathrm{MHz}$; \citealt{2022NatAs...6..984D}) and SARAS 3 ($40$ -- $230~\mathrm{MHz}$; \citealt{2021arXiv210401756N}).

%%%%%%%%%%%%%%%%%%%%%%%%%%%%%%%%%%%%%%%%%%%%%%%%%%
\section{Summary \& Conclusions}
\label{Sec:Conclusions}
%%%%%%%%%%%%%%%%%%%%%%%%%%%%%%%%%%%%%%%%%%%%%%%%%%

Observations of CD and the EoR have the potential to answer long-standing questions of astrophysical interest, including "how massive were the galaxies and halos in which the sources driving reionization formed?" and "how long did it take for luminous sources to reionize the IGM?".
Data from a panoply of experiments spanning the electromagnetic spectrum have been used to place constraints on models of the high-redshift Universe. Jointly analysing these data sets within a self-consistent modelling framework facilitates recovery of a broader, more complete and robust picture of galaxy properties and their impact on the IGM during CD and the EoR.

In this paper, we derived stringent but model-dependent astrophysical constraints from which follow detailed, quantitative answers to these questions. To achieve this, we used a machine-learning-accelerated Bayesian statistical framework to jointly analyse constraints deriving from CMB power spectrum measurements from Planck and SPT, IGM neutral fraction measurements from Lyman-line-based data sets and 21-cm power spectrum upper limits from HERA, LOFAR and the MWA.

We evaluated our full joint analysis results within a Bayesian statistical framework, comparing the prior and posterior densities of astrophysical parameters of the early Universe. This yields strong but model dependent constraints, consistent between data sets, on the duration of the EoR and the mass of dark matter halos hosting the galaxies that were the dominant drivers of reionization. In the context of the considered 21cmSPACE model with mass-independent SFE, the data included in our joint analysis was found to be accurately described a reionization history in which galaxies forming from predominantly atomic-cooled hydrogen gas in dark matter halos with masses $M_\mathrm{min} \gtrsim 2.6 \times 10^{9}~M_{\odot} ((1+z)/10)^{\frac{1}{2}}$ at 95\% credibility ($V_\mathrm{c} \gtrsim 50~\mathrm{km~s^{-1}}$) are the dominant galactic population driving reionization. The data requires that these galaxies rapidly reionized the Universe over a narrow redshift interval $\Delta z_\mathrm{re} < 1.8$ at 95\% credibility (see \Cref{Fig:DerivedParamPosteriors}, right). We find that the midpoint of reionization (when the sky-averaged IGM neutral fraction had dropped to 50\%) can be confidently estimated as $z_{50} = 7.16^{+0.15}_{-0.12}$, and that the IGM was predominantly ionised by $z = 6.2$ ($x_\mathrm{H_{I}}<0.25$ at 95\% credibility).

In addition to this result, we performed Bayesian information theoretic and statistical analyses comparing the prior and posterior densities of astrophysical parameters of the early Universe, given fits of the 21-cm, CMB and Lyman line data sets individually (as well as collectively). Our main inferences and conclusions from these analyses are as follows:
\begin{itemize}
    \item In \Cref{Sec:DKL}, we introduced information triangle plots, which illustrate the EoR information content of the data sets quantified by the marginal KL divergence between the 1D and 2D joint prior and posterior densities of the astrophysical parameters for each analysis. This allowed us to quantify the non-uniformity of the information pertaining to the astrophysical parameters of CD and the EoR in our model of the high-redshift Universe across the data sets included in our analysis. Using this technique, we determined quantitative estimates showing that Lyman line constraints on the IGM neutral fraction, in aggregate, are the most EoR information rich data in our analysis, with an EoR information content of several nats. In comparison, CMB power spectrum constraints provide approximately an order of magnitude lower EoR information (a few tenths of a nat). Finally, the HERA+MWA+LOFAR redshifted 21-cm power spectrum constraints used here provide a factor of a few lower EoR information than the CMB based constraints (approximately a tenth of a nat)\footnote{In contrast to the relatively weak constraining power of current upper limits, the astrophysical parameter sensitivity of the 21-cm power spectrum in our model (see \Cref{Fig:SensitivityPlotxHI_T21_21cmPS}) suggests a detection with upcoming experiments such as the SKA, construction of which has recently begun, would provide a significant increase in the EoR information content.}.
    \item As well as the total information content being unevenly distributed between data sets, we found that so too is its distribution between parameters. The 21-cm power spectrum upper limits constrain $\log_{10}(f_\mathrm{X})$ and $\log_{10}(f_\mathrm{radio})$, providing the most information about the $\log_{10}(f_\mathrm{X})$ - $\log_{10}(f_\mathrm{radio})$ joint PDF. In contrast, the Lyman line and CMB power spectrum data sets both constrain $\log_{10}(V_\mathrm{c})$ and $\tau_\mathrm{CMB}$, providing the most information about the $\log_{10}(V_\mathrm{c})$ - $\tau_\mathrm{CMB}$ joint PDF.
    \item Additionally, we calculated the total EoR information content as the KL divergence between the full five dimensional prior and posterior probability densities of the parameters of our model and found that $72.8 \pm 1.6 \%$, $49.5 \pm 1.7 \%$, $1.4 \pm 0.1\%$ and $0.9 \pm 0.4\%$ of our astrophysical prior volume is consistent with the posterior given the HERA+MWA+LOFAR upper limits, the CMB power spectrum measurements and constraints, the Lyman line measurements and constraints, and their combination, respectively.
    \item Our Bayesian statistical analysis of the sampled parameter and posterior PDs of $T_{21}$ and $\Delta^{2}_{21}$ showed that, in the context of our model, the constraints derived from the 21-cm power spectrum upper limit included in our joint analysis shift the $T_{21}$ and $\Delta^{2}_{21}$ posterior densities to lower amplitudes and that the remaining data sets concentrate the posterior densities at lower redshifts. The posterior PDs of the 21-cm signal from our full joint analysis exhibit a combination of both effects.
\end{itemize}

The shift to lower redshifts of the peak amplitudes of the posterior PDs of both the global 21-cm signal and the 21-cm power spectrum, given the model and data sets considered in this work, is favourable with respect to the reduced brightness of the foreground emission at these redshifts. In the case of the 21-cm power spectrum, in particular, the $\Delta^{2}_{21}(k = 0.36 h~\mathrm{Mpc^{-1}}) \le 3.5\times10^{3}~\mathrm{mK^{2}}$ most recent \citet{2023ApJ...945..124H} upper limits on the 21-cm power spectrum at $z=7.9$ are within approximately an order of magnitude of the $\Delta^{2}_{21}(k = 0.1 h~\mathrm{Mpc^{-1}}) \sim 10^{2}~\mathrm{mK^{2}}$ upper limit on the $68\%$ HPDI of the posterior PD of the 21-cm power spectrum, given our full joint analysis. Furthermore, EDGES high-band, REACH and SARAS 3 have significant sensitivity to the global 21-cm signal in the lower redshift spectral band in which our full joint analysis posterior PD of $T_{21}$ is concentrated. Thus, if the qualitative features of the posterior PDs of the 21-cm signal hold given updated modelling and inclusion of additional data sets in future work, this would be cause for optimism with respect to the detectability of these signals with ongoing 21-cm experiments in this redshift range.

We caution, however, that the tight constraint on $\log_{10}(V_\mathrm{c})$ and its impact on the posterior PDs of the 21-cm power spectrum and global signal are model-dependent. Updates to our analysis using simulations that incorporate more flexible modelling of the galaxy-mass and redshift dependence of the SFE and escape fraction of ionising radiation (see, e.g., \citet{2024arXiv241200799Q, 2025arXiv250321687D}), as well as potential differences between the star formation efficiencies of Population II and III stars (e.g., \citealt{2024MNRAS.531.1113P, 2024MNRAS.529..519G}), are expected to be valuable for refining these results in future work.

Finally, we highlight that there is potential to further enhance our understanding of CD and the EoR by incorporating in a joint analysis additional data sets and measurements that constrain astrophysical processes during these periods. Taking the posterior parameter volume defined by the present joint analysis as the prior volume for a future analysis, the inclusion of additional constraints, has great potential to further enhance and tighten our understanding of CD and EoR astrophysics. Constraints of interest for this purpose derive from data sets including:
\begin{enumerate}
    \item upcoming global 21-cm and 21-cm power spectrum measurements,
    \item measurements of the UV-luminosity function (e.g. \citealt{2017MNRAS.464.1365M, 2019MNRAS.484..933P, 2025arXiv250321687D}),
    \item estimates of the cosmic radio and X-ray background (e.g. \citealt{2011ApJ...734....5F, 2018ApJ...858L...9D, 2024MNRAS.531.1113P, 2024MNRAS.529..519G}), measurements of ionizing photon mean free path (e.g. \citealt{2021ApJ...917L..37C}) and \Lya\ opacity (e.g. \citealt{2024MNRAS.531.1951C, 2024arXiv241200799Q}) constraints from \Lya\ forest data and, in the future,
    \item cross-correlations measurements between CMB, line intensity maps and galaxy observations (e.g. \citealt{2022ApJ...928..162L, 2023ApJ...944...59L, 2023arXiv230906477F, 2023arXiv230800749M}).
\end{enumerate}
Methods such as those employed in this work will be essential for achieving this as we move towards a more robust, comprehensive, multi-probe understanding of these early periods of cosmic history.

% The last numbered section should briefly summarise what has been done, and describe
% the final conclusions which the authors draw from their work.

%%%%%%%%%%%%%%%%%%%%%%%%%%%%%%%%%%%%%%%%%%%%%%%%%%
\section*{Acknowledgements}
%%%%%%%%%%%%%%%%%%%%%%%%%%%%%%%%%%%%%%%%%%%%%%%%%%

% The Acknowledgements section is not numbered. Here you can thank helpful
% colleagues, acknowledge funding agencies, telescopes and facilities used etc.

PHS acknowledges partial support for this work through funding from the Canada 150 Research Chairs Program and a fellowship from the McGill Space Institute. PHS thanks Jiten Dhandha, Adrian Liu, Andrei Mesinger, Jonathan Sievers and Irina Stefan for valuable discussions and Irina Stefan for helpful comments on a draft of this manuscript. HTJB acknowledges support from the Kavli Institute for Cosmology, Cambridge, the Kavli Foundation and of St Edmunds College, Cambridge. Dominic Anstey acknowledges support from the Science and Technologies Facilities Council.

%%%%%%%%%%%%%%%%%%%%%%%%%%%%%%%%%%%%%%%%%%%%%%%%%%
\section*{Data Availability}
%%%%%%%%%%%%%%%%%%%%%%%%%%%%%%%%%%%%%%%%%%%%%%%%%%

% The inclusion of a Data Availability Statement is a requirement for articles published in MNRAS. Data Availability Statements provide a standardised format for readers to understand the availability of data underlying the research results described in the article. The statement may refer to original data generated in the course of the study or to third-party data analysed in the article. The statement should describe and provide means of access, where possible, by linking to the data or providing the required accession numbers for the relevant databases or DOIs.

The data from this study will be shared on reasonable request to the corresponding author. The HERA data is publicly available at \url{https://reionization.org}. The Planck optical depth data is publicly available at \url{https://pla.esac.esa.int/#home}.



\bibliographystyle{mnras}
\bibliography{bibliography} % if your bibtex file is called bibliography.bib

%\begin{thebibliography}{99}
%\end{thebibliography}

%%%%%%%%%%%%%%%%%%%%%%%%%%%%%%%%%%%%%%%%%%%%%%%%%%

%%%%%%%%%%%%%%%%% APPENDICES %%%%%%%%%%%%%%%%%%%%%

\appendix

%%%%%%%%%%%%%%%%%%%%%%%%%%%%%%%%%%%%%%%%%%%%%%%%%%
\section{Derivation of contaminated upper limit (CUL) likelihood}
\label{Sec:UpperLimitLikelihood}
%%%%%%%%%%%%%%%%%%%%%%%%%%%%%%%%%%%%%%%%%%%%%%%%%%

Several of the constraints used in this work derive from $y \sigma$ upper limits on a quantity of interest, $q$, defined such that $q \le d + y\sigma$, where $y$ is a positive integer and $d$ is the measured data. In this appendix, we calculate the likelihood\footnote{For the interested reader, related derivations in the context of 21-cm power spectrum upper-limits can be found in \citet{2019ApJ...887..141L, 2020MNRAS.493.4728G, 2022ApJ...924...51A}.}, $\mathcal{L}_\mathrm{CUL}(\sTheta)$, of measuring data, $d = q + \epsilon + n$, given a model for a subcomponent of the data $q$, which we write as $m(\sTheta)$, where $\sTheta$ is a set of parameters of the model. Here, $q$ is the quantity of interest, $\epsilon$ is a positive nuisance parameter, $n$ is measurement noise and $\mathcal{L}_\mathrm{CUL}(\sTheta) \equiv \mathcal{P}(d \vert \sTheta)$ is the marginal probability of the data given $\sTheta$, having marginalised out $\epsilon$.

Constraints of this type appear in the measurement of the \Lya\ and \Lyb\ forest dark pixel fraction and in the measurement of the ACF of high-redshift LAEs. In these two cases, $q$ corresponds to the IGM neutral fraction. In the former case, $\epsilon$ corresponds to the component of the dark pixel fraction not due to a neutral IGM (for example, due to the presence of damped \Lya\ absorbers or post-reionization ionized gas with sufficient optical depth to produce dark pixels at the signal-to-noise limit of the spectrum, along a particular line of sight; e.g. \citealt{2015MNRAS.447..499M}). In the latter case, it corresponds to the normalisation of the ACF and, in particular, its luminosity-weighted average host halo mass dependence (e.g. \citealt{2015MNRAS.453.1843S}). In both cases, here we assume a uniform prior on $\epsilon$ of the form,
\begin{equation}
    \mathcal{P}(\epsilon)  =
    \begin{cases}
    \frac{1}{\epsilon_\mathrm{max} - \epsilon_\mathrm{min}} , & \text{for $0 \le \epsilon_\mathrm{min} \le \epsilon \le \epsilon_\mathrm{max}$}.\\
    0, & \text{otherwise}.
    \end{cases}
\label{Eq:Palpha}
\end{equation}
Furthermore, we assume zero-mean statistically homogeneous Gaussian distributed noise\footnote{
Thus, in the absence of the nuisance parameter, $\epsilon$, the data likelihood would be Gaussian,
\begin{equation}
    \mathcal{L}(\sTheta)  = \frac{1}{\sqrt{2\pi} \sigma} \exp\left[ -\frac{1}{2}\frac{(d - m(\sTheta))^{2}}{\sigma^{2}} \right] \ . \nonumber
\label{Eq:StandardGaussianUL}
\end{equation}
} on the data, with standard deviation $\sigma$.

Defining a stochastic model for $q + \epsilon$, $m^{\prime}(\sTheta, \epsilon) = m(\sTheta) + \epsilon$,we can write the joint conditional probability of the data given $\sTheta$ and $\epsilon$ as,
\begin{equation}
    \mathcal{P}(d \vert \sTheta, \epsilon)  = \frac{1}{\sqrt{2\pi} \sigma} \exp\left[ -\frac{1}{2} \frac{(d - m^{\prime}(\sTheta, \epsilon))^{2}}{\sigma^{2}} \right] \ .
\label{Eq:JointLikelihood}
\end{equation}
Making use of Bayes' theorem and the independence of $\sTheta$ and $\epsilon$, we can write the posterior density of $\sTheta$ and $\epsilon$ given the data as,
\begin{align}
    \mathcal{P}(\sTheta, \epsilon \vert d) &= \frac{\mathcal{P}(d \vert \sTheta, \epsilon) \mathcal{P}(\sTheta) \mathcal{P}(\epsilon)}{ \mathcal{P}(d \vert m^{\prime})} \ .
\label{Eq:JointPosterior}
\end{align}
Marginalising out the nuisance parameter $\epsilon$, the marginal posterior density of $\sTheta$ is thus,
\begin{align}
    \mathcal{P}(\sTheta \vert d) &= \frac{\mathcal{P}(\sTheta)}{\mathcal{P}(d \vert m^{\prime}) \sqrt{2\pi} \sigma} \int\limits_{\epsilon_\mathrm{min}}^{\epsilon_\mathrm{max}} \exp\left[ -\frac{1}{2} \frac{(d - m^{\prime}(\sTheta, \epsilon))^{2}}{\sigma^{2}} \right] \mathcal{P}(\epsilon) \mathrm{d}\epsilon \\ \nonumber
    &= [\mathcal{P}(d \vert m^{\prime}) 2 (\epsilon_\mathrm{max} - \epsilon_\mathrm{min})]^{-1} \mathcal{P}(\sTheta) \\ \nonumber
    &~\times \left[ \mathrm{erf}\left(\frac{(\epsilon_\mathrm{max} + m(\sTheta) - d)}{\sqrt{2}\sigma}\right) - \mathrm{erf}\left(\frac{(\epsilon_\mathrm{min} + m(\sTheta) - d)}{\sqrt{2}\sigma}\right) \right] \\
\label{Eq:MarginalPosterior}
\end{align}
Applying Bayes' theorem again and making use of the fact that $\mathcal{P}(d \vert m^{\prime}) = \mathcal{P}(d \vert m)$, the marginal likelihood is thus given by,
\begin{multline}
    \mathcal{L}_\mathrm{CUL}(\sTheta) = [2 (\epsilon_\mathrm{max} - \epsilon_\mathrm{min})]^{-1} \\
    \times \left[ \mathrm{erf}\left(\frac{(\epsilon_\mathrm{max} + m(\sTheta) - d)}{\sqrt{2}\sigma}\right) - \mathrm{erf}\left(\frac{(\epsilon_\mathrm{min} + m(\sTheta) - d)}{\sqrt{2}\sigma}\right) \right] \ .
\label{Eq:MarginalLikelihood}
\end{multline}
When using \Cref{Eq:MarginalLikelihood} in the context of modelling constraints derived from the measurement of the \Lya\ and \Lyb\ forest dark pixel fraction, $\epsilon_\mathrm{max} = 1$ corresponds to a scenario in which one is observing a dark spectrum derived entirely from sources other than a neutral IGM during reionization and in the context of and measuring the ACF of high-redshift LAEs to a scenario in which the luminosity-weighted average host halo mass is such that the measured ACF would be consistent with $\overline{x}_\mathrm{H_{I}} = 0$ (see \citealt{2015MNRAS.453.1843S}). Similarly, in the opposite extreme, $\epsilon_\mathrm{min} = 0$ corresponds in the former case to a situation where a measured dark spectrum is due purely to propagation through neutral IGM and in the latter case to a situation in which the luminosity-weighted average host halo mass is sufficiently low for any non-zero ACF to be attributable to non-zero $\overline{x}_\mathrm{H_{I}}$. In this case \Cref{Eq:MarginalLikelihood} simplifies to,
\begin{multline}
    \mathcal{L}_\mathrm{CUL}(\sTheta) = \frac{1}{2} \left[ \mathrm{erf}\left(\frac{(1 + m(\sTheta) - d)}{\sqrt{2}\sigma}\right) - \mathrm{erf}\left(\frac{(m(\sTheta) - d)}{\sqrt{2}\sigma}\right) \right] \ .
\label{Eq:MarginalLikelihood2}
\end{multline}

%%%%%%%%%%%%%%%%%%%%%%%%%%%%%%%%%%%%%%%%%%%%%%%%%%
\section{Marginalisation over redshift uncertainty in \Lya\ EW analyses}
\label{Sec:LyaRedshiftMarginalisation}
%%%%%%%%%%%%%%%%%%%%%%%%%%%%%%%%%%%%%%%%%%%%%%%%%%

As described in \Cref{Sec:SplinePDFlike}, where the redshift uncertainty associated with the LAE-derived $\overline{x}_\mathrm{H_{I}}^\mathrm{d}(z)$ PDF is reported, we add the redshift of the constraint as an additional free parameter of our model and computationally marginalise out the redshift uncertainty. We treat the redshift estimates in these cases as being Gaussian distributed (see e.g. \citealt{2022MNRAS.517.3263B}). Writing the expectation value of the redshift and the redshift uncertainty on the $j$th object as $\hat{z}_{j}$, and $\sigma_{z_{j}}$, respectively, we write the marginal probability of measuring the $j$th PDF in this set as:
\begin{multline}
    \pi_{\mathrm{LAE}_j}(\overline{x}_\mathrm{H_{I}}(\sTheta)) = \\
    \frac{1}{\sqrt{2\pi} \sigma_{z_{j}}} \int\limits_{z} \pi_{\mathrm{LAE}_j}(\overline{x}_\mathrm{H_{I}}(z, \sTheta)) \exp\left[-\frac{1}{2}\left(\frac{\hat{z}_{j} - z}{\sigma_{z_{j}}}\right)^2 \right] \mathrm{d}z\ .
\end{multline}

\begin{figure}
    \centerline{
        \includegraphics[width=0.5\textwidth]{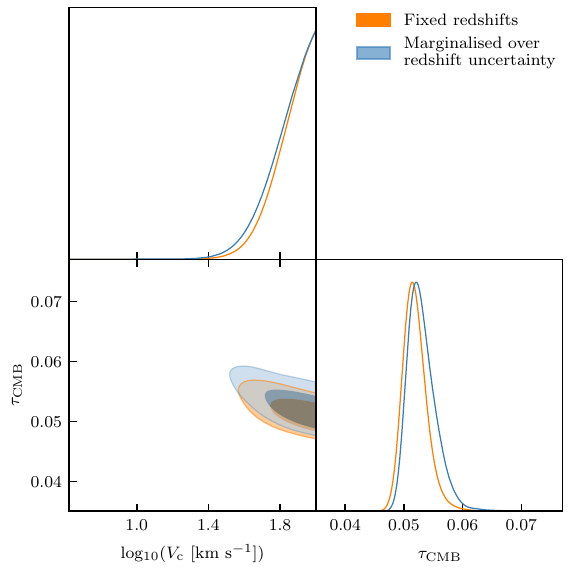}
    }
    \caption{
        One- and two-dimensional posterior probability distributions of the circular velocity, $\log_{10}(V_\mathrm{c})$, and the CMB optical depth, $\tau_\mathrm{CMB}$. Posteriors in orange neglect the uncertainty in the redshift estimates of the $\overline{x}_\mathrm{H_{I}}$ constraints derived from Lyman continuum data (see \Cref{Fig:xHIContour}, bottom and \Cref{Tab:DataSets}). Those in blue marginalise over the redshift uncertainties. The solid and transparent shaded contours plotted in the 2D posteriors contain 68\% and 95\% of the probability, respectively.
        }
    \label{Fig:RedshiftMarginalisation}
\end{figure}

In \Cref{Sec:Results}, we compare the recovered marginal posteriors, for those parameters constrained by the LAE observations, when marginalising over the redshift uncertainty as described above. We find that inclusion of the redshift uncertainty increases the uncertainty recovered parameter constraints by $\sim20\%$. The 1D and 2D marginal posteriors on the parameters in the two cases are shown in \Cref{Fig:RedshiftMarginalisation}.

% \section{Some extra material}

% If you want to present additional material which would interrupt the flow of the main paper,
% it can be placed in an Appendix which appears after the list of references.

%%%%%%%%%%%%%%%%%%%%%%%%%%%%%%%%%%%%%%%%%%%%%%%%%%

% Don't change these lines
\bsp	% typesetting comment
\label{lastpage}
\end{document}